\documentclass[nolinenumbers]{aastex631}
\usepackage{epsfig}
\usepackage{rotating}
\usepackage{mathtools}
\usepackage{comment}
\usepackage{bm}
\usepackage{epstopdf}
\usepackage{rotating}
\usepackage{array}
\newcolumntype{H}{>{\setbox0=\hbox\bgroup}c<{\egroup}@{}}
\usepackage[caption=false]{subfig}
\newcommand{\kms}{\mbox{km~s$^{-1}$}}
\newcommand{\s}{\mbox{$''$}}

\newcommand{\mdot}{\mbox{$\dot{M}$}}
\newcommand{\my}{\mbox{$M_{\odot}$~yr$^{-1}$}}
\newcommand{\ls}{\mbox{$L_{\odot}$}}

\newcommand{\ms}{\mbox{$M_{\odot}$}}

\newcommand{\vexp}{\mbox{$V_{\rm exp}$}}
\newcommand{\vsys}{\mbox{$V_{\rm sys}$}}
\newcommand{\vlsr}{\mbox{$V_{\rm LSR}$}}

\newcommand{\tkin}{\mbox{$T_{\rm kin}$}}

\newcommand{\codos}{$^{12}$CO}
\newcommand{\cotres}{$^{13}$CO}
\newcommand{\dostotres}{$^{12}C/^{13}C$}
\newcommand{\jcinco}{J=5--4}
\newcommand{\jtres}{J=3--2}
\newcommand{\jdos}{J=2--1}
\newcommand{\junos}{J=1--0}

\newcommand{\hctresnhi}{HC${_3}$N\,J=38--37}

\newcommand{\cssiete}{CS\,J=7-6}
\newcommand{\cstres}{$^{13}$CS}
\newcommand{\siodonau}{$^{29}$SiO\,J=8-7\,(v=0)}

\newcommand{\sicdo}{SiC$_2\,15(2,14)-14(2,13)$}

\newcommand{\codosabund}{[$^{12}$CO]/[H$_2$]}

\newcommand{\vdensunit}{cm$^{-3}$}

\DeclareCaptionFormat{cont}{#1 (cont.)#2#3\par}

\shorttitle{ALMA finds a Multi-Ring Circus in V Hya}
\shortauthors{Sahai et al.}

\begin{document}
\title{The Rapidly Evolving AGB Star, V\,Hya: ALMA finds a Multi-Ring Circus with High-Velocity Outflows}
\author{R. Sahai}
\affiliation{Jet Propulsion Laboratory, MS\,183-900, California Institute of Technology, Pasadena, CA 91109, USA}

\author{P-S. Huang}
\affiliation{Academia Sinica Institute of Astronomy and Astrophysics, No. 1, Sec. 4, Roosevelt Road, Taipei 10617, Taiwan}

\author{S. Scibelli}
\affiliation{Steward Observatory, University of Arizona, Tucson, AZ 85721, USA}

\author{M.\,R. Morris}
\affiliation{Department of Physics and Astronomy, UCLA, Los Angeles, CA 90095-1547, USA}

\author{K. Hinkle}
\affiliation{NOAO, Tucson, AZ 85721}

\author{C-F. Lee}
\affiliation{Academia Sinica Institute of Astronomy and Astrophysics, No. 1, Sec. 4, Roosevelt Road, Taipei 10617, Taiwan}

\email{raghvendra.sahai@jpl.nasa.gov}
\begin{abstract}
We have observed the mass-losing carbon star V\,Hya that is apparently transitioning from an AGB star to a bipolar planetary nebula, at an unprecedented angular resolution of $\sim0\farcs4-0\farcs6$ with the Atacama Large Millimeter/submillimeter Wave Array (ALMA). Our \cotres~and \codos~(\jtres~and \jdos) images have led to the discovery of a remarkable set of six expanding rings within a flared, warped Disk structure Undergoing Dynamical Expansion (DUDE) that lies in the system's equatorial plane. We also find, for the first time, several bipolar, high-velocity outflows, some of which have parabolic morphologies, implying wide opening angles, while one (found previously) is clumpy and highly collimated. The latter is likely associated with the high-velocity bullet-like ejections of ionized gas from V\,Hya; a possible molecular counterpart to the oldest of the 4 bullets can be seen in the \codos~images. We find a bright, unresolved central source of continuum emission (FWHM size $\lesssim$165\,au); about 40\% of this emission can be produced in a standard radio photosphere, while the remaining 60\% is likely due to thermal emission from very large (mm-sized) grains, having mass $\gtrsim10^{-5}$\,\ms. We have used a radiative transfer model to fit the salient characteristics of the DUDE's \cotres~and \codos~emission out to a radius of $8{''}$ (3200\,au) with a flared disk of mass $1.7\times10^{-3}$\,\ms, whose expansion velocity increases increases very rapidly with radius inside a central region of size $\sim200$\,au, and then more slowly outside it, from 9.5 to 11.5\,\kms. The DUDE's underlying density decreases radially, interspersed with local increases that represent the observationally well-characterised innermost three rings.

\end{abstract}


\section{Introduction}\label{intro}
As asymptotic giant branch (AGB) stars age over 10$^4$-10$^5$ years, they eject over half or more of their mass in slow winds, and then, in a short 100--1000 year period, are transformed into planetary nebulae (PN) with a dazzling variety 
of morphologies and widespread presence of point-symmetry \citep{2011AJ....141..134S}. 
Recent morphological studies with the Hubble Space Telescope (HST) support the idea that directed, high-speed outflows initiated during the very late AGB phase play a crucial role in the transformation to PNe (\citealt{1998AJ....116.1357S}, \citealt{2002ARA&A..40..439B}, \citealt{2007ApJ...658..410S}). However, evidence for such jets and disks in AGB stars is scant; this phase is presumably so short that few nearby stars are likely to be 
caught in the act. The carbon star, V\,Hya (distance$\sim400$\,pc; GAIA (eDR3) parallax = $2.31\pm0.108$\,mas\,yr$^{-1}$, Lindegren et al. 2018, 2021), is the nearest and best example of this phase, and 
a key object for understanding the formation of bipolar and multipolar PN.

V\,Hya has been observed to have both high-speed, collimated outflows and a ``disk'' (Sahai et al. 2003, Hirano et al. 2004 [HSD04]).  The
spatially integrated mass loss rate observed in rotational CO lines, $>$10$^{-5}$ M$\odot$ yr$^{-1}$ (Knapp et al. 1997), is
at the high end of the range of mass-loss rates found for AGB stars. Observations of CO 4.6\micron~vibration-rotation lines in V\,Hya, using the Kitt Peak National Observatory's 4-m FTS \citep{1988A&A...201L...9S}, followed by a detailed study covering 6 epochs \citep{2009ApJ...699.1015S}, revealed the presence of several high-velocity absorption components with $V_{exp} \approx 70-120$ \kms, along with a slow outflow moving at about 10 \kms. The fast outflows in V\,Hya are very similar to those observed in pre-planetary nebulae, both in terms of the high expansion speeds and their collimation (e.g., Bujarrabal et al. 2001, S{\'a}nchez Contreras \& Sahai 2012, and references therein).

V\,Hya is believed to have an unseen secondary companion surrounded by
an accretion disk.  Using archival photometry Knapp et al. (1999) find that the AGB star
undergoes partial eclipses by the disk every 16.9 $\pm$1.1 yr. 
Sahai, Scibelli \& Morris (2016: SSM16) suggest an orbital period half as long. A  
jet outflow is present consisting of high-speed ($\sim180-240$\,\kms) bullet-like ejections detected in [SII] emission from HST/STIS data spanning 6 epochs (2001--2004 \& 2011--2013) (SSM16; Scibelli, Sahai, \& Morris 2019 [SSM19]). This extended 25\,yr history of the latter has revealed how these bullets emerge and move away from the central star.  Based on these data 
SSM16 have proposed a model in which bullets (with masses $\sim10^{27}$\,g) are ejected
at speeds of $\sim180-220$\,\kms~once
every $\sim$8.5\,yr from V\,Hya, during the periastron passage of a main-sequence binary
companion in an eccentric orbit around V\,Hya. SSM16 find that 
the bullet-ejection axis flip-flops around a roughly eastern direction, both in and perpendicular to the sky-plane, likely due to a wobbling/precessing disk around the companion.  V\,Hya also shows the presence of FUV emission that could arise in a hot, active accretion disk around the companion \citep{2008ApJ...689.1274S}.

The model invoking the role of a binary companion in the ejection of the bullets was embellished by  Salas et al.\ (2019), who argued that the presence of a distant third body in the system would be needed to account for the inferred eccentricity of the inner companion's orbit, given the tendency of tidal dissipation to circularize the inner companion's orbit.

Evidence that the jet engine has been operating in V\,Hya over a much longer time (100--250\,yr) comes from CO mm-line observations (Kahane et al.\ 1996, Knapp et al.\ 1997, HSD04). HSD04 find the presence of an extended high-velocity outflow that is bipolar, blobby and highly collimated from their Submillimeter Array (SMA) observations of CO J=2--1 and 3--2 emission with $\sim3{''}-4{''}$ resolution. This outflow has clumps  that appear to be offset laterally from an overall E--W axis, qualitatively consistent with long-term persistence  of the precessional-resonance producing the flip-flop.
Deconvolved, ground-based mid-IR images at 9.8 and 11.65\,\micron~show a hot-spot offset to the East from the central star by 0\farcs9, attributable to the knotty jet (Lagadec et al. 2005).

However, despite the decades of work done to investigate the circumstellar structures of V\,Hya, there was still much that remained unknown about the precise configuration of the molecular material, which harbors the bulk of the ejected mass. We report here the results of a new study of V\,Hya using ALMA to carry out the very high angular resolution observations at (sub)millimeter-wavelengths. The plan of this paper is as follows. In \S\,\ref{obs} we describe the observational setups, and data reduction and calibration procedures. In \S\,\ref{results}, we present our main observational results, which include the discovery of a dense, equatorial ringed-structure, high-velocity outflows, and an unresolved central continuum source. In \S\,\ref{anal}, we use the molecular-line data to derive the physical properties of these structures, in \S\,\ref{discuss} we discuss the origin of the ring structures and outflows, and in \S\,\ref{sumry} we conclude with a summary of our main findings.

\section{Summary of Observations}\label{obs}
We observed V\,Hya with ALMA in order to map the detailed structure of the extended circumstellar envelope, the high-speed outflow and the central disk with high angular resolution.  The phase center for our observations was located at RA, Dec (J2000) = 10\,51\,37.241, -21\,15\,00.280. We obtained data in Band 7 using the ACA (7m-array) and the 12m-array, and in Band 6 using the 12m-array (2018.1.01113.S). The 12m data were taken in configuration C43-3, providing a beamsize of $0\farcs44\times0\farcs36$ ($0\farcs62\times0\farcs52$) and a Maximum Recoverable Scale (MRS) of $\sim4\farcs7$ ($\sim7\farcs0$), in Band 7 (6), at a frequency of 345.8\,GHz (230.54\,GHz). We have used the pipeline\footnote{CASA 5.6.0-60}-calibrated datacubes in our study. In order to estimate flux losses in the 12m data, we have compared the intensities in the 12m-array \cotres\jtres~image convolved to the angular resolution of the ACA \cotres\jtres~image -- we find that there is a modest flux loss in the 12m-array \cotres\jtres~image of about $20-25$\,\%. Since we do not have ACA data for Band 6, we have used only the 12m-array data for our analysis. The molecular-line and continuum observations discussed in this paper are listed in Tables\,\ref{obslog} and \ref{vhya-mmsed-tbl}, respectively.

\section{Observational Results} \label{results}
We present our main observational results here, which include the discovery of multiple expanding rings (\S\,\ref{diskring}) within a flared, warped, expanding disk structure (\S\,\ref{warped}), several bipolar, high-velocity outflows (\S\,\ref{hivelflows}) and a bright, unresolved central source of continuum emission (\S\,\ref{cont-src}).

\subsection{A Central Disk with Rings}\label{diskring}
\begin{figure}[ht!]
\vspace{-0.1cm}
\includegraphics[width=0.85\textwidth]{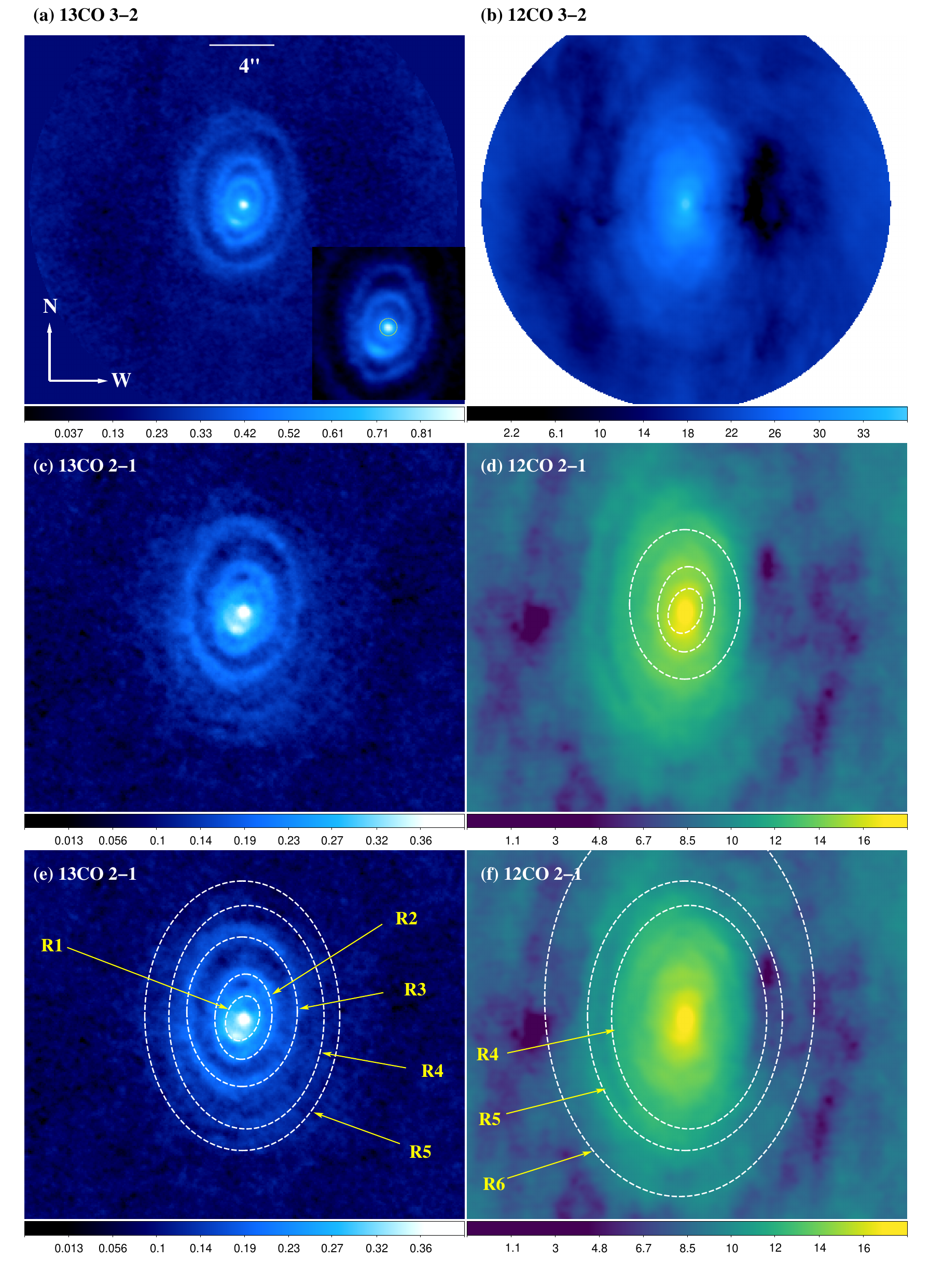}
\caption{Moment 0 maps of \codos~and \cotres~emission lines towards V\,Hya, covering the velocity range  \vlsr=-12 to -27 \,\kms: (a) \cotres~J=3--2; inset shows a magnified view of the center, with a 0\farcs8 diameter (dashed-red) circle around the 
central emission peak, at (RA, Dec) = 10:51:37.24 -21:15:00.3, (b) \codos~J=3--2, (c) \cotres~J=2--1, and (d) \codos~J=2--1. Panel e (f) shows the same image as panel c (d), but with ellipses (dashed white curves) that represent the different rings (R1--R6), overlaid. Beamsizes are as follows: \cotres~J=3--2: $0\farcs463\times0\farcs375$ ($PA=74.24\arcdeg$),  \cotres~J=2--1: $0\farcs643\times0\farcs541$ ($PA=69.25\arcdeg$), \codos~J=3--2: $0\farcs443\times0\farcs361$ ($PA=74.47\arcdeg$), and \codos~J=2--1: $0\farcs620\times0\farcs521$ ($PA=70.6\arcdeg$).  All panels have the same field-of-view with North up and East to the left. The intensity scale at the bottom of each panel is in Jy\,\kms\,beam$^{-1}$.
}
\label{all-mom0}
\vskip -0.1in
\end{figure}

The \cotres~and \codos, \jtres~and \jdos~moment-zero maps of V\,Hya (Fig.\,\ref{all-mom0}) show multiple elliptical ring-like structures. The central star is located within the bright emission peak at the center of these images. The innermost three of these (R1, R2, R3 in order of increasing size), are seen completely in the \cotres~\jtres~and \jdos~images. These rings are partly visible in the \codos~images, but are not as well-defined as in the \cotres~images. The simplest explanation for this is that the \codos\,\jtres~emission is optically thick (therefore sensitive to temperature), whereas \cotres\,\jtres~is optically thin (therefore sensitive to density), and that the temperature contrast between ring and inter-ring regions is not as high as the density contrast.

A small arc structure, A3, is seen to the north of R3 that appear connected to the latter. A large arc-like structure, representing an incompletely manifested fourth ring (R4), is also seen towards the south. The R4 ring, and a larger one, R5, can be seen in the \codos\,\jdos~image, mostly on the eastern side of the central star. In addition, the \codos\,\jdos~image also shows a diffuse arc structure mostly to the south of the central star, that may be a segment of a sixth ring (R6). 
 
The major axes of the three rings seen completely, R1, R2 and R3,  rotate steadily eastward, with $PA\sim-10$ to $-20\arcdeg$ for R1,  $PA\sim-5\arcdeg$ for R2, and $PA\sim0$ to $-5\arcdeg$ for R3. 
By fitting ellipses to the rings in the \cotres\,\jtres~image (by eye; shown as dotted curves in Fig.\,\ref{all-mom0}), we find that the semi-major axis lengths of rings R1, R2, and R3 are $1\farcs4$, $2\farcs6$, and $4\farcs35$, respectively (with uncertainties of $\lesssim5$\%) and the corresponding major to minor axis ratios are 1.4, 1.5, and 1.3, respectively (with uncertainties of $\lesssim10$\%). The major-to-minor axis ratio of R3 is significantly smaller than that of R1 and R2. The centers of the rings are located at the same value of the RA (10:51:37.24), but differ in declination (-21:15:00.27, -21:15:00.17, and -21:14:59.87 for R1, R2, R3). The center of R3 is significantly north of R1 (by 0\farcs4) and R2 (by 0\farcs3).
Rings R4, R5 and R6, are best seen in the \codos\,\jdos~image. We have therefore made elliptical fits (by eye; shown as dotted curves in Fig.\,\ref{all-mom0}) in this image to these rings, but these fits are significantly more uncertain than for rings R1--R3. We find semi-major axis lengths for these to be 6\farcs8, 8\farcs2, and 12\farcs0, respectively (with uncertainties of about $10$\%, $10$\%, and $15$\%) and  major to minor axis ratios of 1.4, 1.4, and 1.4 (with uncertainties of about $15$\%, $15$\%, and $20$\%).

\begin{figure}[ht!]
\includegraphics[width=1.0\textwidth]{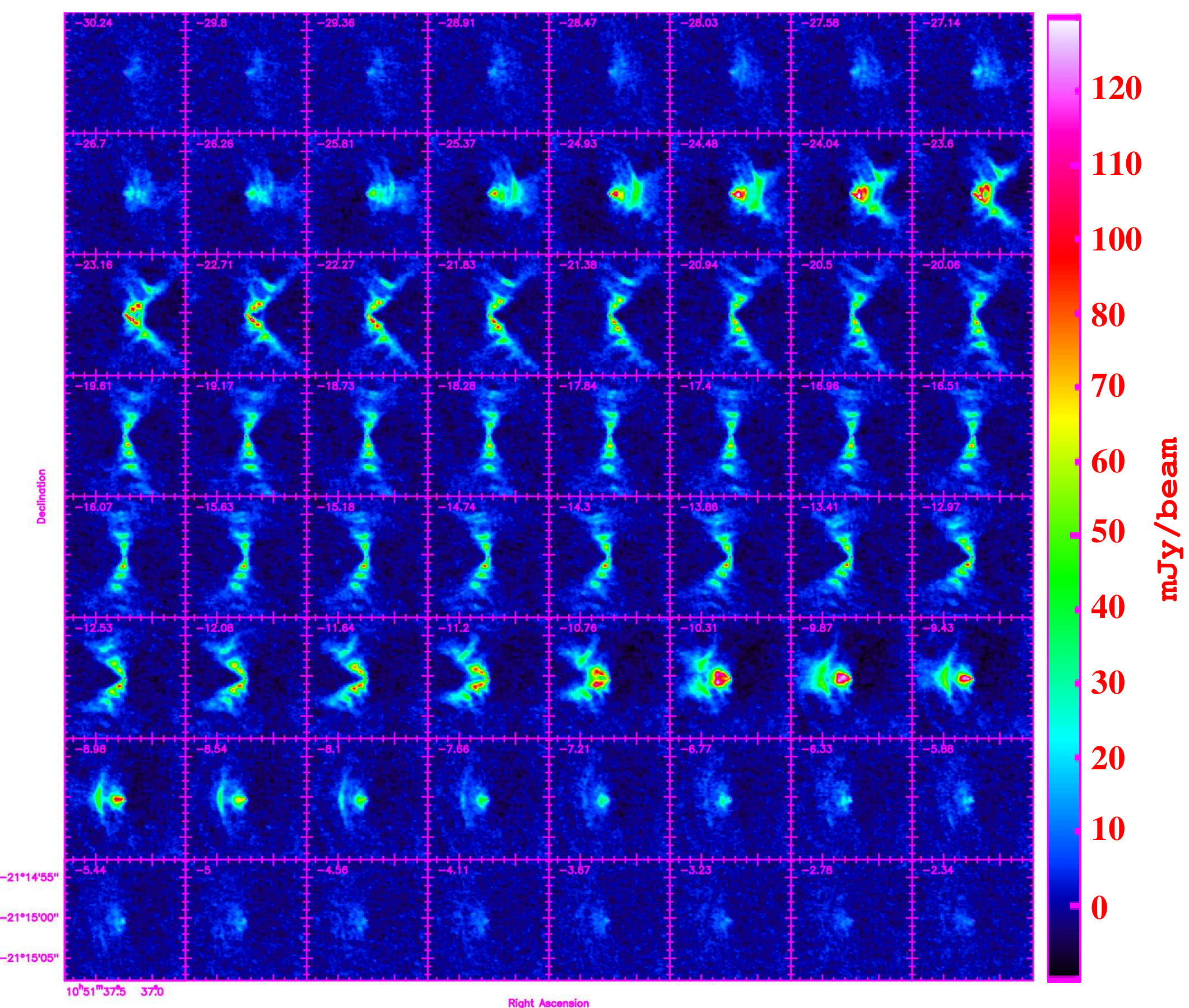}
\vspace{0.5cm}
\caption{Channel/velocity maps of the \cotres~J=3--2 emission from V\,Hya (beam as in Fig. \ref{all-mom0}). Only the central $15{''}\times15{''}$ field-of-view is displayed in order to show the detailed structure of the DUDE. Each channel is 0.44\,\kms~wide.
}
\label{13co32chan}
\end{figure}

The \cotres\jtres~channel maps (Fig.\,\ref{13co32chan}) rule out the possibility that each of the rings is a projection of an expanding ellipsoidal shell structure, since such a geometry would produce a ring-like brightness distribution whose size would be largest at the systemic velocity, and would get progressively smaller towards higher redshifted and blueshifted velocities (e.g., see ellipsoidal shell model for TX\,Psc in Brunner et al. 2019).

These rings are presumably annuli of enhanced density in the equatorial plane of a large (size $\sim25{''}$), expanding central disk-like structure in \codos\,\jdos~emission found by  HSD04. We will refer hereafter to this structure as a {\it Disk Undergoing Dynamical Expansion}, or DUDE.
If we make the reasonable assumption that rings R1--R6 are intrinsically circular, the major-to-minor axis ratios imply that the inclination angles of the ring-planes to the line-of-sight are $46\arcdeg$, $42\arcdeg$, and $50\arcdeg$ for R1, R2, and R3 (with uncertainties of about $\pm3\arcdeg$, respectively), and about $\sim46\arcdeg$ for rings R4--R6 (with uncertainties of about $\pm5\arcdeg$, respectively). These uncertainties in the inclination angles are simply due to the measurement uncertainties in the lengths of the major and minor axes. The flared shape of the DUDE affects these estimates of the inclination angles more significantly (\S\,\ref{dude-model}).
  

HSD04 derive an inclination of $60\arcdeg$ of the DUDE to the line of sight from the ratio of the semi-minor to semi-major axis ($f_{min/maj}$) of the ``low-velocity" ($\pm8$\,\kms) emission in their \codos\,\jdos~($f_{min/maj}=13/15$) and \codos\,\jtres~($f_{min/maj}=7/8$) maps obtained with the SMA. In contrast, after convolving the ALMA maps to the same beam parameters (size, PA) as for the SMA data ($4\farcs4\times3\farcs1$, PA=-28\arcdeg~for \codos\,\jdos; $2\farcs1\times1\farcs9$, $PA=-28\arcdeg$~for \codos\,\jtres), we infer significantly smaller values for $f_{min/maj}$ -- $\sim0.6$ for \codos\,\jtres~and \codos\,\jdos, measured by fitting ellipses to the half-peak intensity contours. It is possible that this discrepancy could be due to differences in the MRS, which is smaller for the ALMA data. However, this is not likely, given that our \codos\,\jtres~ACA map, which has a lower resolution ($4\farcs5\times2\farcs7$, PA=88.5\arcdeg) than the SMA data, shows $f_{min/maj}=0.68$, which is also significantly smaller than the SMA value. Our best guess for this discrepancy is that the SMA data have much more limited UV-coverage and lower signal-to-noise (S/N) compared to the ALMA data.

\begin{figure}[ht!]
\vspace{-0.1cm}
\hspace{2cm}
\includegraphics[width=0.6\textwidth]{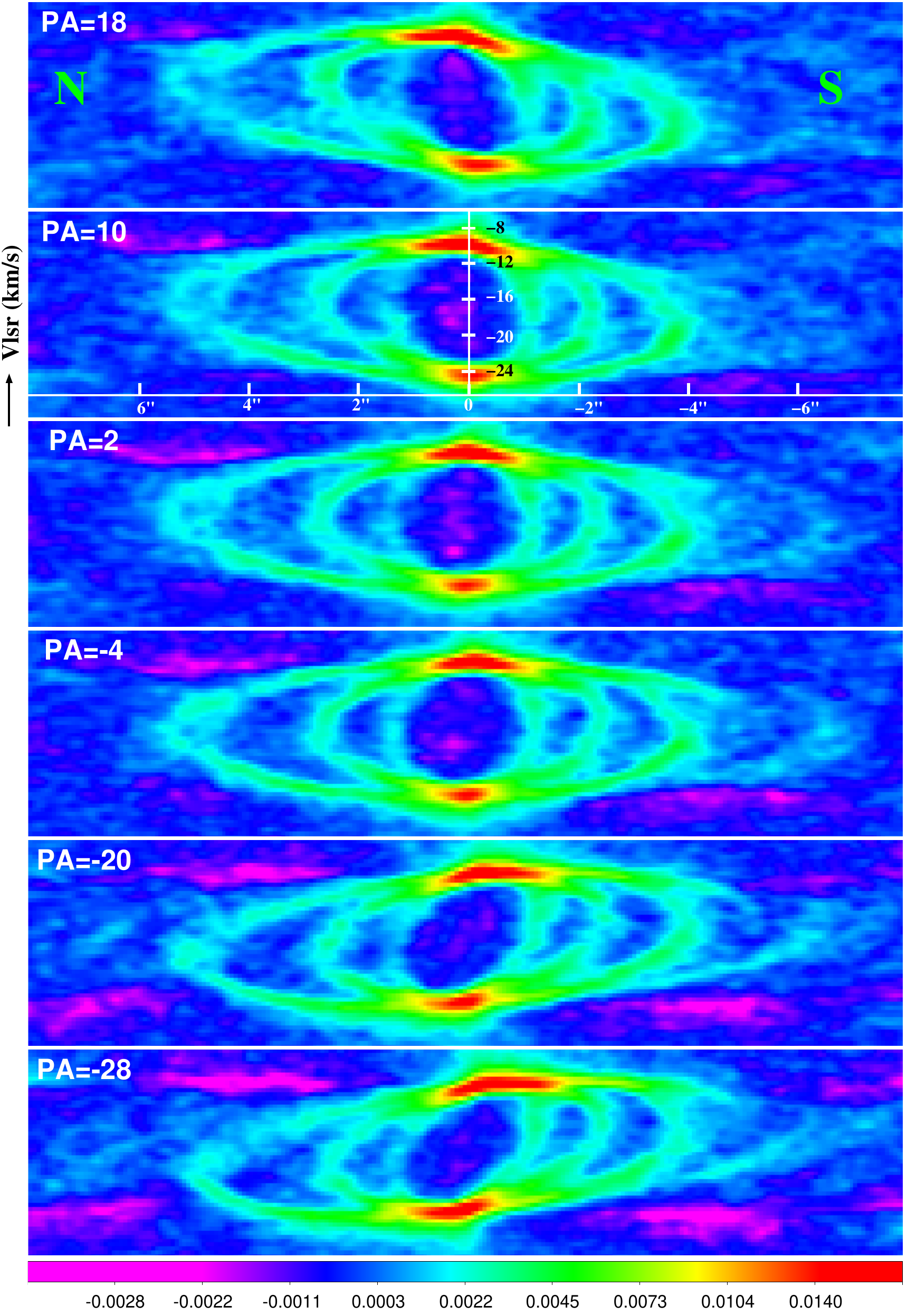}
\vspace{1cm}
\caption{Position ($x$-axis)--velocity ($y$-axis) intensity cuts of the \cotres\,\jtres~emission near and along the major axis of the DUDE. The width of the cut is $16{''}$. The intensity scale at the bottom (Jy\,beam$^{-1}$) applies to all panels. The beamsize is $0\farcs463\times0\farcs375$ ($PA=74.24\arcdeg$).
}
\label{13co32-pv-ns}
\end{figure}

\begin{figure}[ht!]
\vspace{-0.1cm}
\includegraphics[width=0.55\textwidth]{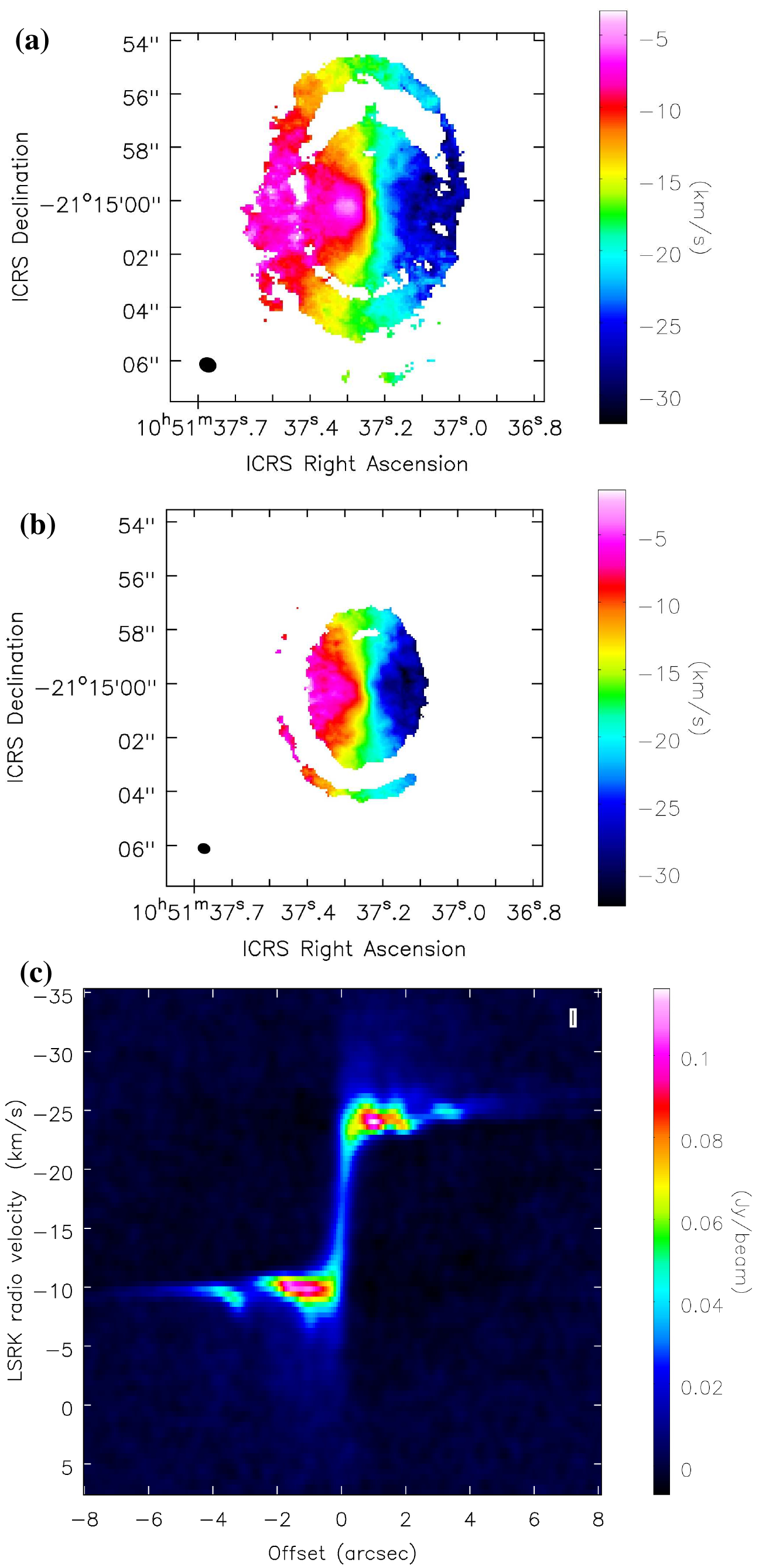}
\caption{(a,b) Moment 1 maps of the \cotres\,\jdos~and \jtres~emission from V\,Hya, (c) Position-velocity intensity cut of the \cotres\,\jtres~emission along the minor axis of the DUDE, i.e., at $PA=86\arcdeg$ (the width of the cut is $1{''}$).
}
\vspace{1cm}
\label{13co32-mom1-pvew}
\end{figure}

Position-velocity intensity cuts of the \cotres\,\jtres~emission along the major-axis (Fig.\,\ref{13co32-pv-ns}), and the moment 1 maps of \cotres~\jdos~and \jtres~(Fig.\,\ref{13co32-mom1-pvew}a,b), reveal the characteristic signature of expansion for a disk-like structure that is tilted roughly along a N--S axis (e.g., Chiu et al. 2006), and consistent with the orientation as inferred from the SMA study by HSD04. The expansion velocities of the rings appear to be the same (or very similar), since in the N--S cut (Fig.\,\ref{13co32-pv-ns}) the ring structures appear to overlap at their maximum velocity points at zero spatial offset, and in the E--W cut (Fig.\,\ref{13co32-mom1-pvew}c), the ring (and inter-ring) emission is contained within two horizontal stripes at maximum and minimum (projected) expansion velocities. However, a very strong velocity gradient can be seen in the emission at offsets $\lesssim0\farcs25$.

\begin{figure}[ht!]
\rotatebox{270}{\includegraphics[width=0.7\textwidth]{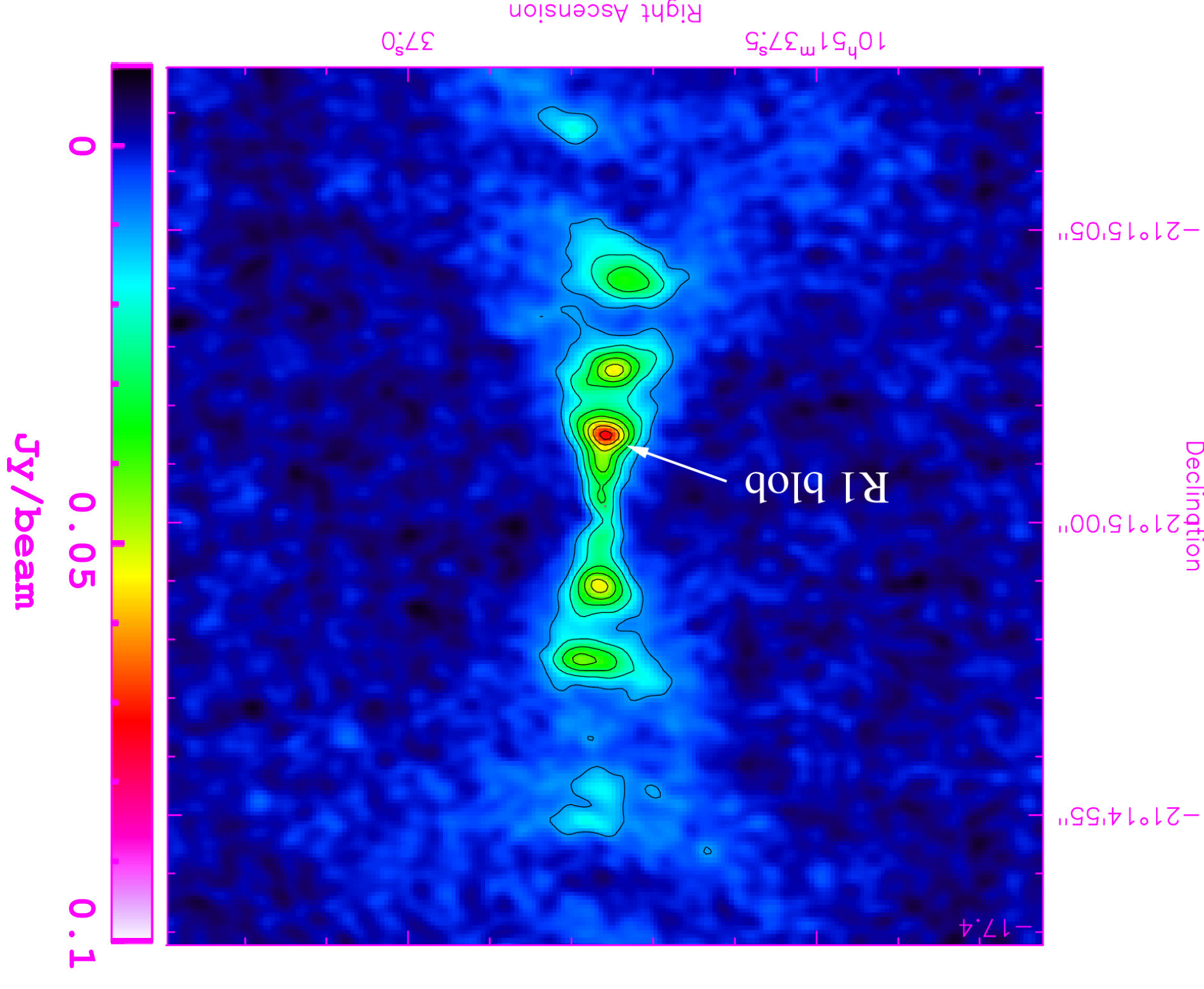}}
\caption{Map of the \cotres~J=3--2 emission from V\,Hya at the systemic velocity, \vlsr=-17.4\,\kms~(beam as in Fig. \ref{all-mom0}. The contour peak (spacing) is 100\,mJy\,beam$^{-1}$ (10\,mJy\,beam$^{-1}$). 
}
\label{13co32cench}
\end{figure}

The \cotres\jtres~emission at a velocity near/at the systemic velocity, $V_{sys}=-17.4$\,\kms~(\vlsr), is shown in Fig.\,\ref{13co32cench} -- the image shows a local minimum in the center, both in size and intensity. 
The bright blobs seen north and south of the center in Fig.\,\ref{13co32cench} (local maxima in size and intensity of emission) show a cross-sectional view of rings R1, R2 and R3. The flared shape of the emission represents a cross-sectional view of the DUDE in which the rings are located -- the DUDE's thickness increases with radius. 

The central peak in the \cotres\,\jtres~moment-zero map is located at J2000 coordinates (RA, Dec) 10:51:37.24, -21:15:00.3. This is 0\farcs28 west of the J2000 SIMBAD coordinates of V\,Hya, 10:51:37.26, -21:15:00.3, implying a proper motion of $\Delta$\,RA,\,$\Delta$\,dec=$-0.0145$,\,$0.0$\,arcsec\,yr$^{-1}$, over the 19.3\,yr period between epoch 2000.0 and the epoch when the 12m ALMA observations of \cotres\,\jtres~were made (see Table\,\ref{obslog}). This result is consistent with V\,Hya's proper motion of $\Delta$\,RA,\,$\Delta$\,dec=$-0.016$,\,$0.002$\,arcsec\,yr$^{-1}$ as measured with GAIA (eDR3 catalog: Lindegren et al. 2020), given the uncertainties in measuring the coordinates of the \cotres\,\jtres~moment-zero map peak (few$\times0\farcs01$).

\begin{figure}[ht!]
\resizebox{0.55\textwidth}{!}{\includegraphics[trim={0cm 0cm 0cm 0cm 0cm},clip]{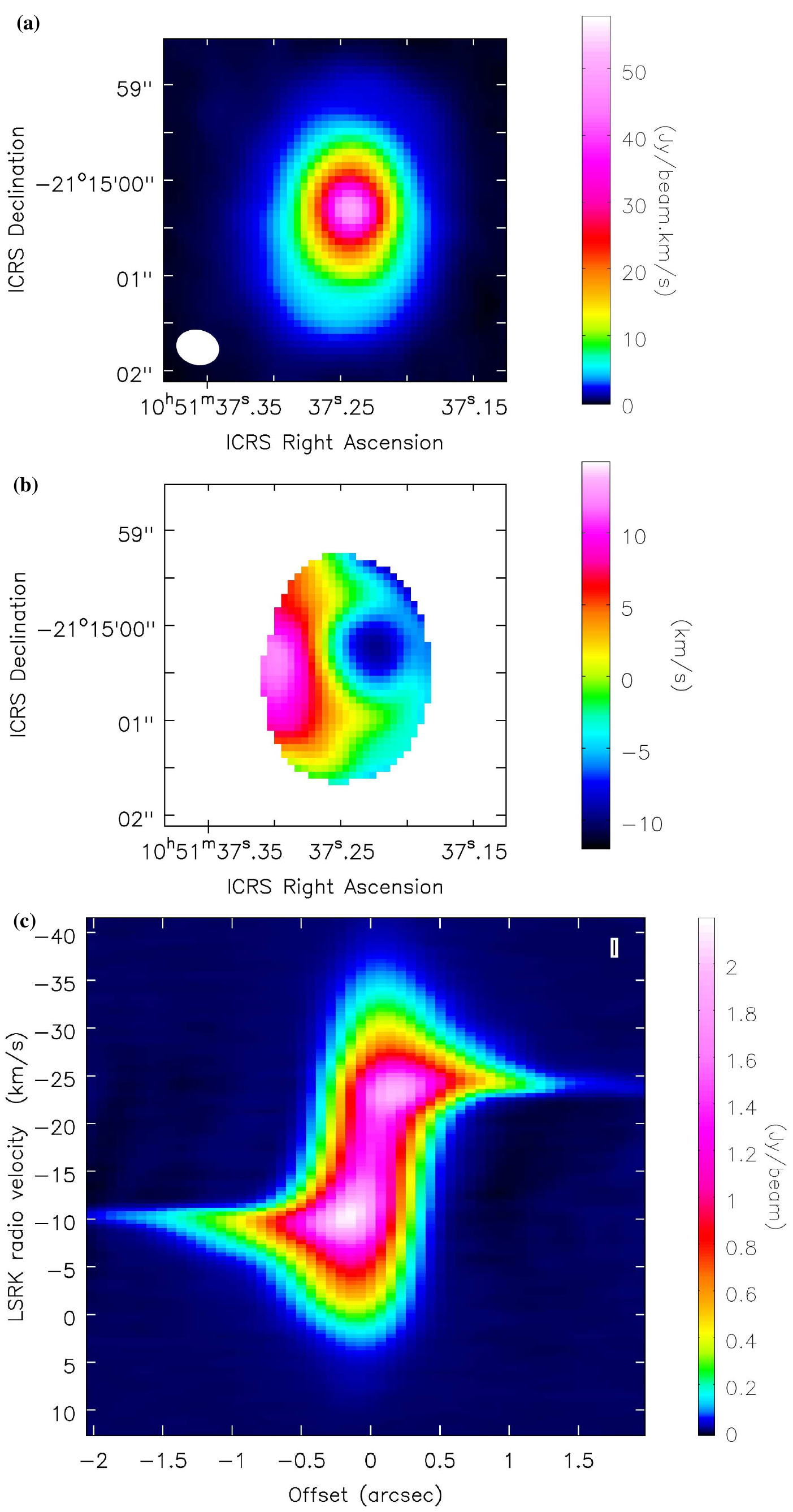}}
\caption{(a, b) Moment 0, 1 maps of the \cssiete~emission from V\,Hya, (c) Position-velocity plot of the \cssiete~emission from V\,Hya, along PA=91\arcdeg~through the stellar position (the width of the cut is 0\farcs36.)}
\label{cs76mom1pv}
\end{figure}

The third-strongest emission line in the band 7 data, after \codos~and \cotres\,\jtres, is \cssiete, and its spectrum also shows a double-peaked profile. However, in contrast to the \cotres\,\jtres~and \codos\,\jtres~emission at the systemic velocity, the \cssiete~emission peaks at the center of the DUDE. The moment 0 map of the \cssiete~emission (Fig.\,\ref{cs76mom1pv}a) is centrally peaked, and the moment 1 map  (Fig.\,\ref{cs76mom1pv}b) shows red (blue)-shifted emission to the east (west) of center, roughly similar to the central region of the moment 1 maps of \codos~and \cotres\,\jtres). However the \cssiete~moment 1 map shows some departures from the latter, with the blue (red)-shifted peak located slightly to the north (south) of center. The \cssiete~position-velocity plot along the E-W direction (Fig.\,\ref{cs76mom1pv}c) shows the characteristic S-shape of an expanding disk.

\subsection{Asymmetries and Warping in the DUDE}\label{warped}
The ALMA data provide several observational indicators that the observed DUDE in V\,Hya is asymmetric and possibly warped. 

First, as described in \S\,\ref{diskring}, the position angles and inclination angles of the best-characterized rings (R1, R2, R3) are not identical, nor are the locations of their centers.

Second, the spectrum of the \cotres\,\jtres~spatially-integrated emission shows that the redshifted emission is slightly stronger than the blueshifted one, most likely due to an intrinsic (mild) asymmetry in the DUDE (Fig.\,\ref{mod-13co32-spec}). 

Third, in the \cotres\,\jtres~and \cotres\,\jdos~maps at the systemic velocity, the symmetry center of the DUDE (assumed to be the local minimum of emission that we see in the center), is located at Dec=-21:15:00.03 (Fig.\,\ref{13co_co_cens}a,b). We identify this minimum as the center of the density distribution of the DUDE in the equatorial plane (hereafter DUDE center). The \codos\,\jtres~and \jdos~maps, like \cotres\,\jtres, show a central minimum at the systemic velocity, but the minimum is located at Dec=-21:15:00.36, i.e., roughly at the location of V\,Hya's central star, but 0\farcs33 south of the corresponding \cotres~minimum (Fig.\,\ref{13co_co_cens}c,d). 

Note also that the emission from several other molecular lines that are prominent in the V\,Hya spectrum, at the systemic velocity, is shown in Fig.\,\ref{othercens}. We find that for all of these, the symmetry centers are co-located with that of \codos, irrespective of whether they are centrally peaked (\siodonau, \cssiete) or have central minima (\sicdo, \hctresnhi) -- presumably the emission in these lines come predominantly from surface layers of the DUDE.


Fourth, the center-of-symmetry of the DUDE in \cotres\,\jtres~channel maps shifts in declination, as a function of velocity.
Closer to the line wings, the emission regions are roughly cone-shaped with their axes aligned east-west, and the apex of each cone points towards the center. 
For a non-warped symmetric DUDE, the north-south location of the apex of each cone should be the same as that of the center of the DUDE (see \S\,\ref{dude-model} of our model DUDE). However, we find that for the blue- and red-shifted channels, the cone apex is located at Dec=-21:15:00.30 and Dec=-21:15:00.41, respectively, which is south of the DUDE's center-of-symmetry at the systemic velocity (Dec=-21:15:00.03) (Fig.\,\ref{13co_symmcens_red_blu}).

Fifth, we find significant mismatches in the position angles and inclination angles of the blue- and red-shifted components of the high-velocity outflows (see \S\,\ref{hivelflows}). If these are channeled by magnetic fields anchored to the DUDE, then the symmetry axes of the field must be similarly  misaligned on either side the DUDE, which we attribute to a warp in the DUDE.

Both the \codos\,\jtres~and \cssiete~lines are significantly more optically-thick than \cotres\,\jtres~(see \S\,\ref{anal-physcond}), hence these emissions are produced in a region that is closer to the DUDE surface, compared to \cotres\,\jtres. The shift in the symmetry center of the DUDE in these lines therefore indicates that the location of the symmetry center changes as one moves from the DUDE mid-plane to the surface. It is likely that the other molecular lines for which the emission is centrally peaked are either also optically thick, or there is a chemical stratification in the DUDE with height above the equatorial plane, such that the abundances of these molecules are significantly enhanced at or near the DUDE surface.



\subsection{High-Velocity Outflows}\label{hivelflows}
The \codos~\jdos~emission shows extensive structure at velocity offsets significantly larger than those where the DUDE emission is prominent (Fig.\,\ref{12co21-hivel}b), i.e., $\gg20$\,\kms~from the systemic velocity (Fig.\,\ref{12co21-hivel}a,\ref{12co21-hivel}c), extending to a velocity offset of roughly 175\,\kms~on either side of the sytemic velocity. Similar structures are seen in \codos~\jtres~emission as well, but are fainter and less well demarcated (Fig.\,\ref{12co32-hivel})\footnote{The bright compact emission that appears around the central star in the velocity range \vlsr$\sim133-146$\,\kms, is due to the \hctresnhi~line.}. 

The high-velocity emission is complex, and its detailed structure is seen more clearly in maps of the faint \codos~\jdos~emission integrated over blue (red) --shifted velocities in the range \vlsr$\sim-61$ to -191\,\kms~(\vlsr$\sim27$ to 148\,\kms) as well as in individual velocity channels (Fig.\,\ref{12co21-arc-hivel}a--d). These maps reveal several outflows with very wide opening angles (labelled HVW-b1, HVW-b2, HVW-r1, HVW-r2 etc)\footnote{The HVW arc structures have been traced by eye, using panels a, b, c, and d; and the symmetry axis vector of the HVW outflows has been estimated by bisecting chords that join symmetric points on the N and S segments of the arcs}. In addition, a highly collimated, clumpy outflow that is oriented roughly E-W (HVC), is also visible. The highest blue (red) --shifted gas, with $\vexp\sim175$\,\kms~($\sim165$\,\kms) is found in the HVC outflow (Fig.\,\ref{12co21-arc-hivel}e,f). Although there is rough (qualitative) symmetry between the blue- and red-shifted sides of the HVW outflows, it is not exact. For example, the symmetry axes of the blue- and red-shifted outflows are not co-linear. Similarly, the HVC is much more prominent on its blue-shifted side, and the ouflow axes of its blue- and red-shifted components are not colinear. 

In position-velocity cuts of the \codos~\jdos~emission, taken along the E-W axis, the high-velocity emission appears as an S-shaped structure, extending to about \vlsr$\sim-170$\,\kms~in the east, to about \vlsr$\sim140 $\,\kms~in the west (Fig.\,\ref{12co21-pv-ew}).  Each of the arms of the S-structure splits up into two or more separate filaments, terminating in shock-like structures at their tips. East of center, we can see 4 blue-shifted shock structures, bs(1)--bs(4), at spatial (velocity) offsets from the center (systemic velocity) of $5{''}$, $8\farcs5$, $10\farcs5$, and $13\farcs5$ ($-170$, $-150$, $-160$, and $-170$\,\kms). West of center, we can see 2 red-shifted shock structures, rs(1)--rs(2), at spatial (velocity) offsets from the center (systemic velocity) of $-5\farcs5$ and $-10{''}$ ($155$ and $145$\,\kms)\footnote{since these shock-like structures are diffuse and extended, the spatial and velocity offsets above are rough estmates, and their values are given to an accuracy of 0\farcs5 and 5\,\kms.}

The bright (saturated) low-velocity central emission in this PV plot arises from the DUDE: its S-shape is oriented in an opposite sense to that of the HV outflows, which is expected since the material in the disk is expanding in directions that are roughly orthogonal to that in the HV outflows. The emission from these outflows at velocities approaching the systemic velocity appears to split up into thinner filaments with ``footprints" that are connected to small protuberances at the periphery of the S-shaped DUDE emission. Some of these filaments appear to be nearly horizontal at low offset velocities ($\lesssim62\pm5$\,\kms), indicating extreme acceleration, followed by a more gentle increase (roughly linear).

\subsubsection{Proper Motion in the High-Velocity Collimated Outflow}\label{propmotion}
The high velocities of the gas in the HVC suggest that we should be able to detect proper motion over the $\sim16.1$\,yr period between the SMA observations of HSD04 (taken during 2003 February and May) and our ALMA observations (taken on 2019 April 27) -- e.g.,  the expected total proper motion is $1\farcs4$ for gas moving with a radial expansion velocity of $\sim170$\,\kms, along an axis with an inclination assumed to be orthogonal to the ring planes, $i\sim45\arcdeg$. We use the compact blob in the HVC outflow seen at a radial velocity offset of $-154\,\kms$ from the systemic velocity and located at an angular offset of $-12\farcs6$ from the center in the SMA PV map of \codos\jdos~in Fig.\,2 of HSD04 (hereafter HVC-b(SMA)), to measure the proper motion. We find that the offset of HVC-b(SMA) in a similar PV map extracted from our ALMA \codos\jdos~map, convolved to the beam in the SMA data, is $-14\farcs1$ (Fig.\,\ref{blob-propmotion}a), implying a proper motion of $1\farcs5$/16.1\,yr -- and thus a tangential\footnote{i.e., orthogonal to the line-of-sight} velocity of 178\,\kms~and a total (3D) velocity of 235\,\kms. 

On the red side of the HVC outflow, there is no isolated blob in HSD04 available for the above analysis. However, if we assume that the local peak centered at a radial velocity offset of $124\,\kms$ from the systemic velocity and an angular offset of $9\farcs9$ from the center, corresponds to the local peak in the HVC outflow in our ALMA map at $\sim125$\,\kms~at an offset of $11\farcs5$ (Fig.\,\ref{blob-propmotion}b, HVC-r(SMA)), then the estimated proper motion is $1\farcs6$. 
Using these measured proper motions, we find that the inclinations of the outflow axes (to the line-of-sight) for the  HVC-b(SMA) and HVC-r(SMA) blobs, respectively, are $49\arcdeg$\,($\pm3.5\arcdeg$) and $57\arcdeg$\,($\pm3.5\arcdeg$) (assuming $\pm$15\% uncertainties in measuring the proper motions). The difference between these inclinations is small, but significant, and comparable to the differences in the position angles of these outflows. Assuming that their proper motions have been roughly uniform over most of their lifetimes, the ages of HVC-b(SMA) and HVC-r(SMA) are 150\,yr and 115\,yr, respectively.

\subsubsection{Ionized-Gas Bullets and the High-Velocity Collimated Outflow}\label{bullets}
The clumpy collimated structure of the HVC suggests that it is closely associated with the high-velocity ionized-gas bullets that have been observed (with HST/STIS) being ejected from V\,Hya every 8.5\,yr over a $\sim$25\,yr period (SSM16, SSM19). Since the gas in these bullets will expand and cool, and/or entrain molecular material, it is plausible that the four bullets detected in the HST/STIS study (numbered $0-3$, with bullet $0$ being the oldest, and bullet $3$ being the youngest) may be directly detectable in our ALMA \codos\jdos~images, taken $\sim6-17$\,yr after the HST observations. 

The tangential velocity is about 100\,\kms~between the last pair of observational epochs for bullets 1 and 2, and 76\,\kms~for bullet 3\footnote{No proper motion could be determined for bullet 0 as it was detected only during one epoch}. If we assume that bullets 0--2 continued to move with tangential velocities of 100\,\kms, and bullet 3 with 76\,\kms, from the last epoch in which they were detected to the epoch of our ALMA observations (2019 Apr-May), then the expected locations of bullets $0-3$ in the ALMA epoch are 1\farcs9, 1\farcs2, 0\farcs8, and 0\farcs3. If we further assume that these bullets  continued to move with the same radial velocities as measured in the last observational epoch when they were detected (\vlsr=-162, -173, -159, -181\,\kms~for bullets $0-3$\footnote{derived from the radial velocities relative to the systemic velocity, as given in Figs. 11 and 12 of SSM19}), then we might expect to see evidence of compact clumps of molecular gas at or near these velocities in the \codos\,\jdos~channel maps, assuming the bullets entrain sufficient ambient molecular gas. Examining the \codos\,\jdos~emission in the velocity range $\vlsr=-166$ to $-191$\,\kms, we find a compact clump at a radial offset of 1\farcs5 from the center, i.e. in the 1\farcs2 and 1\farcs9 region (Fig.\,\ref{blob-bullets}) -- it is plausible that this clump is bullet 0, moving with a tangential velocity of 53\,\kms. Another compact blob is seen at a radial offset of 2\farcs4. Assuming this blob represents another bullet ejected $n\times8.5$\,yr prior to bullet 0, where 8.5\,yr is the period between bullet ejections (SSM16), then assuming a tangential velocity of $\sim$50\,\kms, we get n=4. There is faint emission in the 0\farcs8--1\farcs2 region, but it is not centrally peaked, so it is difficult to associate it with bullets 1--3, that were ejected after bullet 0. The \codos\,\jtres~emission in the same velocity range shows a very similar structure as \codos\,\jdos.

We note that the tangential velocity of the HVC-b(SMA) clump, as derived from its proper motion, is about 180\,\kms, significantly higher than those derived for the ionized-gas bullets. This may be because the overall ejection velocity of bullets was higher in the past, and/or the trajectory of the clump has changed since its ejection as a bullet, bending away from the line-of-sight ($los$) and towards the sky-plane.

\subsection{Central Continuum Source}\label{cont-src}
The continuum emission at 0.8\,mm (band 7), observed with a beam-size $0\farcs454\times0\farcs365$ ($PA\sim75\arcdeg.3$), is centered at (J2000) RA,\,Dec=10:51:37.239,\,-21:15:00.32, and is unresolved (Fig.\,\ref{cont-bnd7}). The location of the continuum source is coincident with that of V\,Hya's primary star (see \S\,\ref{diskring}). The properties of the continuum source are given in Table\,\ref{vhya-mmsed-tbl}, including previous measurements (both interferometric and single-dish). The nature and origin of the continuum emission is discussed in \S\,\ref{dust-anal}.

\section{Analysis} \label{anal}
\subsection{Physical Conditions in the DUDE and its Rings}\label{anal-physcond}

\subsubsection{Ring R1}\label{anal-r1}
A comparison of the brightness temperatures (T$_{mb}$) of the \codos~and \cotres~J=2--1 and J=3--2 lines provides constraints on the physical conditions in the DUDE. For this purpose, we have convolved all images to the same angular resolution as that for \cotres\,\jdos, which has the lowest resolution, and we carry out a representative analysis of the bright blob representing R1's cross-section (hereafter ``R1 blob'': Fig.\,\ref{13co32cench}). Since we are analyzing locally compact structures, the flux losses for these are likely to be small and roughly similar for lines in the same band. We find that, for the \codos~J=2--1 and J=3--2 lines, the median T$_{mb}$ values (extracted from an aperture of size $0\farcs7$ from their respective images at the systemic velocity) are essentially equal (50.7 and 51.3\,K), consistent with these lines being optically thick and roughly in LTE, i.e., the kinetic temperature is $\gtrsim51$\,K. 

For the \cotres~J=2--1 and J=3--2 lines, the T$_{mb}$ values for the R1 blob are 1.8 and 2.9\,K -- since these values are much lower than the kinetic temperature estimated above, and different from each other, we conclude that these lines are optically thin. 
Assuming that the \cotres~and \codos~lines are in LTE, with the same excitation temperature, we use the \cotres~to \codos~\jdos~(\jtres) line ratio to infer the optical depth of the \cotres~\jdos~(\jtres) line to be 0.036 (0.056). Applying the same analysis to the average line intensity ratio (in the systemic velocity channel) extracted from a $2\arcdeg$~angular wedge centered at the location of the primary star and oriented along the average PA of the disk, we derive the radial variation of the \cotres\,\jtres~optical depth
(Fig.\,\ref{13co32tau}) -- the latter shows local maxima near, but displaced slightly beyond, the observed brightness peaks due to the rings in the \cotres\,\jtres~image. The increase in optical depth and thus the column density, between the ring peaks and the inter-ring minima between the peaks, is quite substantial, e.g., by factors of 1.4, 2.1, and 4.4 for rings 1, 2 and 3 south of center. 

We now infer the optical depths of the respective \codos~lines using the \dostotres~abundance ratio. 
Published values of the circumstellar \dostotres~ratio based on single-dish \codos~and \cotres~millimeter-wave observations for V\,Hya are 33 (Jura et al. 1988) and $\sim70$  (see Milam et al. 2009 for a compilation). Jura et al. (1988) did not correct their \dostotres~ratio for possible optical depth effects of the \codos\,\junos~line, so we have adopted the larger value (i.e., 70), which is also representative of the photospheric \dostotres~ratio in V\,Hya derived by Lambert et al. (1986). We thus obtain optical depths of the \codos~\jdos~and \jtres~lines to be 2.5 and 3.9 along a line-of-sight that goes through the R1 blob. 

Since the \codos~J=3--2 line is slightly brighter than the J=2--1 line, we require that the former be adequately excited, requiring the volume density to be $n(H_2)\gtrsim1.8\times10^4$\,\vdensunit, which is the critical excitation density for the \cotres\,\jtres~line ($7\times10^4$\,\vdensunit) multiplied by the escape probability $\beta=(1-exp(-\tau))/\tau=0.26$. 
However, it is likely that $n(H_2)$ is significantly higher than $1.8\times10^4$\,\vdensunit~since (as noted earlier) we detect high-density tracers such as the \cssiete~and \hctresnhi~lines in the DUDE as well, and these lines have critical densities for excitation of $10^{\sim(6-7)}$\,\vdensunit~(e.g., Shirley 2015). However, note that IR radiative excitation of the v=0 rotational ladder via pumping of population to excited vibrational levels (followed by cascade down to v=0) is likely to be important for these high dipole-moment molecules (as, e.g., in IRC+10216: Morris 1975). As a result, the \cssiete~and \hctresnhi~lines can be excited at lower densities than required for pure collisional excitation (e.g., Mills et al. 2013). Although similar IR excitation can also affect CO (and its isotopologues), the 4.6\,\micron~vibration-rotation lines responsible for v=0 to v=1 excitation are likely to have high optical depths in the DUDE, reducing the effectiveness of IR excitation. This is likely to be true for the DUDE, for which the effective mass-loss rate is likely $>\sim2\times10^{-6}$\,\my~(the value derived for the ``15\,\kms~wind" by Knapp et al. (1997)), since Morris (1980) showed that collisional excitation dominates IR excitation for $\mdot\gtrsim3\times10^{-8}$\,\my.

The brightness temperature T$_{mb}$=20\,K of \cssiete~in the R1 blob is lower than the estimated kinetic temperature, implying that the line is somewhat subthermally-excited, since it is most likely quite optically-thick, given that we detect the \cstres\,\jcinco~line as well. The latter is detected with a fairly low brightness temperature, T$_{mb}$=0.2\,K, hence this line must be quite optically-thin and subthermally-excited. Unfortunately, since the \cstres\,\jcinco~line was detected in the low-spectral-resolution continuum spectral window (SPW), we cannot directly compare the \cssiete~and \cstres\,\jcinco~line emission for a more quantitative analysis. 

Using the non-LTE 1D code RADEX (Van der Tak et al. 2007), we obtain a consistent solution for the observed brightness temperatures of \codos~and \cotres~(\jdos~and \jtres) and \cssiete~in the R1 blob, with $T_{kin}=61$\,K, $n(H_2)=2\times10^5$\,\vdensunit~and abundance ratios of [\codos]/[CS]$\sim$85, and [\codos]/[\cotres]$\sim$70. 
The optical depth of \cssiete~is 18, which lowers its critical excitation density to $\sim1.7\times10^{6}$\,\vdensunit~consistent with this line being subthermally excited. If [\codos]/[H$_2$]=$10^{-3}$, we get [CS]/[H$_2$]=$1.2\times10^{-5}$, which is within the broad range of CS abundance values ($2.7\times10^{-7}-2.1\times10^{-5}$) derived for carbon-rich CSEs by Massalkhi et al. (2019).


\subsubsection{The Central Region}\label{cendude-anal}
We noted earlier that in the central region ($r\lesssim$200 au), the \jdos~and \jtres~lines of \codos~and \cotres~show an intensity minimum (see \S\,\ref{warped} and Fig.\,\ref{13co_co_cens}). This reduction in the intensities could be due to a reduction in the column density or abundances or both. A reduction in average column density towards the center is unlikely because several high-excitation species such as \siodonau~and \cssiete~show an intensity maximum at the center (Fig.\,\ref{othercens}) at the systemic velocity. It is more likely that the abundances of \codos~and \cotres~(and possibly, other molecular species as well) are lower in the central region, compared to the outer parts of the DUDE.

The average brightness temperature of \cssiete~over the central region within a beam-sized aperture ($0\farcs32$) is $T_{b(\rm CS\,7-6)}\sim59$\,K. We also detect the \cstres\,\jcinco~line at the center with a brightness temperature that is a factor $\sim$50 less than that of \cssiete, about 1.1\,K. Assuming that the [$^{13}$CS]/[CS] ratio is 1/70, we find, using RADEX, that a simultaneous fit to the brightness temperatures of both the \cssiete~and \cstres\,\jcinco~lines, requires T$_{kin}\sim(100-250)$\,K, and relatively high H$_2$ densities, $n\sim(0.15-2)\times10^7$\,cm$^{-3}$, with lower values of T$_{kin}$ associated with higher values of $n$.

In contrast to \cssiete, \codos\,\jdos~and \jtres~have lower brightness temperatures, 33\,K and 47\,K, respectively. The resulting \codos\,\jtres/\codos\,\jdos~brightness temperature ratio, 1.4, requires that $T_{kin}\lesssim100$\,K and optical depths less than unity for these lines. The discrepancy between the \tkin~values derived from CO and CS, can be resolved if there is a vertical temperature gradient in the DUDE, with \tkin~increasing with height above the equatorial plane of the DUDE. Since \cssiete~is optically-thick, it would preferentially sample the hotter surface layers of the DUDE, compared to \codos\,\jdos~and \jtres, which are optically-thin in the central region.

The chemical properties of the central region appear to be different from the outer parts of the DUDE, 
based on the very low CS abundance in it, which we estimate as follows. 
The projected thickness of the DUDE in the central region,  i.e., at r$\lesssim0\farcs25$ or 100\,au, derived from the FWHM of the \cssiete~intensity profiles in E-W intensity cuts in this region, is $0\farcs5$ or $\sim3\times10^{15}$\,cm.  Accounting for the inclination, we derive an intrinsic thickness of $\sim4.3\times10^{15}$\,cm, and a path length along the $los$ of $\sim6.1\times10^{15}$\,cm\footnote{The \cotres\jtres~image at the systemic velocity gives similar results, but the measurement is more uncertain}. From our RADEX modeling, the maximum CS column density achieved within the range of possible solutions, is N$_{CS}\sim2.2\times10^{15}$($\delta$V/5\,\kms)\,cm$^{-2}$, implying a [CS]/[H$_2$] abundance ratio $\sim1.8\times10^{-7}$($\delta$V/5\,\kms), which is substantially lower than that estimated above for ring R1, $1.2\times10^{-5}$. If we use our spatio-kinematic model of the DUDE (see \S\,\ref{dude-model}), we find that the H$_2$ column density in the central region along the $los$, is $\sim5.2\times10^{21}$\,cm$^{-2}$, implying a [CS]/[H$_2$] abundance ratio $\sim4.2\times10^{-7}$($\delta$V/5\,\kms). We conclude that the CS abundance is significantly lower (by a factor $\gtrsim30$) in the central region, relative to the outer parts of the DUDE. The abundances of \codos~and \cotres~in the central region are also lower by at least a factor 2.5, based on our radiative transfer modeling (described below.)

The emission from other molecular lines such as SiC$_2$ 15(2,14)-14(2,13) and \hctresnhi, indicate that there are additional complexities in the central region (see Appendix) which we do not analyze in detail in this paper.  A detailed analysis and modeling of the central region is deferred to a future study.


\subsubsection{Dust Emission from the DUDE center}\label{dust-anal}

All high-angular-resolution continuum images show an unresolved source (Table\,\ref{vhya-mmsed-tbl}). There are four possible sources of this compact mm-wave emission -- (i) the stellar photosphere, (ii) radio photosphere, (iii) free-free emission from ionized gas, and (iv) thermal dust emission. We estimate the stellar photospheric contribution using a blackbody, with a stellar luminosity of $1.36\times10^4$\,\ls~at our adopted distance of D=400\,pc\footnote{scaled from $4.11\times10^4$\,\ls~at D=694\,pc (McDonald 2012)}, and an effective temperature of the star ($T_{eff}=3091$\,K:  McDonald (2012)). We find that the stellar photospheric emission is about a factor 3 lower that the observed ``small-beam" fluxes (Table\,\ref{vhya-mmsed-tbl}, Fig.\,\ref{mod-cont-vhya}) -- e.g., at 225 \& 338\,GHz, the photospheric emission is 7.9 \& 17.9\,mJy\footnote{assuming black-body emission}. Next, we estimate the expected emission from a standard radio photosphere (T$_{eff}(radio)=1590$\,K, radius=$5\times10^{13}$\,cm), using the 
relationship given by Reid \& Menten (1997), $S=0.5(\nu/10\,{\rm GHz})^{1.86}(D/100{\rm pc})^{-2}$\,mJy, and find that this emission, although higher than the stellar photospheric emission, is also much lower (factor $\sim2.4$) than the observed fluxes (Fig.\,\ref{mod-cont-vhya}). Scaling up the radio photospheric emission by a factor 2.4 provides a good fit to the small-beam data, suggesting that either V\,Hya's radio photosphere is larger than average (by a factor $\sim1.5$), or the excess is due to another emission source. However, since the opacity (due to H$^{-}$ and H$_2^{-}$) varies inversely as the square of the frequency (Reid \& Menten 1997), the size of the radio photosphere is likely to be smaller, not larger, at mm/submm wavelengths compared to that in the radio regime. Furthermore, the observed excess above the radio photosphere emission is larger for larger beams, consistent with what we would expect if it is due to thermal dust emission from an extended circumstellar envelope.  Although ionized gas is present around V\,Hya (in the form of ``bullets", see SSM16), the free-free contribution is likely to be optically-thin at (sub)millimeter wavelengths and inadequate for providing the excess flux. We tentatively conclude that dust emission from large grains is the more likely explanation for the excess mm/submm emission from the compact central continuum source, after accounting for the contribution of a radio photosphere. 

Thermal dust emission can provide this excess. The power-law exponent of the excess is $\sim$2 (in Fig.\,\ref{mod-cont-vhya}), implying a grain-emissivity power-law ($\kappa(\nu)\propto\nu^{\beta}$) index of $\beta\sim$0 (in the Rayleigh-Jeans or R-J approximation) -- hence the emitting dust grains would be quite large (size $\gtrsim$ mm). If we drop the R-J approximation, then determining $\beta$ requires a knowledge of the dust temperature, $T_d$. Assuming that the size of the unresolved continuum source is equal to or smaller than the beamsize, the brightness temperature of the excess flux at 338\,GHz  is $T_b$(338\,GHz)$>7.5$\,K -- hence $T_d>7.5$\,K. Fitting the excesses in the wavelength range $0.89-3.1$\,mm within (conservative) uncertainties of $\pm25$\%\footnote{This uncertainty is dominated by the systematic uncertainty arising from the radio photosphere flux estimates, which are based on a standard model for AGB stars; we have made an eyeball estimate of this uncertainty from the scatter of the high S/N flux measurements around the standard model, in Fig.\,9 of Reid \& Menten (1997)}, we find $T_d\,\ge\,7.5$\,K, and $\beta\,\le\,1$.

However, we show that it is much more likely that $T_d$ is significantly greater than 7.5\,K, and $\beta$ is significantly lower than 1. Since the continuum source is unresolved in our Band 7 data, the source size is less than the beam-size i.e., $<0\farcs41$ (164\,au). Using the equation for the radius ($r_d$) at a given value of the dust temperature ($T_d$) from Herman et al. (1986), 
\begin{equation}
r_d=(L_*\,T_{*}^\beta/16\,\pi\,\sigma)^{1/2}\,T_d^{-(2+\beta/2)}, 
\end{equation}
where $\sigma$ is the Stefan-Boltzmann constant, for heating by a central stellar source with luminosity and temperature of $L_*=1.36\times10^4$\,\ls~and $T_*=T_{eff}=3091$\,K, we find that dust grains at a radius of 82\,au should have a temperature of $(330-520)$\,K, for $\beta$ in the range $0-1$. At these high temperatures, a fit to the mm/submm excesses (within $\pm25$\%) requires $\beta\sim0$. The low value of $\beta$ suggests that the grains in the waist producing the observed emission are several mm in size. Similar large-sized grains have been found in the disks of several disk-prominent post-AGB objects (Sahai et al. 2011), and in HD\,101584, believed to be a post-RGB object like the Boomerang Nebula (Olofsson et al. 2019). Sahai et al. (2017) showed that two alternative mechanisms that produce shallow mm-submm spectral indices, namely ($a$) spinning dust grains, and ($b$) ferromagnetic or ferrimagnetic nanoparticles, could not explain the Boomerang Nebula's mm-submm continuum fluxes. We can rule out mechanism $a$ in the case of V\,Hya for exactly the same reasons as in the case of the Boomerang Nebula. For mechanism $b$, we estimate the dust mass in Fe-nanoparticles, M$_{d,Fe}$, required to produce the excess 338\,GHz flux in V\,Hya ($\sim30.9$\,mJy) from Figs.\,3-4 of Draine \&  Hensley (2012) -- these show that F$_{\nu}$(340\,GHz)$\lesssim0.9$\,mJy (D/62\,kpc)$^{-2}$ (M$_{d,Fe}$/1\,\ms), implying that for V\,Hya, 
M$_{d,Fe}\gtrsim1.4\times10^{-3}$\,\ms. This value of the dust mass is implausibly large compared to our estimate of $1.7\times10^{-5}$\,(330\,K/$T_d$)\,\ms~assuming a typical dust absorption coefficient of 1.5 cm$^2$\,g$^{-1}$ at 1.3\,mm. Also, assuming a cosmic abundance of Fe ([Fe]/[H]=$4\times10^{-5}$) and that 100\% of all the available Fe is in these small grains, we find that the associated mass of H is 0.64\,\ms, which is orders of magnitude larger than the estimate from our spatio-kinematic model (see below). Significant extinction and reddening of the starlight by dust close to the central star in V\,Hya will cause the grains to be somewhat cooler  -- see e.g., a model of the disk/torus in the PPN, IRAS\,22036+5306 (Sahai et al. 2006) which also shows the presence of large grains at relatively low temperatures -- in this case, the dust mass in V\,Hya's central region would be somewhat larger than the values derived above.

The gas mass within a radius of 82\,au as derived from the density law in our spatio-kinematic model (\S\ref{dude-model}), is $1.2\times10^{-5}$\,\ms, implying an unprecendent low value of the gas-to-dust ratio of $\lesssim1$ compared to the values typical of the dust outflows from AGB stars, $few\times100$. A possible solution to this discrepancy is that the gas density in this region is substantially higher than in our model, and the \codosabund~abundance ratio substantially lower, which is quite plausible since we already find from our modeling that a decrease in this abundance is required for the central region.






\subsubsection{Spatio-kinematic Model of the DUDE}\label{dude-model}
We have made a quantitative model of the DUDE which extends to a radius of $8{''}$, fitting the observed \cotres\,\jtres~emission with the main goal of providing some basic constraints on the DUDE's physical parameters (e.g., geometry, density, velocity, and temperature). In order to improve the S/N of faint structures in channel maps for model-fitting, the \cotres\,\jtres~datacube was rebinned from a velocity-resolution of 0.443\,\kms~to 1.329\,\kms. A similar model for the HV outflows is deferred to a following paper -- the faint emission from these outflows near the DUDE overlaps the very bright emission from the latter, so a combined model of outflows+DUDE is required and is outside the scope of this paper. 

Our model consists of a compact central component (C$_0$) and three rings (R1, R2, and R3). For simplicity, we assume a common inclination angle of all the ring-planes to the $los$, of $40\degr$ -- this is a few degrees smaller than the inclination angles inferred in \S\,\ref{diskring} for reasons discussed below.  We use an abundance ratio [\cotres]/[H$_2$]=$1.4\times10^{-5}$, assuming a cosmic oxygen abundance [\codos]/[H$_2$]=$10^{-3}$, and [\codos]/[\cotres]=70 (see above) everywhere in the DUDE except in its central region -- for $r<$ 200 au, we find that the abundances of \codos~and \cotres~have to be reduced, i.e., [\codos]/[H$_2$]=$4\times10^{-4}$, in order to match the observed intensities in this region.

We generate \cotres\jtres~emission maps by integrating the local emission from each point in our model along each line of sight, using a radiative transfer code and assuming local thermal equilibrium (see e.g., Huang et al. 2016, 2020). We use the following criteria to find our best-fit model: 
(1) in the \cotres\,\jtres~line profile, the difference of the flux density at the blueshifted and redshifted peaks between our model and observation should be $<$10\%;  
(2) the radii and amplitudes (above the underlying DUDE emission) of the brightness peaks (R1, R2, and R3) at the systemic velocity should be consistent (within 10\%) with the average of those observed in the north and south (e.g., as seen in Fig.\,\ref{13co32tau}); and 
(3) in the E-W PV intensity cut across the center, the difference of the expansion velocity and position of the intensity peaks between the model and observation should be $<$10\%. 

We parameterize the spatio-kinematic structure of the DUDE as follows (see Fig.\,\ref{schem-dude}). We have used both spherical and cylindrical coordinate systems for defining the structure; the radial coordinate in spherical geometry is $r$, and that in cylindrical geometry, $R$. We refer to the latter as the cylindrical radius $R$ to avoid any ambiguity. The radial variation of the fundamental physical parameters characterising the model DUDE is shown in Fig.\,\ref{mod-parms}. 
The number density is given by
\begin{equation}
n(r,\theta,\phi)=n(r)\,s(r,\theta)\,\frac{A(\phi)}{\frac{1}{2\pi}\int A(\phi)d\phi}, 
\end{equation}
where $\phi$ is the azimuthal angle in the equatorial plane of the DUDE, with $\phi=0$ to the south, $A(\phi)$ is the dependence of density on $\phi$, and $s(r,\theta)$ is the dependence of density on radius and the latitudinal angle (and thus the height above the equatorial plane, Z -- see eqn. \ref{sfunc} below.)
The function, $A(\phi)=1+C_1 {\rm sin}^2(\phi/2-\pi/4)$, produces a maximum in the eastern part of the DUDE ($\phi=3\pi/2$), and a minimum in the western part ($\phi=\pi/2$) (Fig.\,\ref{ab-phi}), making the DUDE slightly azimuthally-asymmetric and thus allowing us to make the model redshifted emission peak in the spatially integrated spectrum stronger than the blueshifted one, as observed.
The gas density dependence on radius $r$ is given by:
\begin{equation}
n(r)=n_1 (\frac{r_1}{r})^{\beta_n} \, {\rm{exp}}(-\frac{(r-r_i)^2}{2{\sigma_i}^2}), 
\end{equation}
where $n_1$, $\beta_n$, and $\sigma_i$ are free parameters,
$r_i$ is the radius of the components C$_0$, R1,
R2, and R3 ($i=$0, 1, 2, and 3, respectively). 
For two adjacent rings, the densities are the same at the boundary.
The function $s(r,\theta)=s(R,Z)$ describes the decrease of density as 
a function of the height $Z$ above the DUDE equator and cylindrical radius $R$:  
\begin{equation}\label{sfunc}
s(r,\theta) = s(R,Z) = {\rm{exp}}(-\frac{Z^2}{2{(h(R)/2)}^2}), 
\end{equation}
where $h(R)$ is the height of the DUDE in cylindrical coordinates ($R$, $Z$, $\phi$). 
The FWHM of the density function along the Z-axis is constrained by the projected radius along the E-W direction of the DUDE shown in the intensity map (Fig.\,\ref{all-mom0}). If the FWHM is significantly larger, the projected radius is correspondingly larger, and the moment 0 map is less elliptical than observed.

Here the height of the DUDE is assumed to linearly increase with the cylindrical radius $R$ as
\begin{equation}
h(R) = h_1 + m_{\rm sur} (R-r_1)
\end{equation}
Here $h_{1}$ is the DUDE height at $R=r_{1}$, and $m_{\rm sur}$ is the slope of the surface relative to the equatorial plane.
Therefore, the DUDE thickness at the cylindrical radius $R$ is $2h(R)$.

The velocity is defined as $\bm{v}=v_Z \bm{\hat{Z}} + v_R \bm{\hat{R}}$ in the cylindrical coordinate system that we have used. 
The value of the slope $m_{\rm sur}$ (0.35), and the corresponding opening angle ($39\degr$) of the model DUDE, is estimated from the flared shape of the DUDE as seen in \cotres\,\jtres~emission in Fig.\,\ref{13co32cench}.

The slope of the velocity vector, $m_{\rm vel}=m_{\rm sur}\,(Z/h(R))$, thus on the DUDE's surface (equatorial plane), the velocity vector is parallel to the surface (equatorial plane).
Because $m_{\rm vel}=v_Z/v_R$, the velocity can be calculated by this equation:
\begin{equation}
\bm{v}= \frac{m_{\rm vel} v_a(r)}{\sqrt{1+{m_{\rm vel}}^2}} \bm{\hat{Z}} + \frac{v_a(r)}{\sqrt{1+{m_{\rm vel}}^2}} \bm{\hat{R}},
\end{equation}
where $v_a(r)$ is the expansion velocity. In our best-fit model, 
\begin{equation}
v_a(r) = \left \{ \begin{array}{ll}
    (v_0+(v_D-v_0)\frac{r}{R_D})\frac{r}{r_s}, \quad \mbox{if $r \leq r_s$} \\
    v_0+(v_D-v_0)\frac{r}{R_D}, \quad \mbox{if $r > r_s$},
  \end{array} \right.
\end{equation}
where $v_0$ and $v_D$ are the velocity at $r=0$ and $r=R_D$, and $r_s$ is a small distance
showing that the expansion velocity linearly increases from 0 to $v_0$. 
For the bulk of the gas in/near the equatorial plane ($m_{\rm vel}=0$) and $r>r_s$, we have $v=v_0+(v_D-v_0)r/R_D$, as shown in Fig.\,\ref{schem-dude}. 

In the PV plot with the cut taken along the E-W direction, the velocity range is
relatively large near the source (i.e., within $\pm0\farcs1$~of the center), and decreases with the distance to the center (Fig.\,\ref{13co32-mom1-pvew}c).
In order to fit this feature, we assume a turbulence velocity
\begin{equation}
v_t(r) = \left \{ \begin{array}{ll}
    v_{t0}-v_{t1}\frac{r}{r_1}, \quad \mbox{if $r \leq r_1 $} \\
    v_{t1}\frac{r_1}{r}, \quad \mbox{if $r > r_1$}.
  \end{array} \right.
\end{equation}

The gas temperature is assumed to be given by:
\begin{equation}
T=T_1 (\frac{r_1}{r})^{\beta_T} \, {\rm{exp}}(-\frac{(r-r_i)^2}{2{\sigma_{Ti}}^2}), 
\end{equation}
where $T_1$, $\beta_T$, and $\sigma_{Ti}$ are free parameters. In addition, we assume that the gas temperature has an upper limit of $\sim$100 K (Fig.\,\ref{mod-parms}c) based on our analysis in \S\,\ref{cendude-anal}. Since the \cotres~and\codos~lines becomes optically-thin in the central region, the kinetic temperature or its variation cannot be determined more accurately. The values of $\sigma_{Ti}$ determine the local rises in the kinetic temperature in the vicinity of the rings which are required to produce the 
local peaks in the brightness temperature of the optically-thick \codos\jtres~line. The value of $\beta_T$ is constrained by the overall decrease with radius in the brightness temperature of the \codos\jtres~line (Fig.\,\ref{13co32tau}). It is possible that the localised brightness temperature peaks in the ring regions R2 and R3 are a result of local increases in the excitation temperature due to local increases in density in these regions (since the excitation there is subthermal), but since we are assuming LTE in our model, we do not have a way of accounting for such increases. However, in ring R1, the density (and optical depth) is high enough for \codos\jtres~to be thermalized, so the local increase in the observed brightness temperature in it does require a local increase in the kinetic temperature.

\begin{figure}[ht!]
\includegraphics[width=0.7\textwidth]{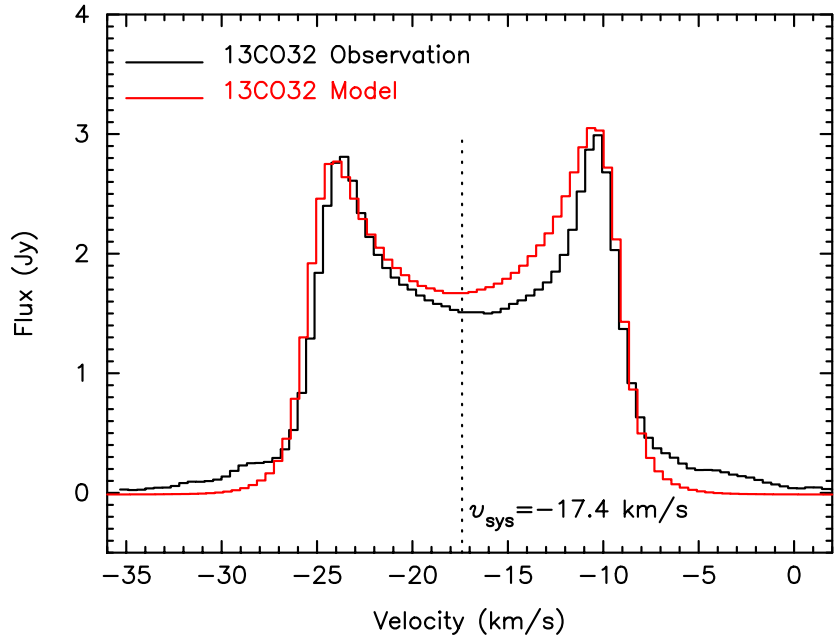}
\caption{Spatially-integrated spectrum of the \cotres\jtres~emission spectrum (black curve) from the DUDE, extracted using an elliptical aperture of $16\farcs5\times15\farcs0$, together with the best-fit model spectrum (red curve). 
}
\label{mod-13co32-spec}
\end{figure}
\begin{figure}[ht!]
\includegraphics[width=1.0\textwidth]{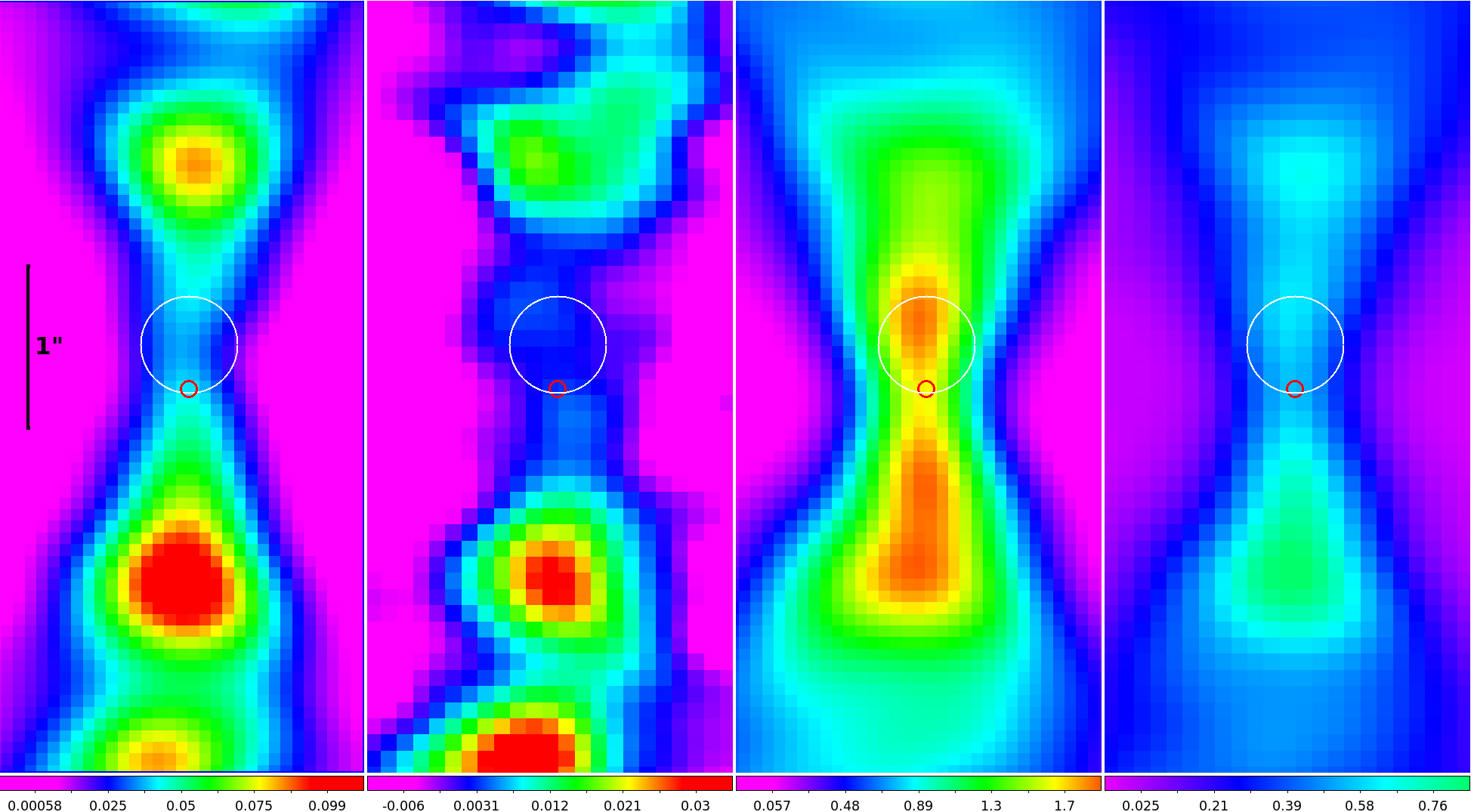}
\caption{The central region of the DUDE as seen in (a) \cotres\,\jtres, (b) \cotres\,\jdos, (c) \codos\,\jtres, (d) \codos\,\jdos~emission at the systemic velocity. The white circle (radius $0\farcs3$) at the center of each image is located at the minimum of the \cotres\,\jtres~image, and the small red circle (radius $0\farcs05$) shows the location of V\,Hya's central star. Images in panels a, c, and d have been convolved to the same beam as in panel b ($0\farcs643\times0\farcs541$, $PA=69.25\arcdeg$). All panels have the same field-of-view with North up and East to the left. The intensity scale at the bottom of each panel is in Jy\,beam$^{-1}$. 
}
\label{13co_co_cens}
\end{figure}

Our best-fit parameters are listed in Table\,\ref{dudeparms}, which is divided into five sections -- geometry, density, temperature, velocity, and abundance -- each including the parameters that govern these characteristics.
For example, if the DUDE has larger values of $h_1$, inclination angle $i$, and opening angle (39\arcdeg), the projected thickness of the DUDE becomes wider.  If $C_1$ is larger, then, in the line profile (Fig.\,\ref{mod-13co32-spec}), the ratio of the flux density of the redshifted peak to that of the blueshifted, is larger. If $n_1$ increases, the total flux increases, and the optical depth of the \cotres\,\jtres~line (Fig.\,\ref{13co32tau}) increases, too.  If the power-law index $\beta_n$ increases, the decrease of the \cotres\,\jtres~optical depth (Fig.\,\ref{13co32tau}) with radius becomes more steep. If $\sigma_i$ and $\sigma_{Ti}$ are smaller than the best-fit values, the local minima of the optical depth and brightness temperature become lower.  If $r_s$ is larger than 100 au, the velocity increases more slowly with offset from the center, and the intensity peaks in the PV plot in Figure\,\ref{mod-13co32-pv} shift to a larger offset. 

The remaining discrepancies between the model and data are mostly in the (i) emission in the line-wings (i.e., at expansion velocities $>10$\,\kms) of the spatially-integrated spectrum, and (ii) ring locations (Fig.\,\ref{13co32tau}).  Discrepancy $i$ is due to the fact that we have not included the high-velocity parabolic outflows in our model, even though their bases which are connected to the DUDE, likely contribute to the emission seen in the line-wings. Discrepancy $ii$ is due to the model assumption that all three rings have the same center, whereas the actual centers of these are not the same.

The total mass of the DUDE is $1.7\times10^{-3}$\,\ms~-- this is a lower limit because the DUDE extends to (at least) the radius of ring R6 ($12{''}$), beyond the model outer radius of $8{''}$. If we assume that the slope for h(R) and the density power-law index remains unchanged at radii $>8{''}$, then, for (say) an outer radius of $14{''}$, the total mass increases to $2.9\times10^{-3}$\,\ms. The average mass-loss rate in the DUDE is $\sim1.1\times10^{-6}$\,yr.

The model (spatially-integrated) spectrum, channel map, and PV plot are shown, together with the data for comparison, in Figs.\,\ref{mod-13co32-spec}, \ref{mod-13co32-rings}, and \ref{mod-13co32-pv}.

We also computed the moment 0 map of the \cotres~\jtres~emission, and found that, because the DUDE is flared, the inclination angle of the ring-planes to the $los$ needs to be 40\arcdeg~(i.e., a few degrees smaller than the inclination angles inferred in \S\,\ref{diskring}) in order to roughly match the average of the observed major-to-minor axis ratios for rings R1--R3, as derived from the observed \cotres~\jtres~moment 0 map.

\section{Discussion}\label{discuss}
The molecular circumstellar environment of V\.Hya, has, as noted in \S\,\ref{intro}, been observed both with single-dish and interferometric facilties, and three main kinematic components have been identified, summarized in HSD04 as -- (1) a slowly expanding disk, (2) a medium-velocity wind, and (3) a high-velocity jet, with their respective emissions spread over radial-velocity offsets (from \vsys) of $-8<\Delta\,V\,(\kms)<8$, $-60<\Delta\,V\,(\kms)<60$~and $-(60 to 160)<\Delta\,V\,(\kms)<(60 to160)$, respectively. In addition to the discovery of the rings and the wide-angle high-velocity outflows, our ALMA observations have now both considerably extended and refined this picture.

\subsection{Equatorial Rings and High-Velocity Outlows}
The previously identified slowly expanding disk is an equatorially-dense, flared structure with six ring-like density enhancements -- the DUDE. The (de-projected) expansion speed is roughly constant, $10.4-11.4$\,\kms~over most of the DUDE, except for a central region (size $\sim$200\,au) with a strong linear velocity gradient. The innermost four rings of the DUDE, which are reasonably well characterised, have modestly different PA's and inclination angles. A plausible explanation for these differences is that (a) the rings represent episodic mass-ejections from the primary star, (b) the orbit of the close companion is not exactly coplanar with the equatorial plane of V\,Hya, (c) over time, the resulting torques cause a precession of V\,Hya's rotation axis\footnote{Barnbaum et al. (1995) find that V\,Hya is rotating quite fast}, and (d) the plane in which the episodic mass ejection occurs (the latter producing each of these rings) is the same as the instantaneous equatorial plane of V\,Hya. 

The ages of the best characterized innermost three rings (R1, R2, R3), inferred by dividing their observed radii by their expansion velocities (as derived from our model), are 270, 485, and 780\,yr. Our model does not extend to rings R4, R5 and R6 but assuming that the expansion velocities of these are similar to that of R3, i.e., 10.6\,\kms, their ages are 1220, 1470, and 2150\,yr.
So what is the physical mechanism for producing these episodic ejections spaced apart by a $\sim$few$\times$100-700 years? It may be indirectly related to the more distant tertiary star that was proposed by Salas et al. (2019) in order to keep the near-companion's orbital eccentricity high. If so, then the tertiary must also be in an eccentric orbit, provoking the enhanced mass ejections in V\,Hya's equatorial plane at periastron. But in such a scenario, one would expect the time-intervals between succesive rings ejections to be equal to the tertiary orbital period and therefore the same -- however, these time-intervals are observed to be different from each other. A plausible explanation of this discrepancy is that (a) the rings result from enhanced mass-loss from cool-spots\footnote{because dust will form more efficiently above such spots} in/near the equatorial plane of V\,Hya that are formed due to enhanced dynamo magnetic activity (e.g., Soker 1998, 2000, Soker \& Zoabi 2002) during periastron passage of the tertiary, and (b) that the dynamo is more stable at some times than others.

The previously identified high-velocity jet corresponds to the HVC outflow in our study, and represents a bipolar outflow that is highly-collimated and is thus qualitatively similar to the fast jet-like outflows that are seen commonly during the AGB-to-PN transition, i.e., during the PPN phase. 

The previously identified medium-velocity wind corresponds to the HVW outflows. The nested hourglass structure of these outflows is a puzzle -- such outflows have not been seen before in AGB stars. We consider two models for the HVW outflows -- (a) radiation-pressure-driven disk winds, and (b) accretion-disk driven winds.  Model $a$ is motivated by the fact that many of the footprints of the parabolic high-velocity outflows appear to be anchored at relatively large offsets from the disk axis. The accelerating force could be provided by radiation pressure from V\,Hya's large luminosity acting on small grains, which drag the gas with them via dust-gas friction. These small grains could be produced by grain-grain collisions of the large grains in the DUDE. Such a model has been proposed to explain the submillimeter fluxes from the giant SS\,Lep by Jura et al. (2001). However such a model would have difficulty explaining the very high-velocities of the parabolic outflows, given that radiation pressure on dust -- the standard model for outflows in AGB stars -- produces expansion velocities of typically $\sim10-20$\,\kms. 

But since we cannot spatially separate the very faint HVW outflow emission from the disk emission at low velocity offsets, we cannot rule out the possibility that the footprints of these parabolic outflows are really anchored at locations near the central binary system (e.g., in an accretion disk around the close companion) and that the gas in these outflows first moves outwards at relatively small angles to the plane of the disk, and then further out undergoes a very strong curvature towards larger angles (as observed). This possibility leads us to Model $b$ which is motivated by the presence of both wide-angle outflows and jets in young stellar objects (YSOs), such as in the Class 0 protostar Cep E (Velusamy et al. 2011). In these objects, the wide-angle outflows are believed to be driven by the magnetocentrifugal wind mechanism along hourglass-like field lines, whereas the highly collimated outflow
is driven by magnetic pressure and guided by straight field lines near the YSO. The difficulty with applying this model to V\,Hya is that the wide-angle outflows in YSOs are of relatively low-velocity.

Soker \& Rappaport (2000) have proposed a mechanism in which a bipolar high-velocity outflow can compress pre-existing material expelled in a slow outflow towards the equatorial plane. Akashi et al. (2015) have made numerical simulations of this mechanism, employing a short-lived bipolar high-velocity (1000\,\kms) outflow with a wide opening angle, $\sim$60\arcdeg--100\arcdeg (referred to as a jet in their study) that expands into a geomerically-thin, slowly expanding (10\,\kms) massive dense spherical shell (resulting from a 3\,yr episode of mass-loss at the rate of $\sim0.03$\,\my~possibly due to an Intermediate-Luminosity Optical
Transient event or ILOT); the shell is surrounded by a slowly expanding (also at 10\,\kms) much more tenuous circumstellar medium resulting from a wind with a mass-loss rate of $10^{-5}$\,\my. This ``jet" compresses gas towards the equatorial plane and generates an equatorial ring structure. This model appears attractive at first glance, but the physical requirements (ejection of $\sim$0.1\,\ms~in a few years, 1000\,\kms~outflow) are rather extreme to be directly applicable to V\,Hya. Most importantly, although V\,Hya's light curve has been monitored by the AAVSO since 1961 (e.g., Knapp et al. 1999) it has never shown any evidence of even a single ILOT.

We offer a speculative but comprehensive model that may explain the HVWs, the HVC, and the DUDE and its rings. In this model, high-speed bullets interact with a circumstellar medium created by a wind from the primary AGB star that experiences periods of enhanced mass-loss every few hundred years. The bullets are ejected during periastron passage of the secondary. The bullet engine resides in an accretion disk around the secondary (SSM16), possibly as a result of Grazing Envelope Evolution, in which the companion grazes the primary star's envelope while both the orbital
separation and the giant radius shrink simultaneously (Soker 2015)\footnote{the binary system avoids experiencing CE evolution, instead evolving in a constant state of ``just enetering a CE phase" for up several hundred years}. The interaction creates a bipolar cavity, and the bullets, whose ejection axis flip-flops, bounce back and forth between the walls of this cavity (SSM16, SSM19). The gas in the cavity walls at intermediate and high-latitudes is thus compressed and accelerated to high velocities. As these walls expand and move outwards, the bullet-wall interaction weakens, with the bullets (presumably with molecular material entrained) being channeled along the symmetry axis of the system, forming the HVC. The cavity walls continue to move outwards ballistically, forming the HVWs. The DUDE and its rings are either a result of the mass-loss from the primary being intrinsically concentrated towards the equatorial plane, or gas at low latitudes in a spherical primary wind being compressed towards the equatorial plane via the mechanism discussed by Soker \& Rappaport (2000).

\subsection{Comparison with other Evolved Stars}
The morphology of the CO emission from the S-star, $\pi^1$\,Gru, imaged with ALMA (Doan et al. 2020, Homan et al. 2020), does bear rough resemblance to some of the structures we see in V\,Hya. Doan et al. (2020) present a model for $\pi^1$\,Gru CO emission that consists of a slowly expanding equatorially-dense expanding structure (similar to the DUDE in V\,Hya), and a high-velocity bipolar flow in the form of thin-walled bubbles expanding radially at 60\,\kms~(similar to the parabolic high-velocity outflows in V\,Hya). Their low-velocity channel maps show partial arc structures that the authors infer to be part of an Archimidean spiral pattern, induced by the presence of a close companion. The main differences between the circumstellar morphology in V\,Hya and $\pi^1$\,Gru are (a) we see complete or partial rings in V\,Hya, not a spiral structure, (b) there are three high-velocity outflows (two wide-angle ones and a collimated one), and (c) the velocities of the expanding material in these outflows is significantly higher than in $\pi^1$\,Gru.

There are also notable similarities between V\,Hya and R\,Aqr, a Mira with a white dwarf companion embedded in an accretion disk, and classified as a D-type symbiotic star. R\,Aqr is an eccentric binary (e=0.25) with a period of $\sim$43.6\,yr (Gromadzki \& Mikolajewska 2009), and it has a bulleted outflow and a complex bipolar nebula (Liimets et al. 2018). For V\,Hya, the period is 8.5 (SSM16) or 17\,yr (Knapp et al. 1999), and 
the photospheric radius of the primary, while highly wavelength dependent, is on the order of several AU (Woodruff et al\, 2009). The period of the orbit and the large size of the AGB star suggest that the AGB star nearly fills its Roche lobe (Barnbaum et al. 1995). In R\,Aqr, the stars are too widely separated for Roche lobe overflow, but the primary's wind likely leads to enhanced accretion near periastron, similar to what has been inferred for V\,Hya (SSM16).

The significant commonality between V\,Hya and R\,Aqr, thus appears to be the presence of a binary in an eccentric orbit, with enhanced activity in an accretion disk around the companion at presiastron passage leading to enhanced mass ejection. Bulleted high-velocity outflows have also been seen in the pre-PN/binary post-AGB object CRL\,618 (Trammell \& Goodrich 2002, Balick et al. 2014), two D-symbiotic systems with extended bipolar nebulae and jets, the Southern Crab (He\,2--104) and BI\,Cru, and the well-known planetary nebulae, the Etched Hourglass Nebula (MyCn\,18) (Corradi and Schwarz 1993, O'Conner et al 2000). We speculate that the central stars of CRL\,618, He\,2--104, BI\,Cru, and MyCn\,18 are also eccentric binary systems.

Both He\,2-104 and MyCn\,18 display a nested hourglass morphology like V\,Hya (Sahai et al. 1999).
MyCn 18 also shows the presence of circumstellar rings in the equatorial plane (Sahai et al. 1999), and has a binary nucleus that is likely post-CE (Miszalski et al 2018). It is quite plausible that V\,Hya will evolve into a PN with a double-hourglass morphology like MyCn\,18 and He\,2-104.

The DUDE may become visible as a large circumbinary disk, while V Hya passes through a post-AGB phase after exhausting its nuclear fuel but before becoming a white dwarf. During this phase mass loss will have ceased and the unseen companion will no longer have an accretion disk.  \citet{oomen_et_al_2018} review orbital elements in a sample of post-AGB binaries, and find that these systems are similar to an evolved V\,Hya binary.  The orbital periods, in the range 0.3 to 8.2 years, require a contact binary or a common envelope phase while on the AGB.    Large circumbinary disks, with outer radii as large as $\sim$few$\times1000$\,au (de Ruyter et al. 2006, Gielen et al. 2011, Bujarrabal et al. 2013), are a common feature.  Given the number of systems known and the very brief post-AGB supergiant lifetime, V Hya does not appear to represent an unusual evolutionary path.  The role of angular momentum transfer in driving the binary eccentricity and shaping the circumbinary disk in these systems remains an open question (Dermine et al. 2013, Rafikov 2016) -- our current study of V\,Hya and future studies with higher-angular resolution that resolve the central region adequately, possibly uncovering evidence of rotation, should help in addressing these issues.


\section{Summary}\label{sumry}
We have observed V\,Hya and its circumstellar environment with ALMA in continuum and molecular-line emission, notably the \jtres~and \jdos~lines of \codos~and \cotres~with an unprecedented angular resolution of $\sim0\farcs5$~(200\,au). Our main findings are as follows:
\begin{enumerate}
\item V\,Hya has a remarkable set of six expanding molecular rings in the system's equatorial plane. These rings represent regions of enhanced density within an equatorially-dense, expanding, structure (dubbed the DUDE). The three innermost rings, which are seen fully, have modestly differing position angles and inclination angles, with radii 560, 1040, 1740\,au and ages of 270, 485, and 780\,yr (with $<10$\% errors). For rings R4, R5 and R6, which are seen partially and less clearly, the radii and ages are about 2720, 3280, 4800\,au and 1220, 1470, and 2150\,yr (with $15-20$\% errors).

\item The ALMA data provide several observational indicators that the DUDE is asymmetric and possibly warped, including a mismatch in the locations of the symmetry centers of the \cotres~and\codos~images at the systemic velocity, and shifts in the symmetry center of the \cotres\,\jtres~emission as a function of velocity offset. In addition, the blue- and red-shifted components of the collimated outflow show small but significant differences between their inclinations and position angles, and the position angles and inclination angles of the best-characterized rings (R1, R2, R3) are not identical, nor are the locations of their centers.

\item In addition to the DUDE, V\,Hya shows the presence of three bipolar, high-velocity outflows, with expansion velocities up to $\sim$175\,\kms. Two of these have parabolic shapes implying very wide opening angles, while a third is clumpy and highly collimated. The velocity gradient in these three high-velocity (HV) outflows is opposite to that seen in the DUDE, suggesting that they are directed along the DUDE's axis.

\item The clumpy, collimated outflow is likely associated with the high-velocity bullet-like ejections of ionized gas from V\,Hya that have been found (using HST) to occur every 8.5\,yr over at least 25\,yr. A possible molecular counterpart to the oldest of the 4 bullets (bullet 0) can be seen in the \codos\jdos~and \jtres~images at a radial offset of 600\,au from center, with a tangential velocity of 53\,\kms. 

\item From a comparison of previous \codos\jdos~images and our ALMA ones, we detect proper motion of blue- and red-shifted clumps in the collimated outflow, that are much more distant than bullet 0. For these, located at radial offsets of 5600\,au and 4600\,au, we find tangential velocities of 178 and 190\,\kms, respectively, much higher than that of bullet 0. The inclination angles of the outflows represented by these distant clumps are different from each other, and their values indicate that they are moving along axes that are roughly, but not exactly, along the DUDE's axis.

\item We find a central, bright, unresolved source of continuum emission (FWHM size $\lesssim$165\,au). Although about 40\% of the mm- and submm- emission from this source can be produced in a standard radio photosphere, the remaining 60\% most likely comes from very large (mm-sized) grains, with a mass of $\gtrsim10^{-5}$\,\ms. Alternatively, the radio photosphere in V\,Hya is a factor 1.5 larger than the standard value. 

\item Using a radiative transfer model, we have fitted the salient characteristics of the \cotres~emission out to a radius of $8{''}$ (3200\,au) with a model disk-like structure whose scale height increases with radius, and an expansion velocity that increases linearly from 9.5 to 11.5 km s$^{-1}$\,\kms~outside of a central region of size $\sim200$\,au. Within the central region the expansion velocity increases very rapidly with radius. The DUDE's underlying density decreases radially, interspersed with local increases that represent the rings R1, R2 and R3. The mass of the DUDE within a radius of 3200\,au is $1.7\times10^{-3}$\,\ms.

\item Within a central region of radius $\lesssim$200\,au, the abundance of CS shows a very significant decrease (by a factor $\gtrsim30$). The gas mass within a radius of 82\,au as derived from the density law in our model is $1.2\times10^{-5}$\,\ms, which implies an unprecendently low value of the gas-to-dust ratio of $\lesssim1$ -- a factor $\sim100$ lower than the typical value for mass ejecta of AGB stars. A plausible resolution of this discrepancy is that the \codosabund~abundance ratio is significantly lower in this region than in the outer parts of the DUDE, and the gas density in this region is correspondingly higher than in our model. 

\item The presence of a pair of hourglass shaped wide-angle outflows and a collimated outflow in V\,Hya makes it remarkably similar to the D-type symbiotic object He\,2--104 and the PN, MyCn\,18.

\end{enumerate}

No existing model can explain the rich diversity of organized structures that we have discovered in V\,Hya. We offer a speculative but comprehensive model in which high-speed bullets interact with a circumstellar medium created by a wind from the primary AGB star that experiences periods of enhanced mass-loss every few hundred years. Our model can potentially explain the origin of the DUDE and its rings, as well as the multiple high-velocity bipolar outflows.

\section{acknowledgements} We thank an anonymous referee for his/her review which has helped us improve the paper. This paper makes use of the following ALMA data: ADS/JAO.ALMA\#2018.1.01113.S. ALMA is a partnership of ESO (representing its member states), NSF (USA) and NINS (Japan), together with NRC (Canada), MOST and ASIAA (Taiwan), and KASI (Republic of Korea), in cooperation with the Republic of Chile. The Joint ALMA Observatory is operated by ESO, AUI/NRAO and NAOJ. The National Radio Astronomy Observatory is a facility of the National Science Foundation operated under cooperative agreement by Associated Universities, Inc.
 
R.S.’s contribution to the research described here was carried out at the Jet Propulsion Laboratory, California Institute of Technology, under a contract with NASA, and funded in part by NASA via ADAP awards, and multiple HST GO awards from the Space Telescope Science Institute. P.-S.H. and C.-F.L acknowledge grants from the Ministry of Science and Technology of Taiwan (MoST 107-2119-M-001-040-MY3) and the Academia Sinica (Investigator Award AS-IA-108-M01). S.S. is supported by National Science Foundation Graduate Research Fellowship (NSF GRF) Grant DGE-1143953. KH acknowledges partial support from USRA award SOF-06-0093 to AURA. This work has made use of data from the European Space Agency (ESA) mission {\it Gaia} (\url{https://www.cosmos.esa.int/gaia}), processed by the {\it Gaia} Data Processing and Analysis Consortium (DPAC, \url{https://www.cosmos.esa.int/web/gaia/dpac/consortium}). Funding for the DPAC has been provided by national institutions, in particular the institutions participating in the {\it Gaia} Multilateral Agreement.

\appendix
The \hctresnhi~and \sicdo~lines show two faint interesting ring-like structures in position-velocity intensity cuts taken along the major axis of the DUDE (Fig.\,\ref{sic2-hc3n-r0}). The 
outer one in these corresponds to R1, but the inner one ~(hereafter R0) has no counterpart in \codos~or \cotres. The radius of R0 is about 200\,au, and corresponds to the edge of the central region, where the \codos~intensity starts going down. R0 probably corresponds to a region of enhanced abundance for \hctresnhi~and \sicdo, suggesting that this region is chemically different from the outer parts of the DUDE. 

\begin{figure}[ht!]
\includegraphics[width=1.0\textwidth]{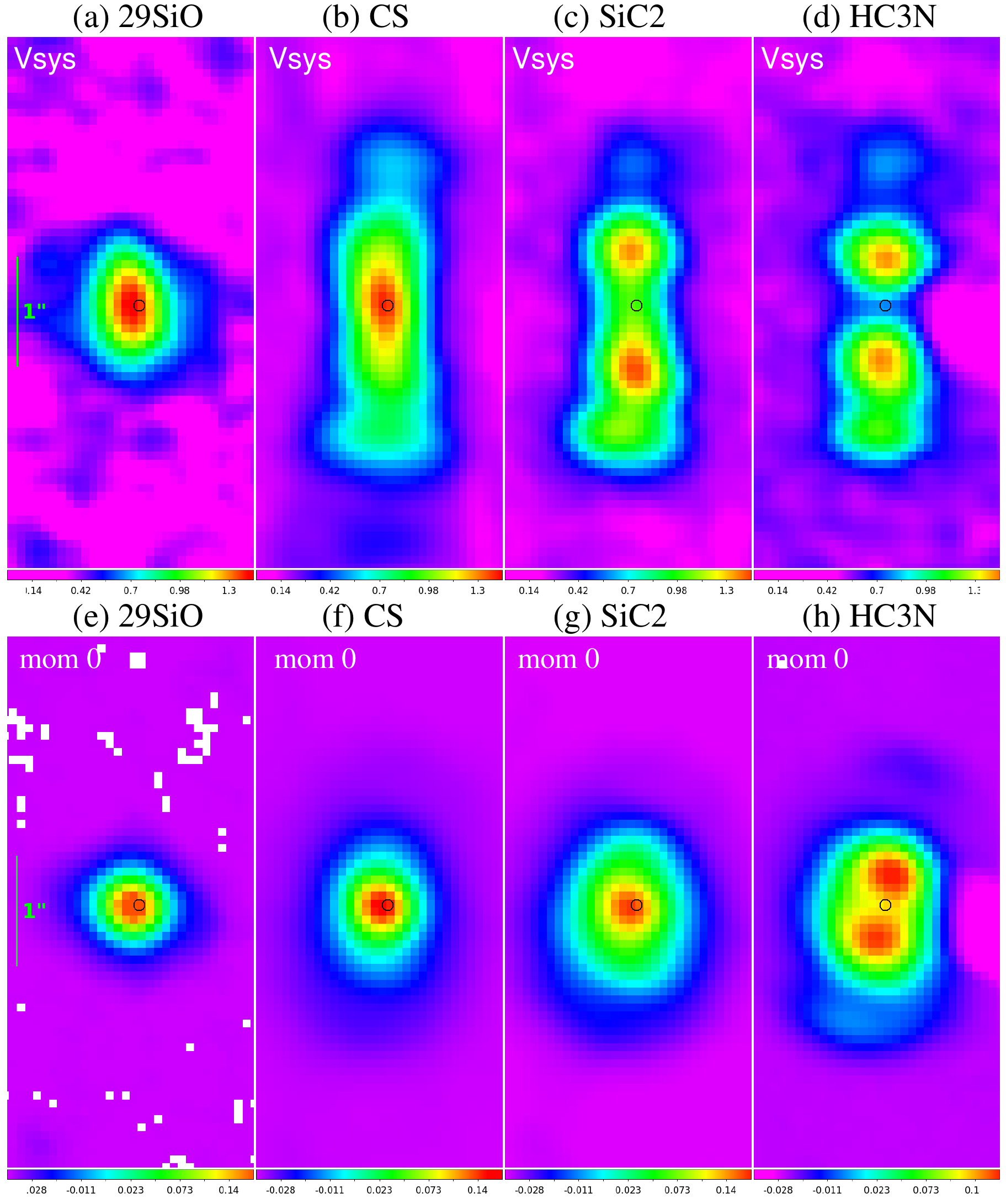}
\caption{The central region of the DUDE as seen in (a) \siodonau, (b) \cssiete, (c) SiC$_2$ 15(2,14)-14(2,13), (d) HC$_3$N\,$38-37$ emission at the systemic velocity. The black circle (radius $0\farcs05$) shows the location of V\,Hya's central star. Panels e--h show the moment 0 maps of the same lines. All panels have the same field-of-view. All panels have the same field-of-view with North up and East to the left. The intensity scale (i) beneath panels a--d is in Jy\,beam$^{-1}$, and (ii) beneath panels e--h is in Jy--\kms\,beam$^{-1}$. The beam sizes are $0\farcs451\times0\farcs363$ ($PA=75.653\arcdeg$) for \siodonau,  \cssiete, and SiC$_2$ 15(2,14)-14(2,13), and $0\farcs443\times0\farcs361$ ($PA=74.473\arcdeg$) for HC$_3$N\,$38-37$.
}
\label{othercens}
\end{figure}

\begin{figure}[ht!]
\includegraphics[width=1.0\textwidth]{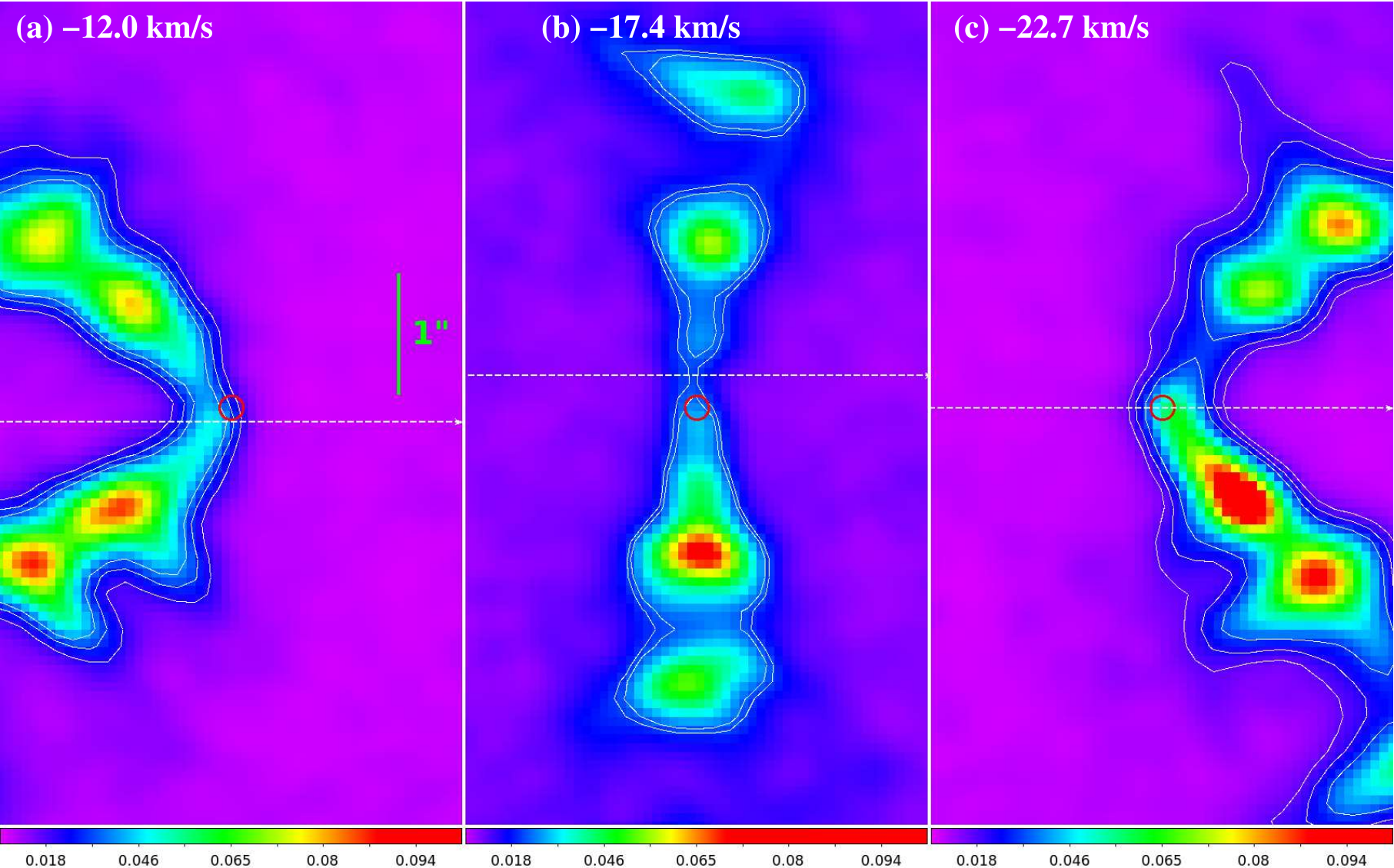}
\caption{The central region of the DUDE as seen in \cotres\,\jtres~at (a) redshifted (\vlsr=-12.1\,\kms) (b) systemic (\vlsr=-17.4\,\kms), and (c) blueshifted (\vlsr=-22.7\,\kms), velocity channels (each channel is 0.44\,\kms~wide).  All panels have the same field-of-view with North up and East to the left. The red circle (radius $0\farcs1$) shows the location of V\,Hya's central star. The horizontal white lines mark the N-S location of the symmetry center in each panel. These lines are respectively located at declinations of -21:15:00.41, -21:15:00.03, -21:15:00.30 in panels a,b, \& c, and the contour levels used to help locate them are, for panel a: 0.022, 0.030, 0.037 Jy\,beam$^{-1}$, panel b: 0.021, 0.024 Jy\,beam$^{-1}$, \& panel c: 0.016, 0.023, 0.032 Jy\,beam$^{-1}$ (see \S\,\ref{warped} for details). The intensity scale at the bottom of each panel is in Jy\,beam$^{-1}$. Beam is $0\farcs443\times0\farcs361$, $PA=74.47\arcdeg$.
}
\label{13co_symmcens_red_blu}
\end{figure}

The maps of the SiC$_2$ 15(2,14)-14(2,13) and \hctresnhi~line intensities, at the systemic velocity, show a central minimum. We infer that, at the systemic velocity, the SiC$_2$ 15(2,14)-14(2,13) and \hctresnhi~lines are optically-thin, in contrast to the \siodonau~and \cssiete~lines, which are optically-thick. The moment 0 maps of \siodonau, \cssiete, and SiC$_2$ 15(2,14)-14(2,13) show intensity maxima at the center (Fig.\,\ref{othercens}, e--g). We infer that the optical depth of SiC$_2$ 15(2,14)-14(2,13) lines increases sufficiently towards the line-wings, such that it becomes optically-thick. But for \hctresnhi, the moment 0 map shows a central intensity minimum, indicating that this line remains optically thin even in the line wings -- surprisingly, though, the moment 0 map of this line shows two compact bright blobs located north and south of the center (Fig.\,\ref{othercens}h).

\clearpage

\scriptsize
\begin{longtable}{p{0.4in}p{1.1in}p{0.4in}p{0.4in}p{0.3in}p{0.5in}p{1.0in}p{0.9in}}
\caption[]{Log of ALMA Molecular Line Observations of V\,Hya}\\
\label{obslog} \\
\hline \\[-2ex]
   \multicolumn{1}{l}{\textbf{Bnd.}} &
   \multicolumn{1}{l}{\textbf{Line}} &
   \multicolumn{1}{l}{\textbf{Freq.}\tablenotemark{(1)}} &
   \multicolumn{1}{l}{\textbf{Range}} &
   \multicolumn{1}{l}{\textbf{$\delta\,\nu$}\tablenotemark{(2)}} &
   \multicolumn{1}{l}{\textbf{Array}} &
   \multicolumn{1}{l}{\textbf{Beam}}  &
   \multicolumn{1}{l}{\textbf{Date}}  \\[0.5ex]
   \multicolumn{1}{l}{\textbf{}} &
   \multicolumn{1}{l}{\textbf{}} &
   \multicolumn{1}{l}{\textbf{GHz}} &
   \multicolumn{1}{l}{\textbf{GHz}} &
   \multicolumn{1}{l}{\textbf{MHz}} & 
   \multicolumn{1}{l}{\textbf{}} &
   \multicolumn{1}{l}{\textbf{${''}\times{''}$,\,(PA)$\arcdeg$}} &
   \multicolumn{1}{l}{\textbf{yyyy-mm-dd}} \\[0.5ex] \hline
   \\[-1.8ex]
\endhead
6  & $^{13}$CO\,J=2--1 & 220.3987  & 220.176--220.644 & 0.244  & 12m  & $0\farcs64\times0\farcs54$,\,69.3  & 2019-04-27 \\
   & $^{12}$CO\,J=2--1 & 230.5380  & 230.081--231.018 & 0.244  & 12m  & $0\farcs62\times0\farcs52$,\,70.6 & 2019-04-27  \\
7  & $^{13}$CO\,J=3--2 & 330.5880  & 330.340--332.213 & 0.488 & 12m   & $0\farcs46\times0\farcs38$,\,74.2 & 2019-04-20  \\
   & $^{13}$CO\,J=3--2 & 330.5880  & 330.355--332.198 & 0.488 & ACA   & $4\farcs72\times2\farcs75$,\,87.8 & 2019-03-08  \\
   & $^{12}$CO\,J=3--2 & 345.7960  & 345.579--346.047 & 0.122 & 12m   & $0\farcs44\times0\farcs36$,\,74.5 & 2019-04-20  \\
   & $^{12}$CO\,J=3--2 & 345.7960  & 345.563--346.063 & 0.122 & ACA   & $4\farcs50\times2\farcs67$,\,88.5 & 2019-03-08  \\
   & \hctresnhi        & 345.6090  & 345.563--346.063 & 0.122 & 12m   & $0\farcs44\times0\farcs36$,\,74.5 & 2019-04-20 \\
   & \cssiete          & 342.8829  & 342.288--344.146 & 0.488 & 12m   & $0\farcs45\times0\farcs36$,\,75.7 & 2019-04-20 \\
   & \sicdo            & 342.8050  & 342.288--344.146 & 0.488 & 12m   & $0\farcs45\times0\farcs36$,\,75.7 & 2019-04-20 \\
   & \siodonau         & 342.9808  & 342.288--344.146 & 0.488 & 12m   & $0\farcs45\times0\farcs36$,\,75.7 & 2019-04-20 \\
\hline
\tablenotetext{1}{Line rest frequency}
\tablenotetext{2}{Frequency width per channel}
\end{longtable}

\newpage
\begin{table}[!t]
\caption{V\,Hya Continuum Emission}
\label{vhya-mmsed-tbl}
\begin{tabular}{ccccccl}
\hline
$\nu$  & Flux\,Density\tablenotemark{(1)} & MajAxis & MinAxis & Beam(Maj) & Beam(Min) & Reference\\
(GHz)    &(mJy)                        &${''}$   &${''}$   &${''}$    & ${''}$   &          \\
\hline
95.0791 & 5.944 & 1.811 & 1.493 & 1.77 & 1.47 & ALMA-12m/2015.1.01271.S,  \\
95.0828 & 4.911 & 0.632 & 0.50 & 0.62 & 0.49 &  ALMA-12m/2015.1.01271.S  \\
225.157 & 26.81 & 0.627 & 0.532 & 0.62 & 0.52 & ALMA-12m/2018.1.01113.S  \\
241 & 63 & -- & -- & 3.0 & 2.2 &  SMA/Hirano et al.\,2004 \\
274.6 & 108 & -- & -- & 18 & 18 & JCMT-UKT14/van der Veen\,1995 \\ 
335 & 105 & -- & -- & 2.0 & 1.3 & SMA/Hirano et al.\,2004  \\
338.194 & 52.18 & 0.460 & 0.374 & 0.45 & 0.36 & ALMA-12m/2018.1.01113.S  \\
337.927 & 83.23 & 4.629 & 2.781 & 4.62 & 2.62 & ALMA-ACA/2017.1.00917.S  \\
338.208 & 79.89 & 4.74 & 2.8 & 4.67 & 2.67    & ALMA-ACA/2018.1.00682.S \\
352.941 & 310 & -- & -- & 13 & 13 & JCMT--SCUBA/Di Francesco et al.\,2008  \\
666.666 & 650 & -- & -- & 7 & 7   & JCMT--SCUBA/Di Francesco et al.\,2008 \\
\hline
\end{tabular}
\end{table}
\tablenotetext{1}{The non-ACA ALMA observations are missing flux at scales signficantly larger than the beam.}


\begin{table}[!h]
\vskip 1in
\caption{Parameters (and best-fit values) of the model DUDE\label{dudeparms}}
\centering
\begin{tabular}{cccc}
\hline\hline
& Parameter & value & unit \\
\hline
geometry & $r_0$, $r_1$, $r_2$, $r_3$ & 160, 560, 1040, 1740 & au \\
 & $h_1$ & 350 & au \\
 & $i$ & 40\arcdeg \\
 & opening angle & 39\arcdeg \\
\hline
density & $C_1$ & 0.1& \\
& $n_1$ & $1.35\times10^5$ & cm$^{-3}$ \\
 & $\beta_n$ & 1.6 & \\
 & $r_s$ & 100 & au \\
 & $\sigma_0$, $\sigma_1$, $\sigma_2$, $\sigma_3$ & 51, 102, 153, 204 & au \\
\hline
temperature & $T_1$ & 63 & K \\
 & $\beta_T$ & 1 & \\
 & $\sigma_{T1}$,  $\sigma_{T2}$,  $\sigma_{T2}$,  $\sigma_{T3}$& 120, 240, 360, 480 & au \\
\hline
velocity & $v_{\rm {sys}}$ & \vlsr=17.4 & km s$^{-1}$\\
 & $v_0$, $v_D$ & 9.5, 11.5 & km s$^{-1}$ \\
 & $v_{t0}$, $v_{t1}$ & 4, 2 & km s$^{-1}$ \\
\hline
abundance &$\left[^{12}{\rm {CO}}\right]$/$\left[{\rm H_2}\right]$ & $10^{-3}$ & \\
& $\left[^{13}{\rm {CO}}\right]$/$\left[{\rm H_2}\right]$ & $1.4\times10^{-5}$ & \\
\hline

\end{tabular}
\end{table}

\clearpage


%

\begin{figure}[ht!]
 \centering
 \subfloat{\includegraphics[width=1.0\linewidth]{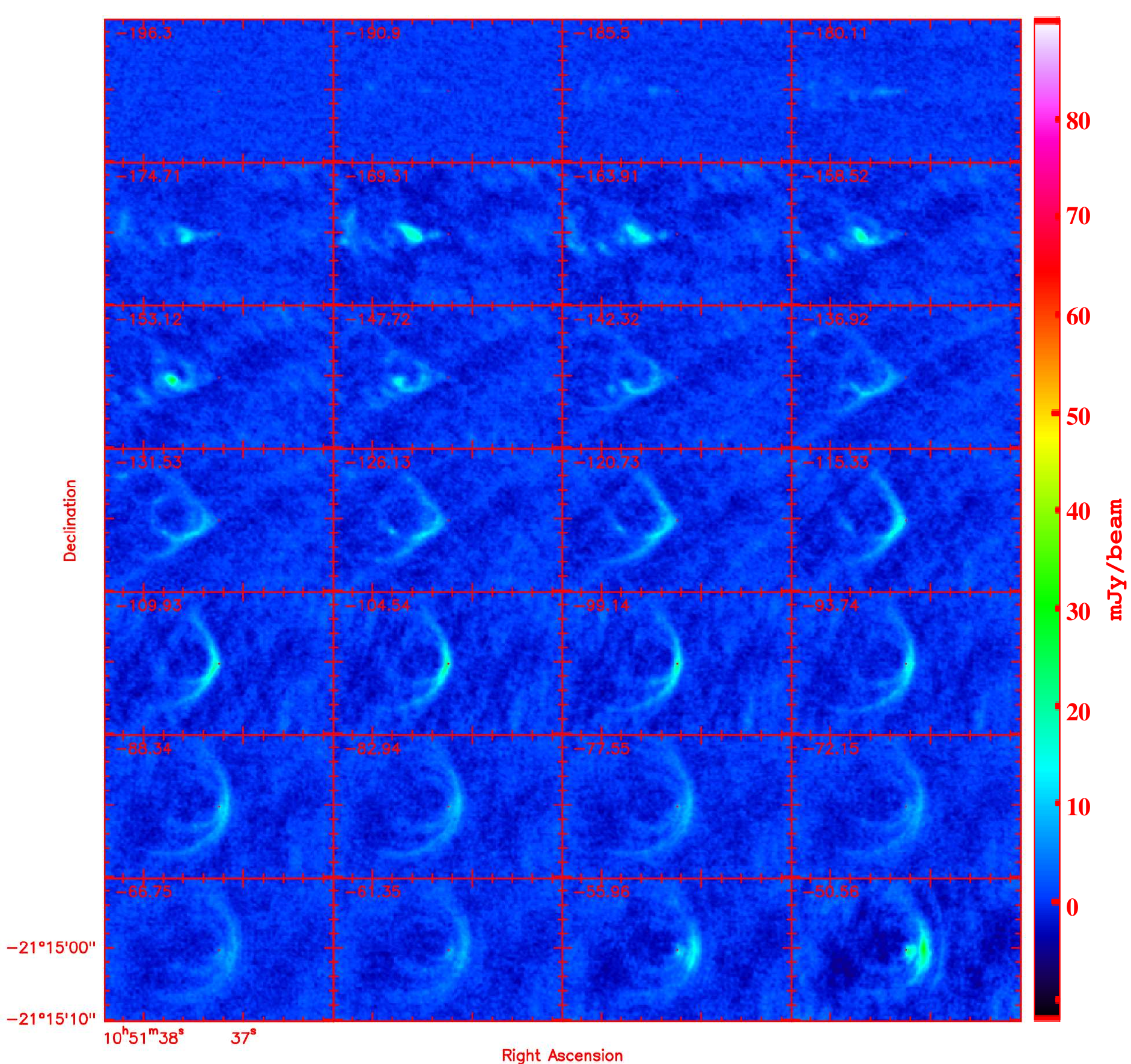}}
 \caption{(a) Channel/velocity maps of the blue-shifted (\vlsr=-196 to -51\,\kms) \codos\,\jdos~emission from V\,Hya. The spectral resolution has been reduced in these maps, such that each channel is $5.4\,\kms$~wide, in order to reveal the very faint emission.} 
\end{figure}

\begin{figure}
 \ContinuedFloat
\captionsetup{list=off,format=cont}
 \centering
 \subfloat{\includegraphics[width=1.0\linewidth]{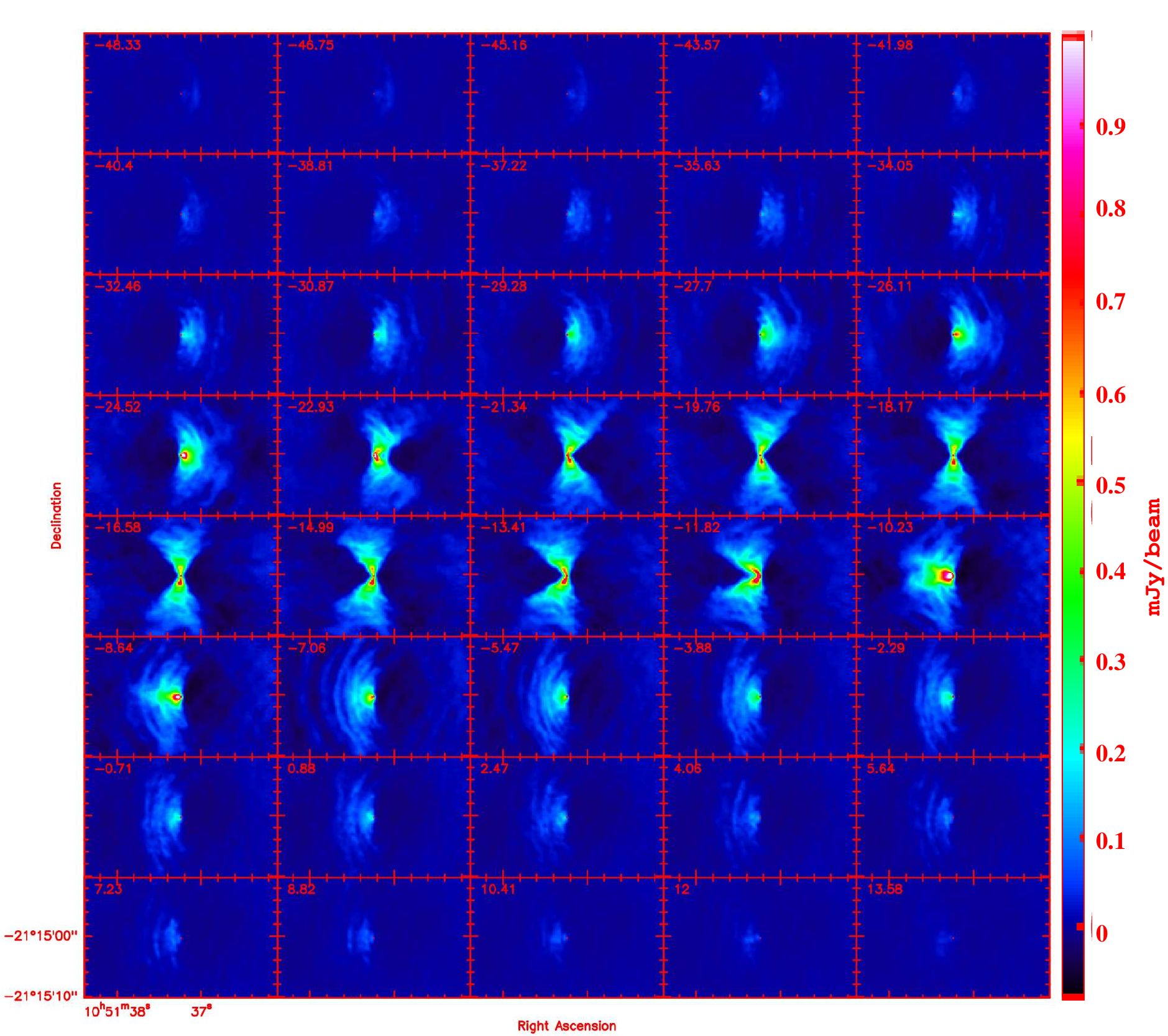}}
 \caption{(b) Channel/velocity maps of the \codos\,\jdos~emission from V\,Hya at and around the systemic velocity (\vlsr=-48 to 14\,\kms). The spectral resolution has been reduced in these maps, such that each channel is $1.6\,\kms$~wide.}
 \end{figure}

\begin{figure}
 \ContinuedFloat
 \captionsetup{list=off,format=cont}
 \centering
 \subfloat{\includegraphics[width=1.0\linewidth]{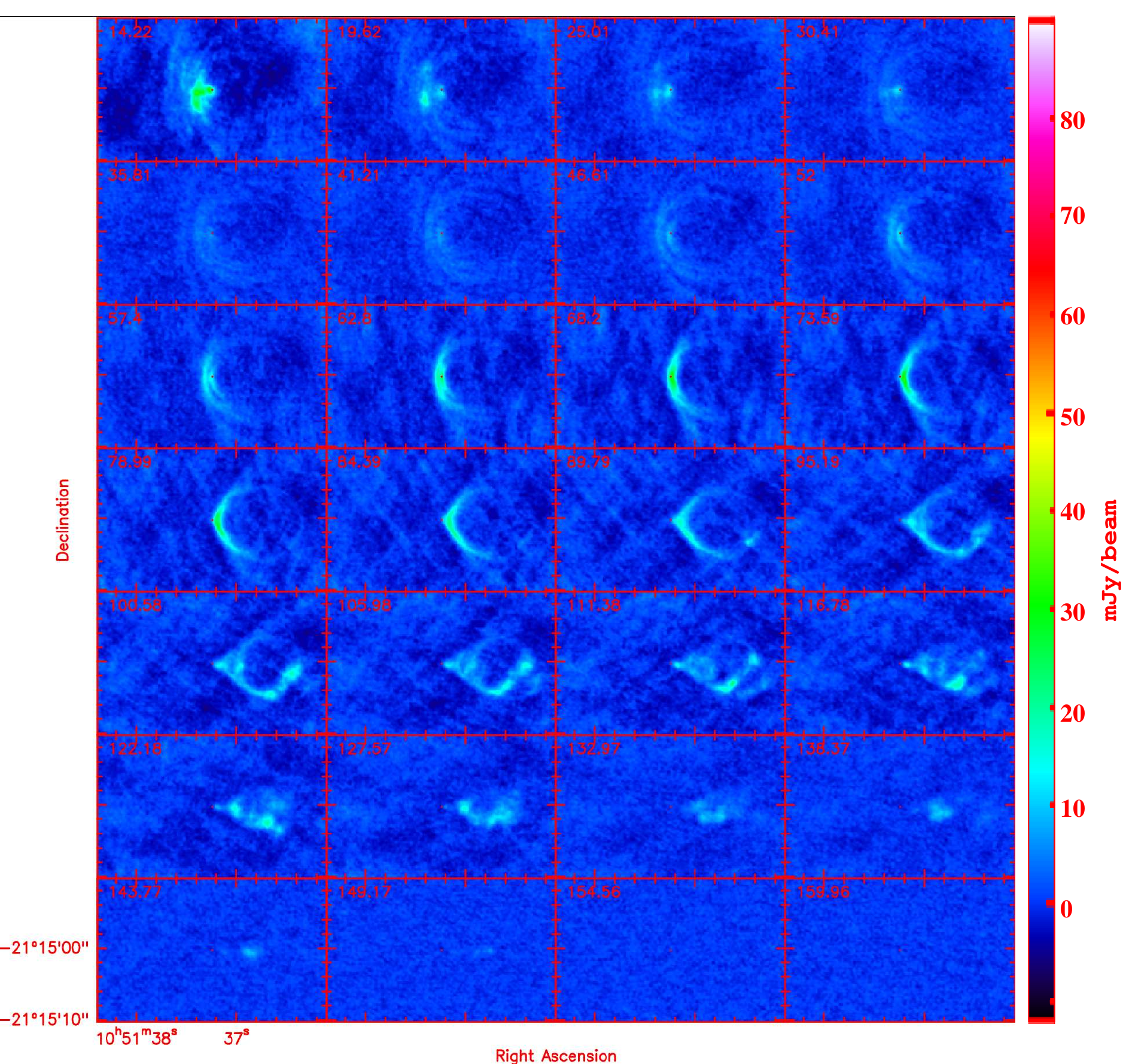}}
 \caption{(c) Channel/velocity maps of the red-shifted (\vlsr=14 to 160\,\kms)\codos\,\jdos~emission from V\,Hya. The spectral resolution has been reduced in these maps, such that each channel is $5.4\,\kms$~wide, in order to reveal the very faint emission.}
 \label{12co21-hivel}
 \end{figure}

\begin{figure}[ht!]
 \subfloat{\includegraphics[width=1.0\textwidth]{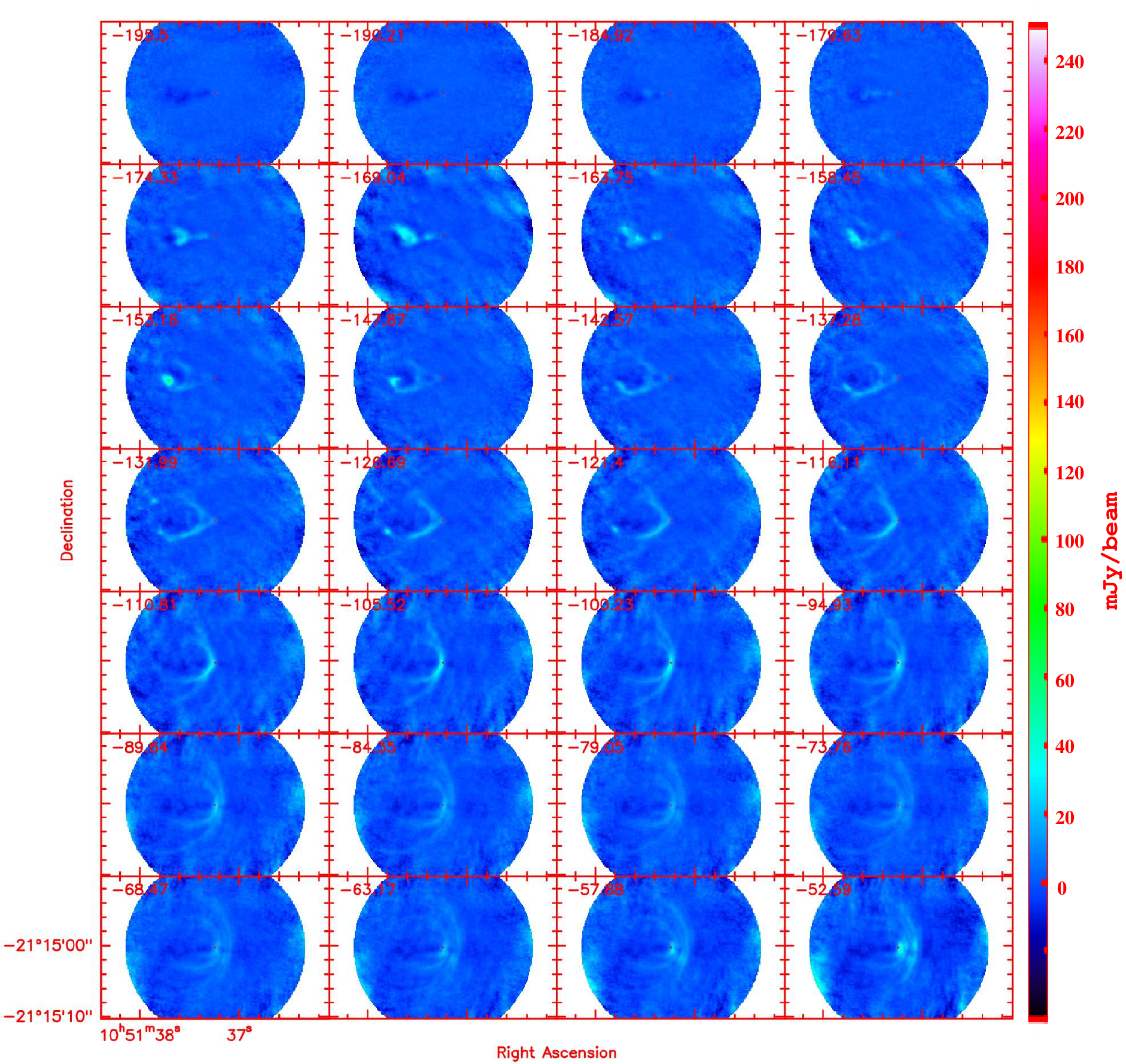}}
 \caption{(a) Channel/velocity maps of the blue-shifted (\vlsr=-196 to -53\,\kms) \codos\,\jtres~emission from V\,Hya. The spectral resolution has been reduced in these maps, such that each channel is $5.3\,\kms$~wide, in order to reveal the very faint emission.  The field-of-view is the same as in Fig.\,\ref{12co21-hivel}.} 
\end{figure}

\begin{figure}[ht!]
 \ContinuedFloat 
 \captionsetup{list=off,format=cont}
 \subfloat{\includegraphics[width=1.0\textwidth]{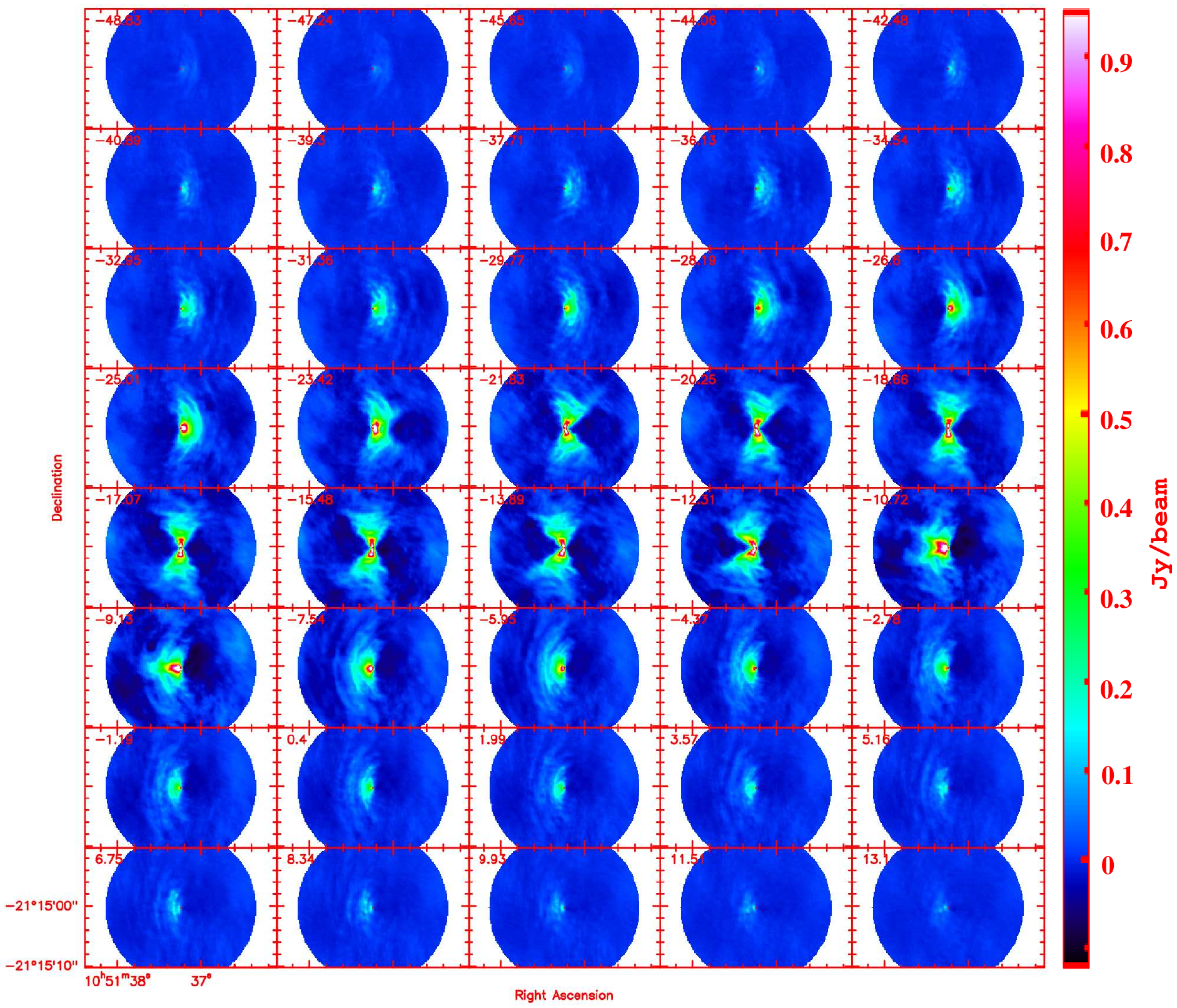}}
 \caption{(b) Channel/velocity maps of the \codos\,\jtres~emission from V\,Hya at and around the systemic velocity (\vlsr=-49 to 13\,\kms). The spectral resolution has been reduced in these maps, such that each channel is $1.6\,\kms$~wide.  The field-of-view is the same as in Fig.\,\ref{12co21-hivel}.}
\end{figure}

\begin{figure}
 \ContinuedFloat
 \captionsetup{list=off,format=cont}
 \centering
 \subfloat{\includegraphics[width=1.0\linewidth]{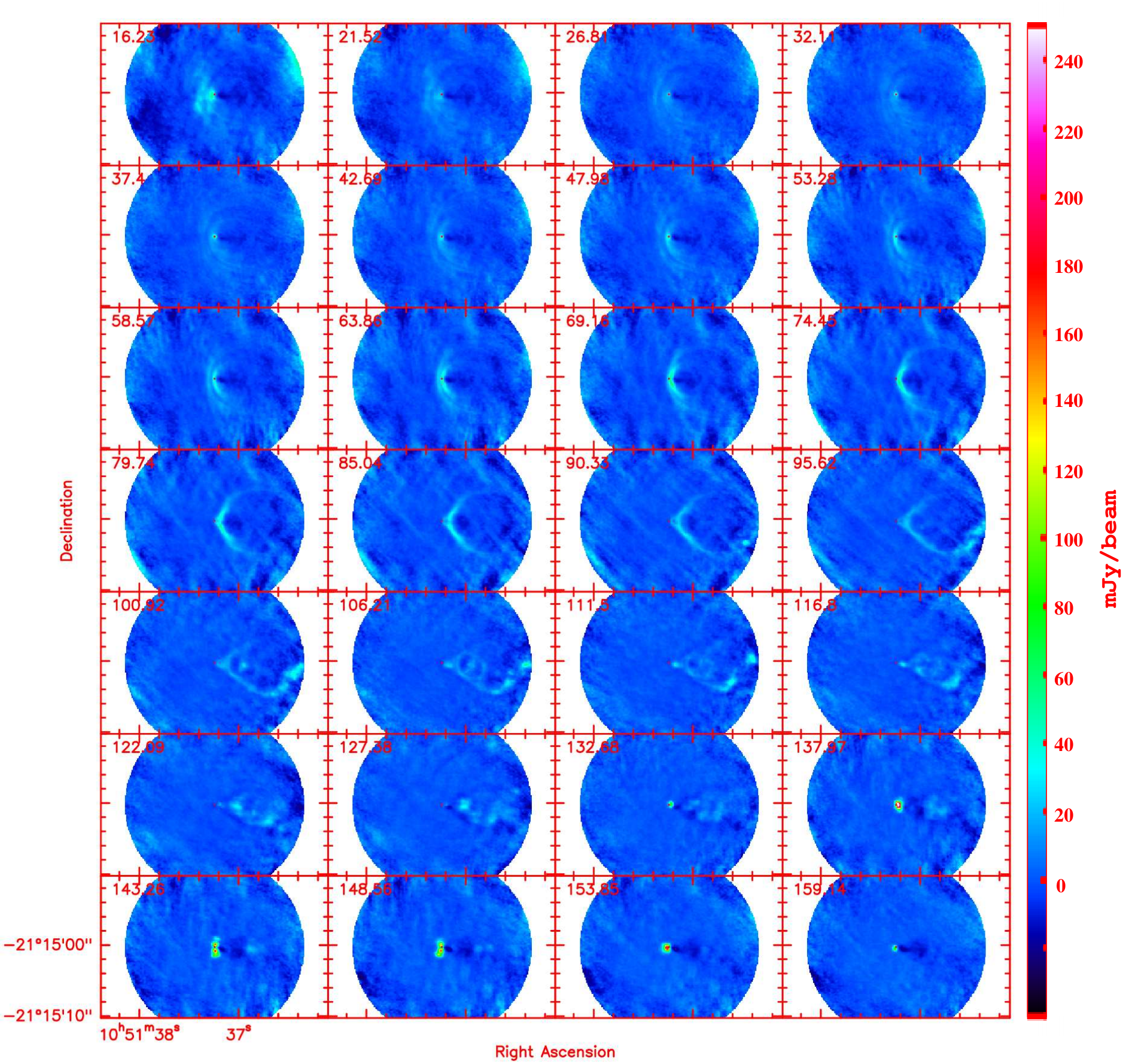}}
 \caption{(c) Channel/velocity maps of the red-shifted (\vlsr=16 to 159\,\kms)\codos\,\jtres~emission from V\,Hya. The spectral resolution has been reduced in these maps, such that each channel is $5.4\,\kms$~wide, in order to reveal the very faint emission.  The field-of-view is the same as in Fig.\,\ref{12co21-hivel}.}
 \label{12co32-hivel}
 \end{figure}

\begin{figure}[ht!]
\includegraphics[width=0.6\textwidth]{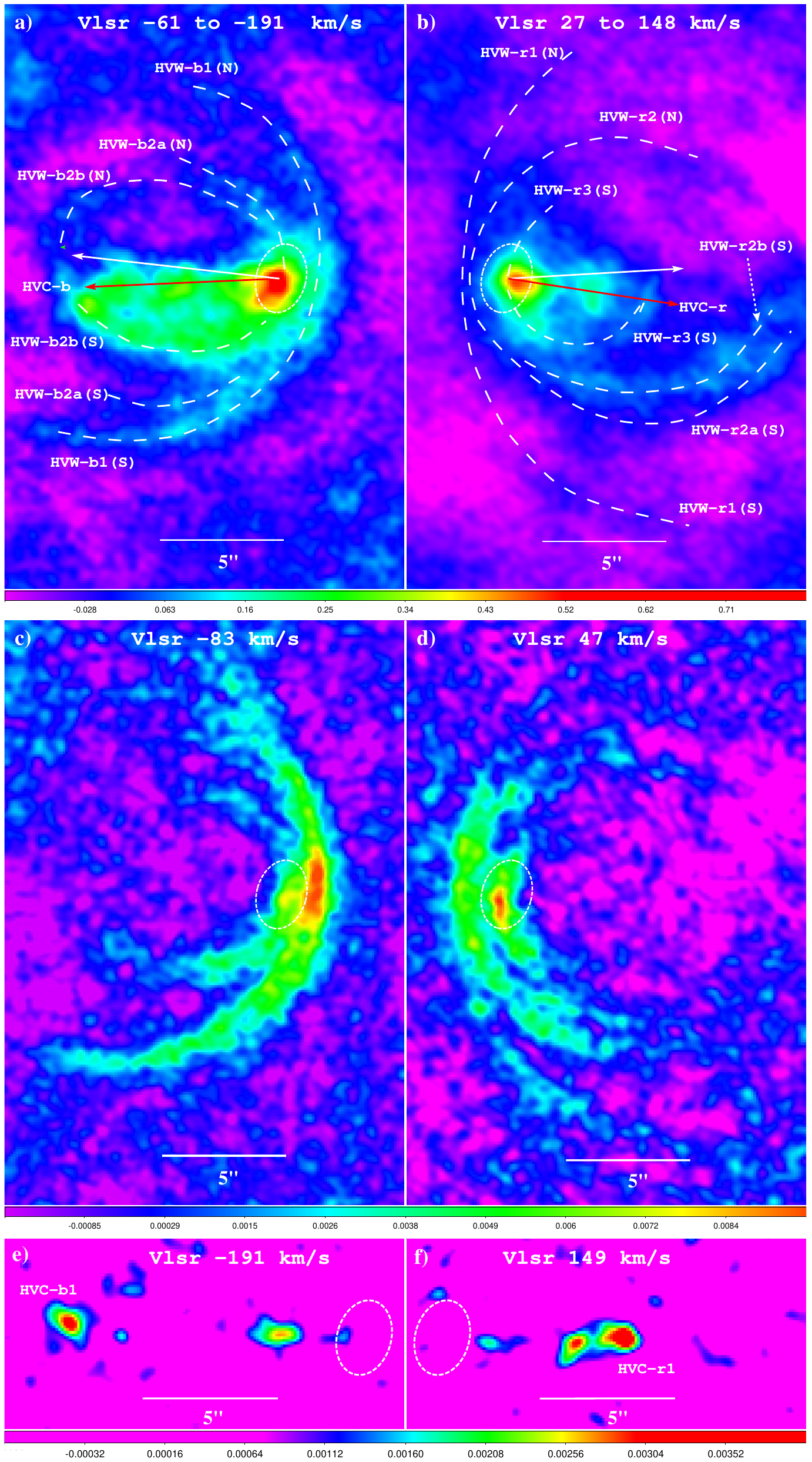}
\vspace{-0.01cm}
\caption{Maps of the \codos\jdos~emission from V\,Hya at specific blue- and red-shifted velocities, chosen (roughly) symmetrically about the systemic velocity, showing the structure of the HV outflows. The blue-shifted and red-shifted emission, integrated over a wide range of expansion velocities (panels a,b) shows two types of outflows, several with wide-opening angles (labelled with prefixes HVW-b and HVW-r) and two that are highly collimated (HVC-b and HVC-r). The HVW outflows are dominant at intermediate expansion velocities (panels c,d), whereas the HVC outflows are dominant at the highest expansion velocities (panels e,f). Dashed white arcs delineate the multiple components in the HVW outflows -- these are fits-by-eye to the observed structures using the images in panels a, b, c, and d. White (red) vectors show the symmetry axes for the parabolic HVW outflows (the collimated HVC outflows.) The spectral resolution of the orginal \codos\jdos~datacube was first reduced to $5.4\,\kms$, and then the images  in panels a,b,c,d and panels e,f were smoothed to reduce the noise (using ds9, with Gaussian kernels of r=1 and 9 pixels, respectively; each pixel is $0\farcs1$). The dashed white ellipse in each panel represents ring R1, for reference. The intensity scale at the bottom of each pair of side-by-side panels is in Jy\,\kms\,beam$^{-1}$ for panels a,b and Jy\,beam$^{-1}$ for panels c--f. The size of panels a--d is  $16\farcs1\times23\farcs5$, and the size of panels e and f is $14\farcs7\times6\farcs9$.
}
\label{12co21-arc-hivel}
\end{figure}

\begin{figure}[ht!]
\includegraphics[width=1.0\textwidth]{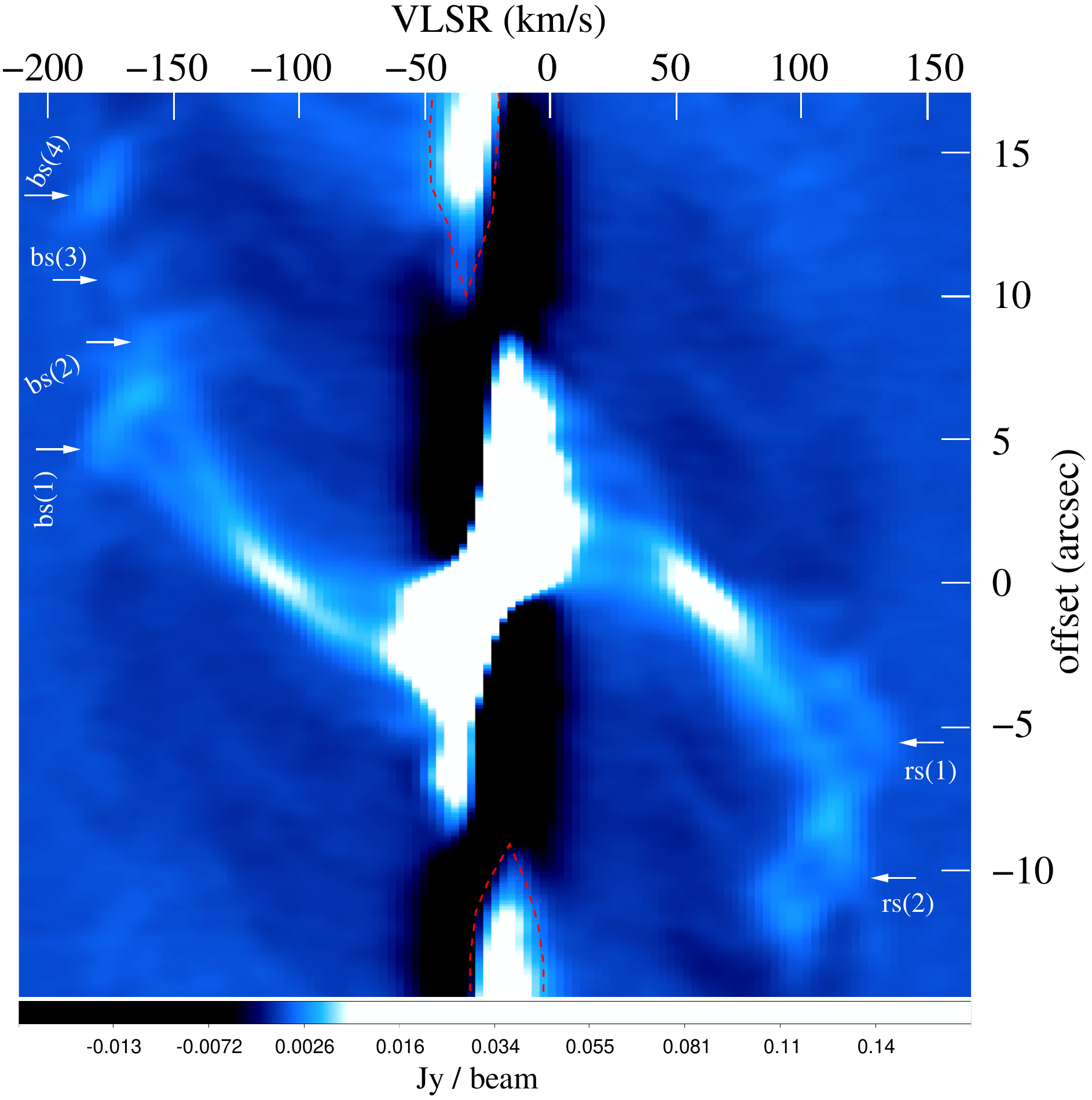}
\vspace{-0.5cm}
\caption{Position-velocity intensity cut of the \codos~\jdos~emission, taken along the E-W axis (positive offsets are towards the East). The width of the cut is $8\arcsec$. The spectral resolution has been reduced to 3.2\,\kms, in order to reveal the faint high-velocity outflow emission. 
Blue and red-shifted shock-like structures, (bs(1)--bs(4) \& rs(1)--rs(2), respectively) are marked with white arrows.
The saturated S-shaped structure at low velocities in and near the center (around the systemic velocity) is due to emission from the expanding DUDE. The dashed red polygons show regions with emission artefacts due to incomplete $uv$-coverage. The beam is $0\farcs62\times0\farcs52$, $PA=70\arcdeg.6$.
}
\label{12co21-pv-ew}
\end{figure}

\begin{figure}[ht!]
\includegraphics[width=0.5\textwidth]{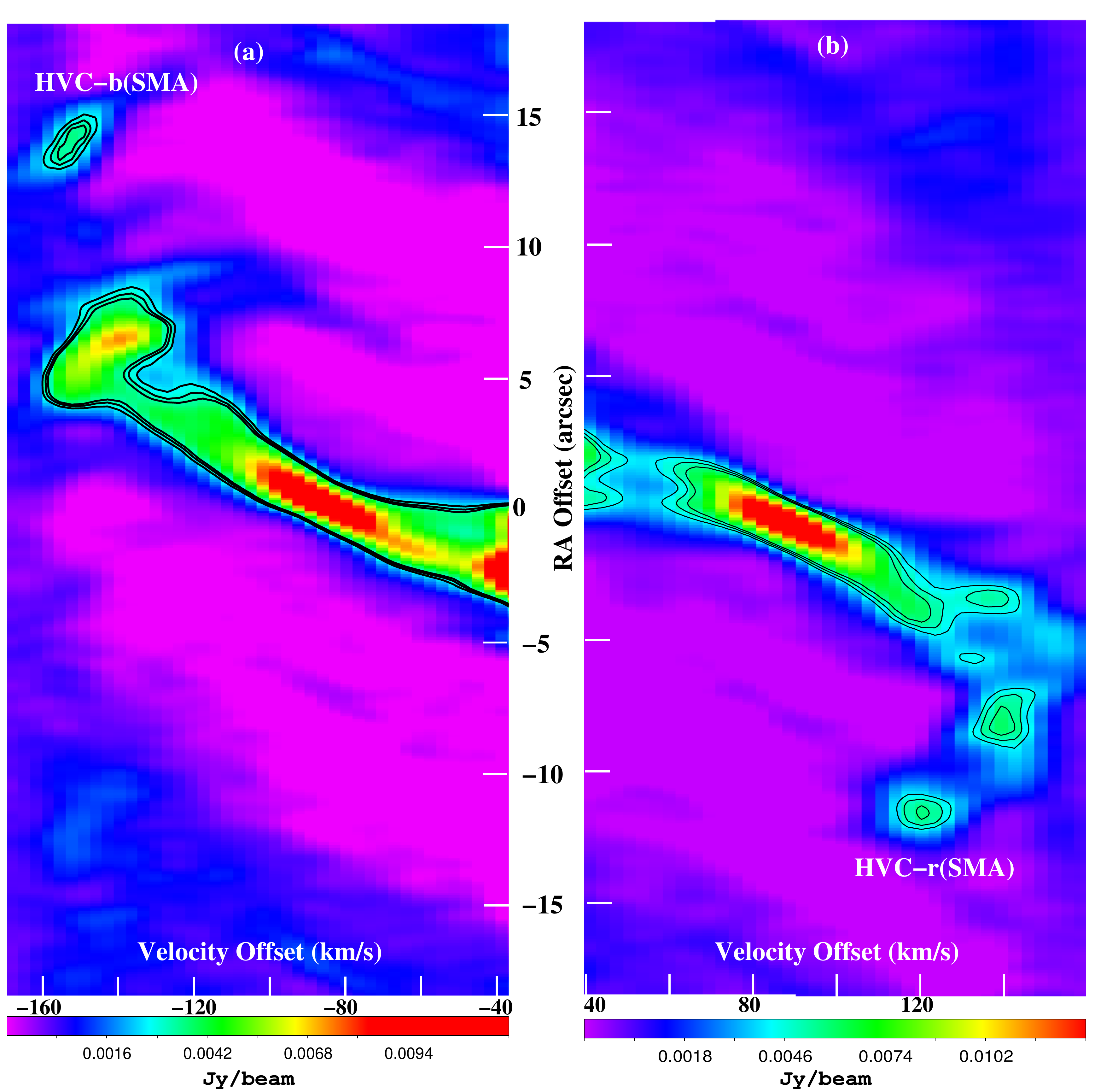}
\vspace{0.01cm}
\caption{Position-velocity intensity cut of the \codos~\jdos~emission, taken along the E-W axis (positive offsets are towards the East), showing the blobs HV-b1 (panel a) and HV-r1 (panel b), used to determine proper motion by comparison with the images of these blobs obtained with the SMA in 2003 (see Fig.\,2 of Hirano et al. (2004)). The width of the cut is $5\arcsec$. Each spaxel is $3.2$\,\kms$\times0\farcs1$, and the image has then been smoothed (in ds9 with a Gaussian kernel r=3 pixels). Velocity (spatial) offsets are relative to the systemic velocity of V\,Hya. Contour levels are 0.0026, 0.0030, and 0.0033, Jy\,beam$^{-1}$ in panel a, and 0.0041, 0.0048 and 0.0055 Jy\,beam$^{-1}$ in panel b.
}
\label{blob-propmotion}
\end{figure}

\begin{figure}[ht!]
\includegraphics[width=0.5\textwidth]{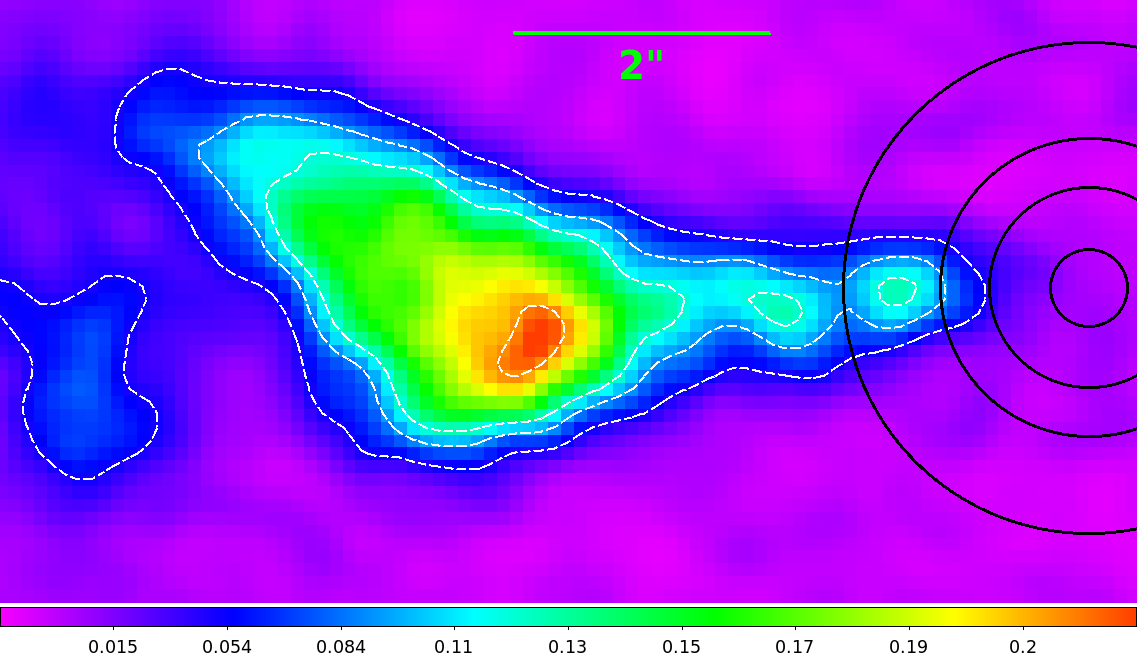}
\vspace{0.01cm}
\caption{Map of the \codos\jdos~emission from V\,Hya over the velocity range $\vlsr=-166$ to $-191$\,\kms, showing the presence of compact clumps that may correspond to the bullet-like high-velocity ejections observed in V\,Hya. The black circles are centered at the location of V\,Hya's primary star, and have radii of 1\farcs9, 1\farcs2, 0\farcs8, and 0\farcs3, which are the expected radial offsets of bullets 0, 1, 2, and 3 at the epoch of the ALMA observations, assuming that they have been moving with the tangential velocities last measured for them (in the SSM16 HST study). Contour levels are  0.05, 0.09, 0.12, and 0.21 Jy\,\kms\,beam$^{-1}$.
}
\label{blob-bullets}
\end{figure}

\begin{figure}[ht!]
\vspace{0.1in}
\includegraphics[width=1.0\textwidth]{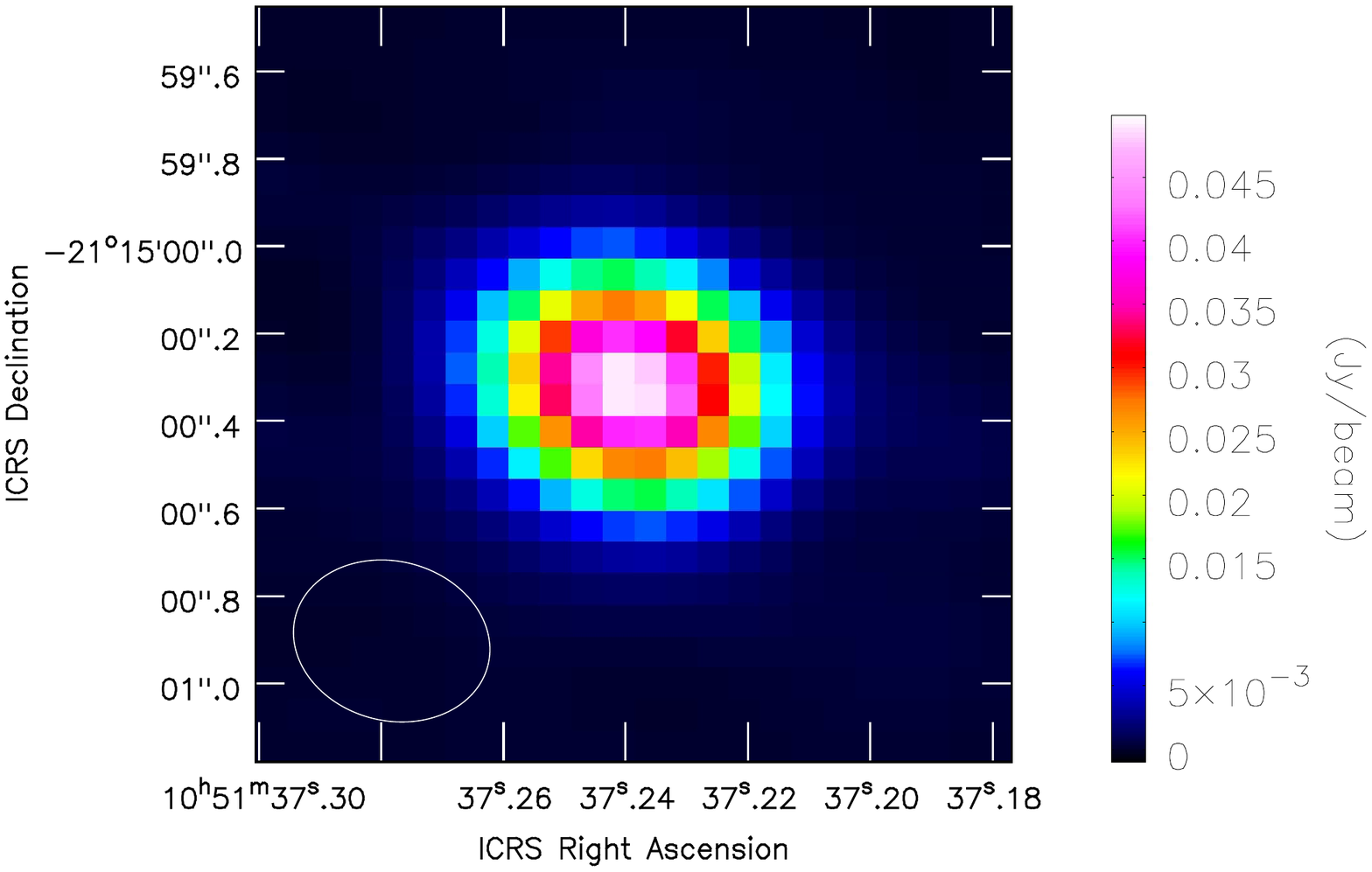}
\vspace{-5.0in}
\caption{The band 7 continuum emission from V\,Hya (beam is $0\farcs454\times0\farcs365$, $PA=75\arcdeg.3$)
}
\label{cont-bnd7}
\end{figure}

\begin{figure}[ht!]
\includegraphics[width=0.5\textwidth]{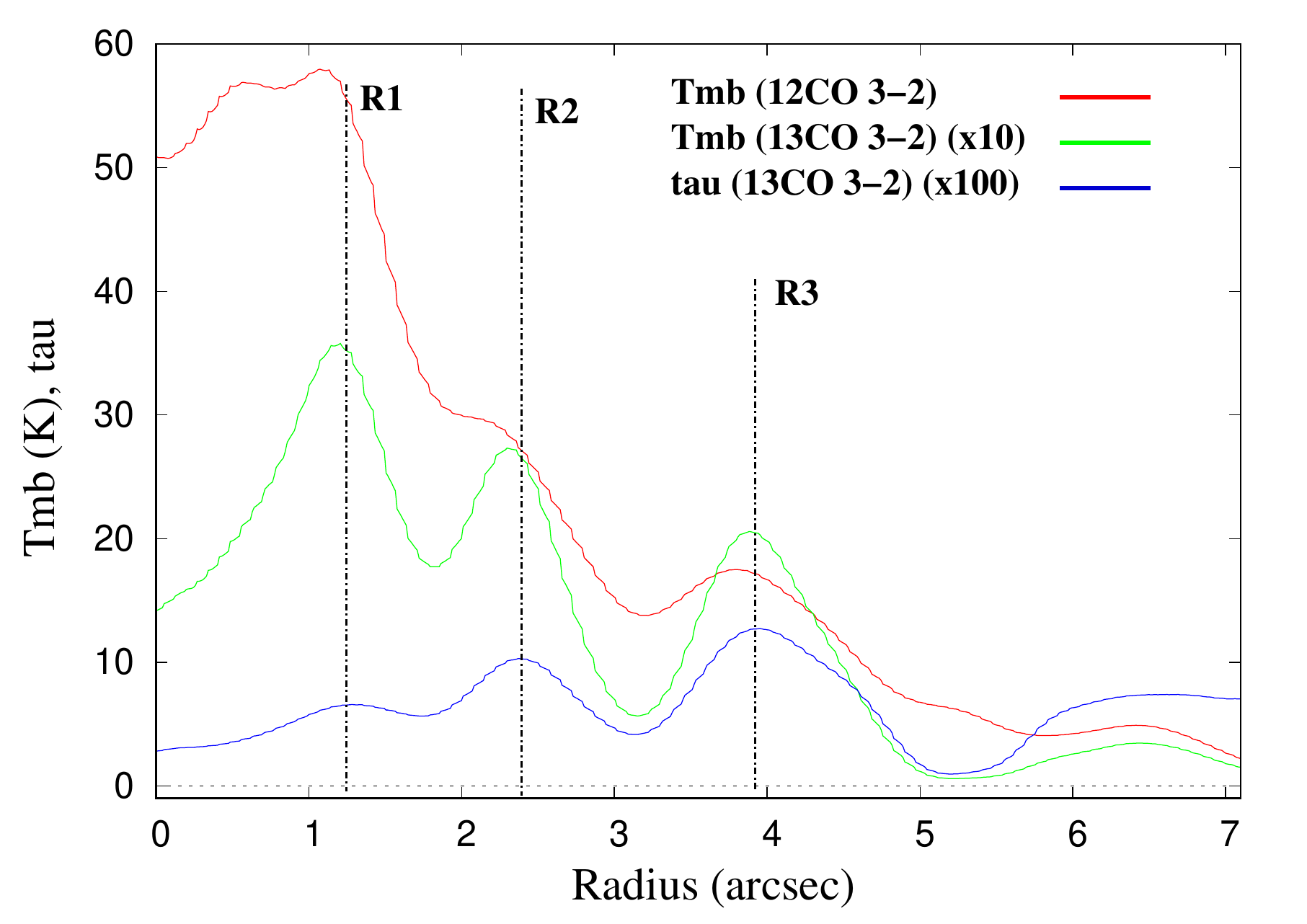}
\includegraphics[width=0.5\textwidth]{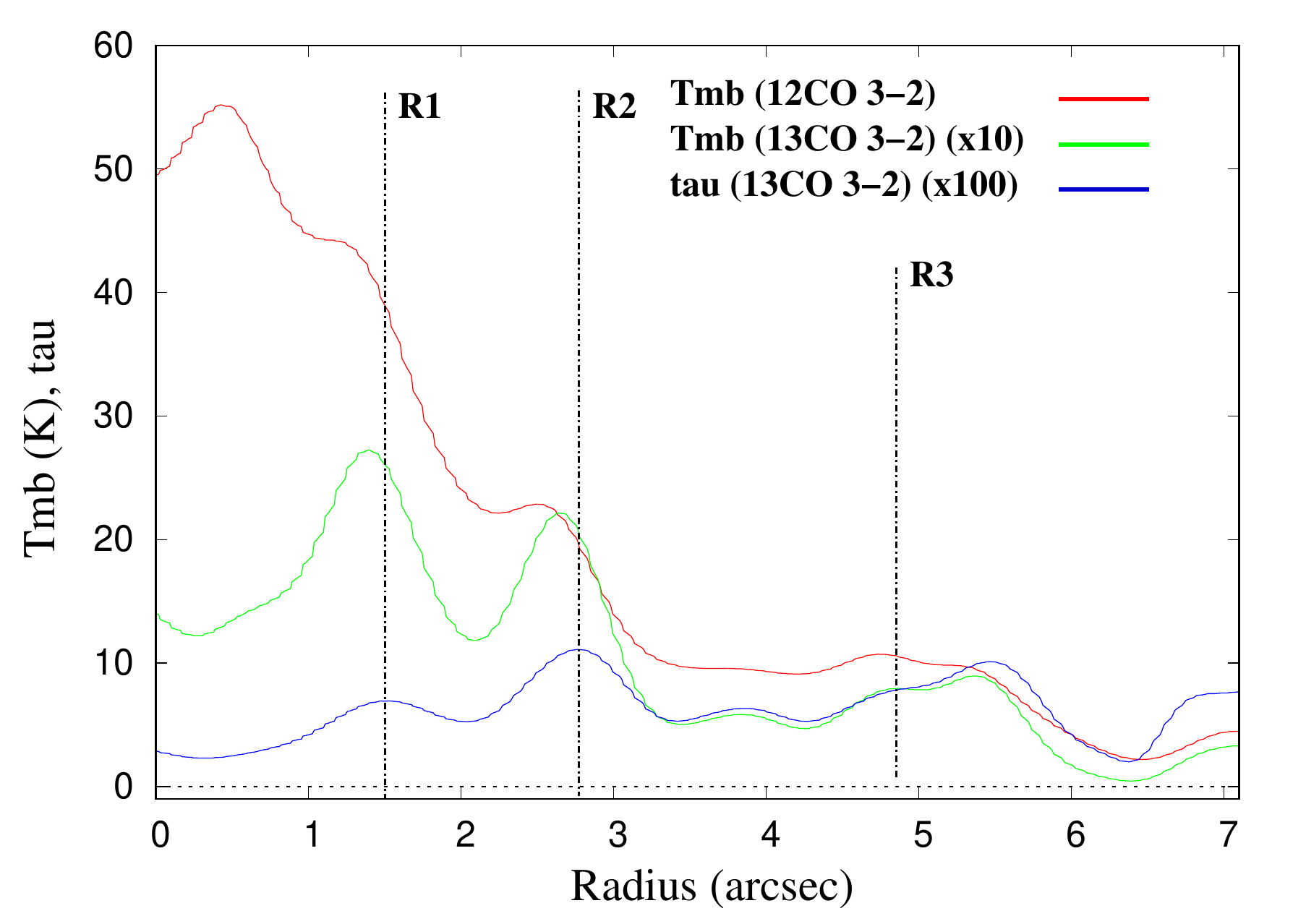}
\vspace{0.0cm}
\caption{Observationally-derived optical depth of the \cotres\,\jtres~line  at the systemic velocity, as a function of radius in the DUDE, and the brightness temperatures of the \cotres~and \codos\,\jtres~lines, averaged over a $2\arcdeg$~angular wedge centered on the primary star in V\,Hya, and oriented along the average PA of the disk (left) to the south, $PA=174\arcdeg$ to $176\arcdeg$, (right) to the north, $PA=-4\arcdeg$ to $-6\arcdeg$. The local peaks in the optical depth represent the radial locations of the centers of rings R1, R2, and R3. The maximum optical depth due to each ring is located at different radial offsets in the two panels, because the center for the cuts (the location of V\,Hya's central star) does not coincide with the centers of the rings. In the north cut, R3, does not display a clear peak because of confusion with the nearby arc structure, A3, which has a slightly larger radius than R3. The \cotres\,and \codos~images have been convolved to a common beam.
}
\label{13co32tau}
\end{figure}
\begin{figure}[ht!]
\includegraphics[width=1.0\textwidth]{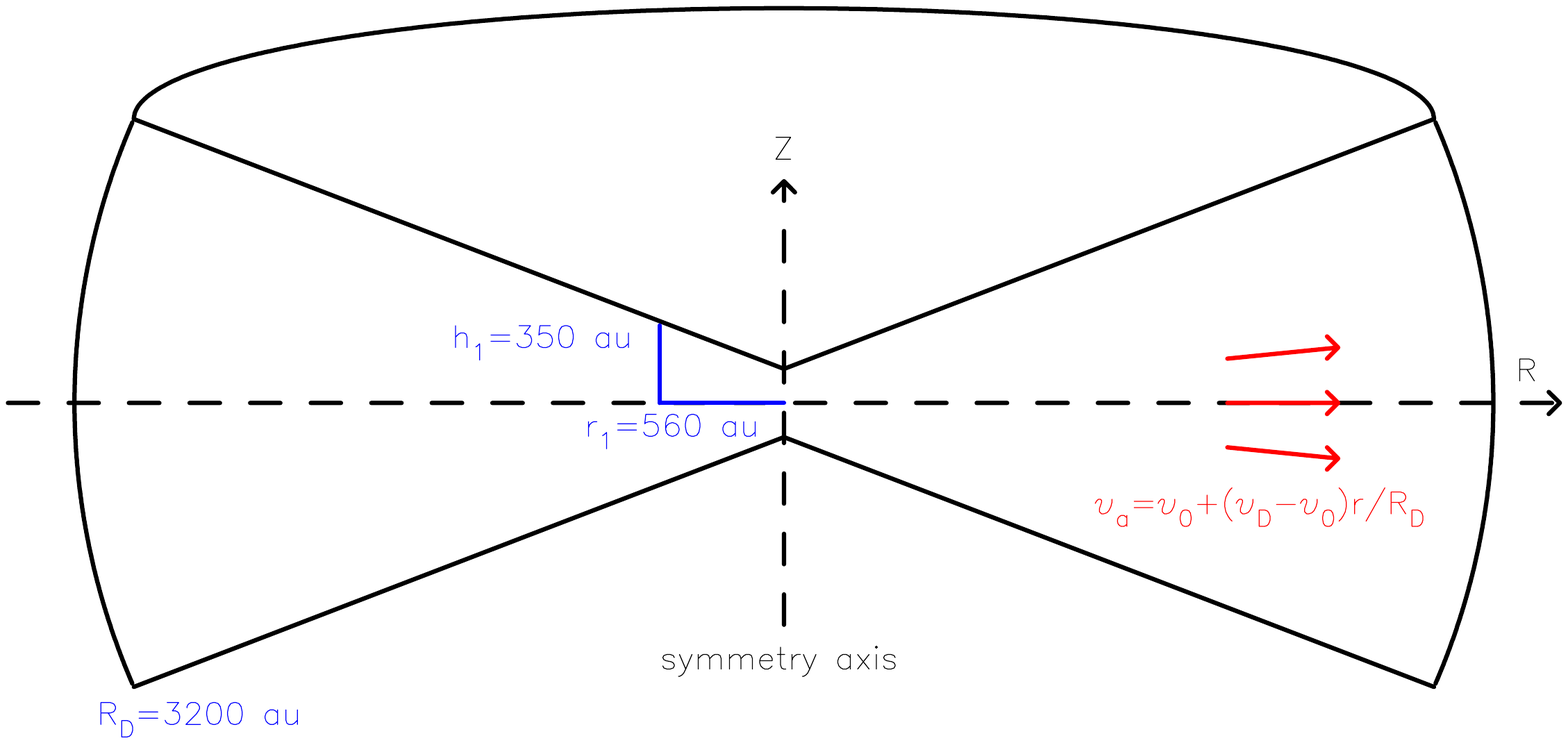}
\caption{Schematic of the model DUDE, showing its spatio-kinematic structure.
}
\label{schem-dude}
\end{figure}

\begin{figure}[ht!]
\includegraphics[width=1.0\textwidth]{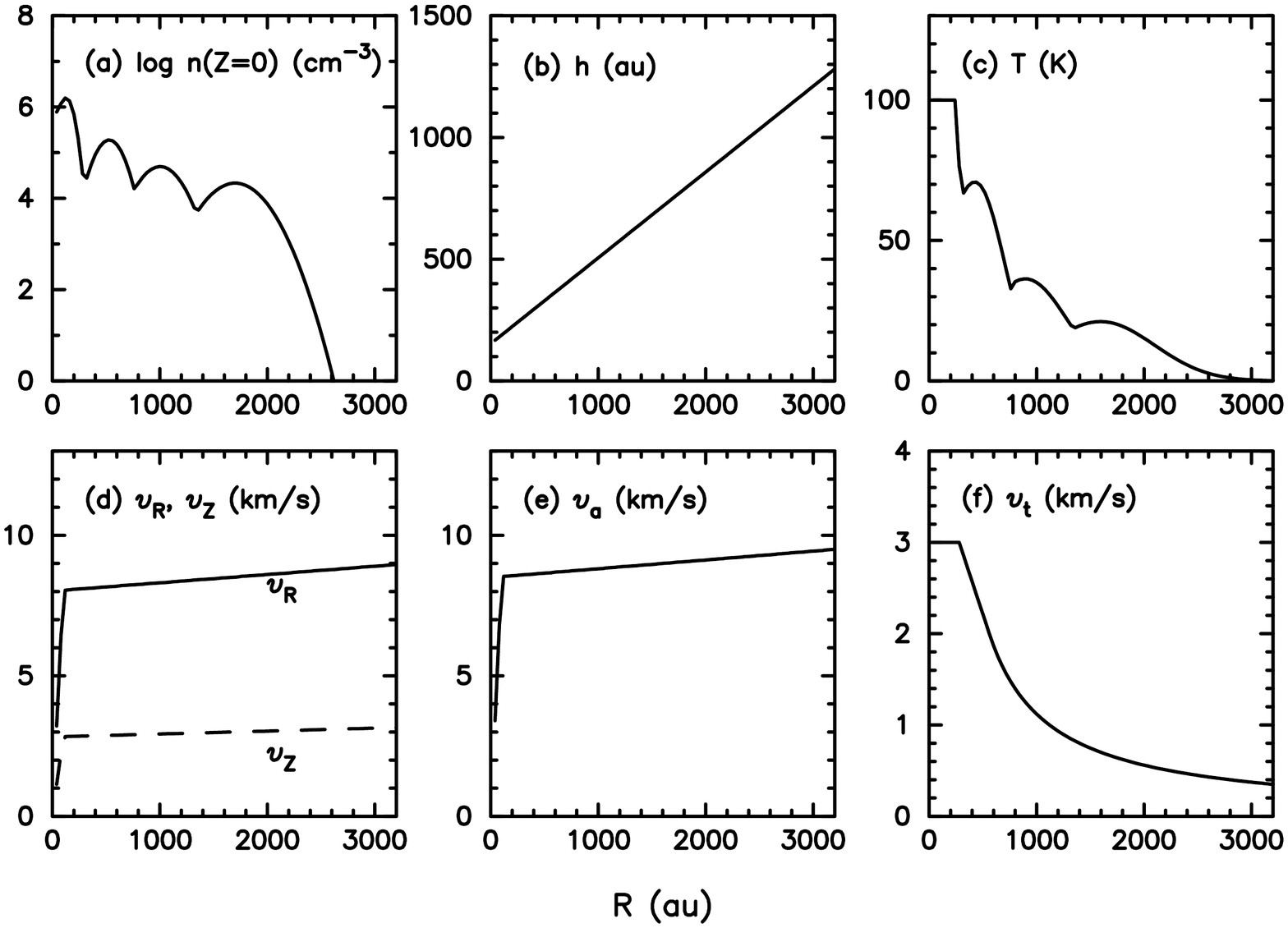}
\caption{The radial variation of the fundamental physical parameters characterising the model DUDE.
}
\label{mod-parms}
\end{figure}

\begin{figure}[ht!]
\includegraphics[width=1.0\textwidth]{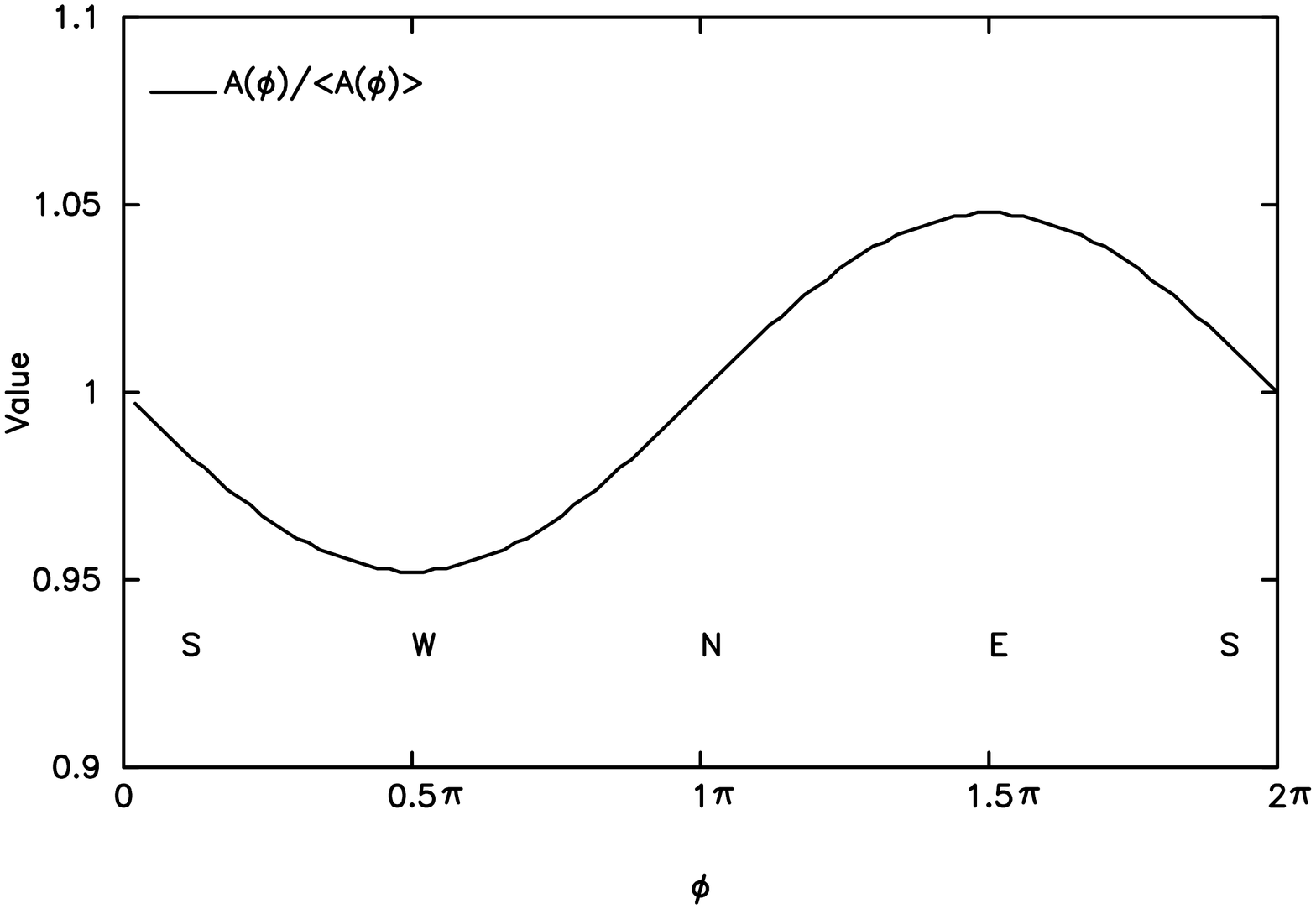}
\caption{The anisotropy function $A(\phi)$ in Eqn. 1 (see \S\,\ref{dude-model}).}.
\label{ab-phi}
\end{figure}

\begin{figure}[ht!]
\hskip -1.0in
\includegraphics[width=1.1\textwidth]{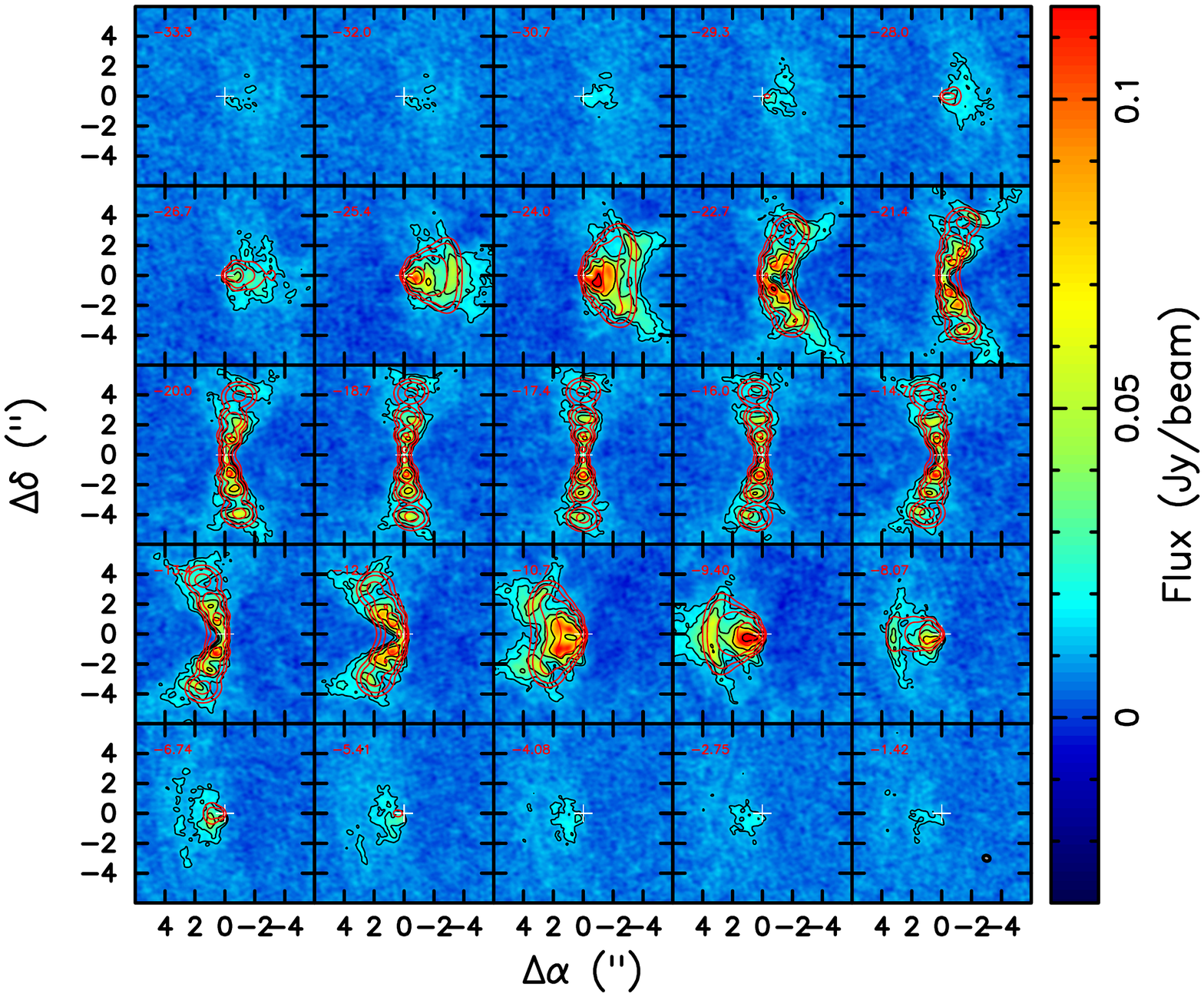}
\caption{Channel/velocity maps of the \cotres\jtres~emission derived from our best-fit spatio-kinematic model of the DUDE (orange contours) overlaid on the observed (colorscale) maps of \cotres\jtres~from the DUDE in V\,Hya. The model contours are 6, 12, 24, 48, ... mJy\,beam$^{-1}$ (beam is $0\farcs64\times0\farcs54$).
}
\label{mod-13co32-rings}
\end{figure}

\begin{figure}[ht!]

\includegraphics[width=0.75\textwidth]{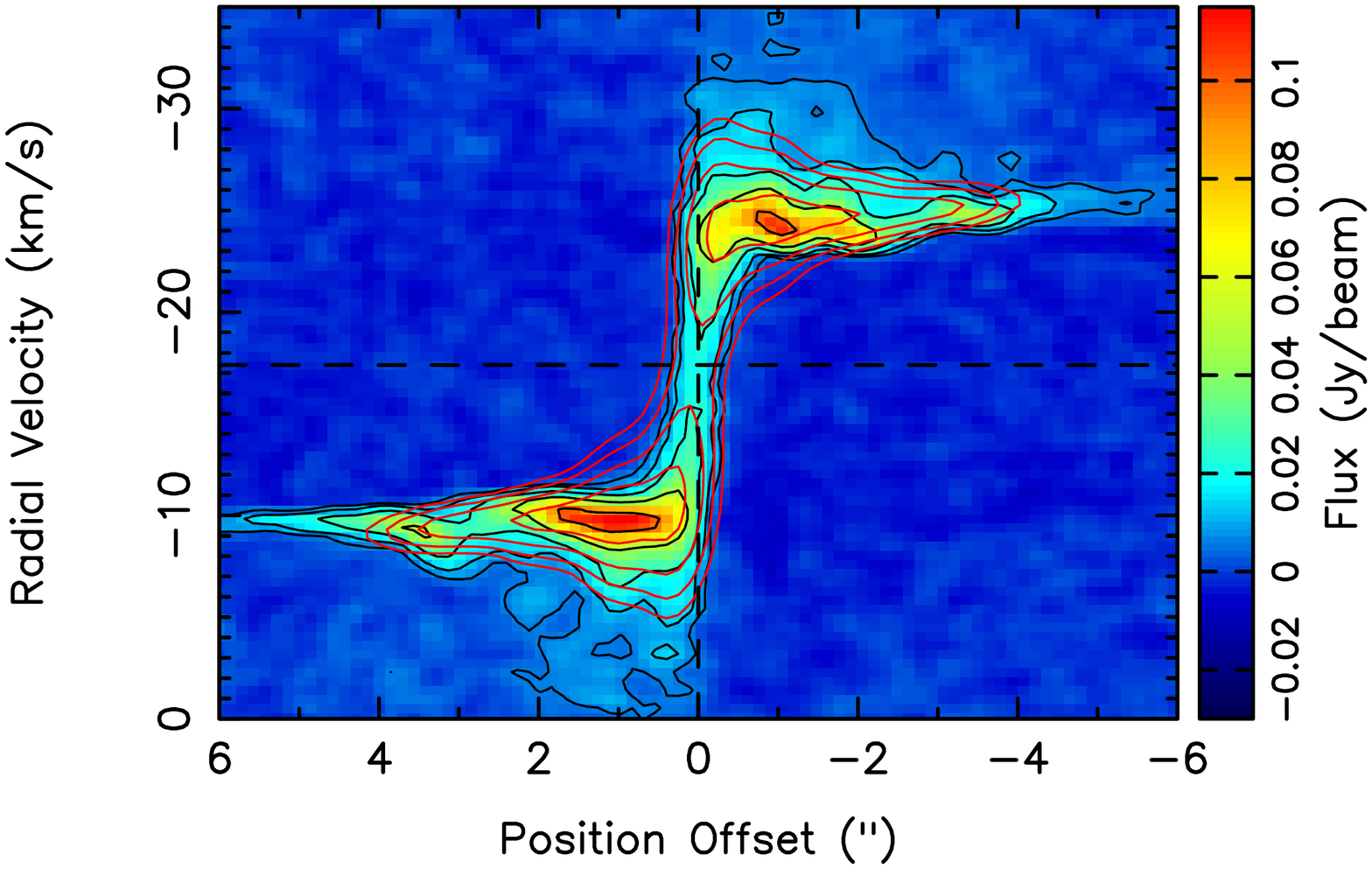}
\caption{Model position-velocity intensity (red contours) taken along the minor axis of the DUDE, overlaid on the observed (colormap and blue contours) \codos~\jtres~emission. The width of the cut is $0\farcs1$.
}
\label{mod-13co32-pv}
\end{figure}

\begin{figure}[ht!]
\includegraphics[width=0.75\textwidth]{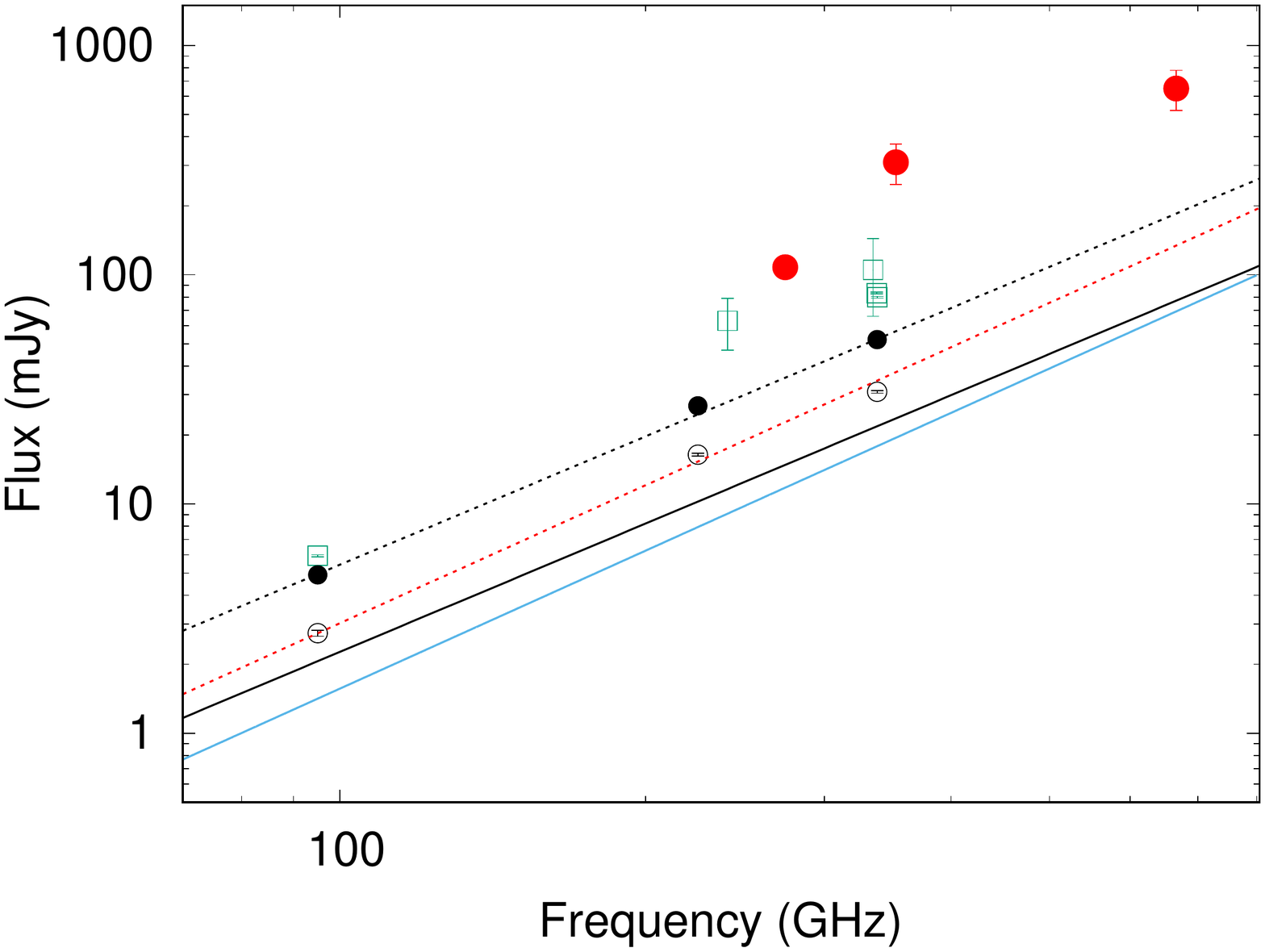}
\caption{Observed millimeter-wave flux densities for V\,Hya as measured with large beams ($7{''}-18{''}$: red circles), medium beams ($\sim1\farcs6-3\farcs5$: green squares) and small beams ($\sim0\farcs4-0\farcs6$: black circles), together with the expected flux density from (i) a standard radio photosphere (solid black line), (ii) radio photosphere scaled up by factor 2.4 (dashed black line) and (iii) the stellar photosphere (light blue line). The excess flux over the emission from a standard radio photosphere for the small-beam measurements (open black circles) is consistent with a $\nu^2$ power-law (dashed red line).}
\label{mod-cont-vhya}
\end{figure}

\begin{figure}[ht!]
\includegraphics[width=0.9\textwidth]{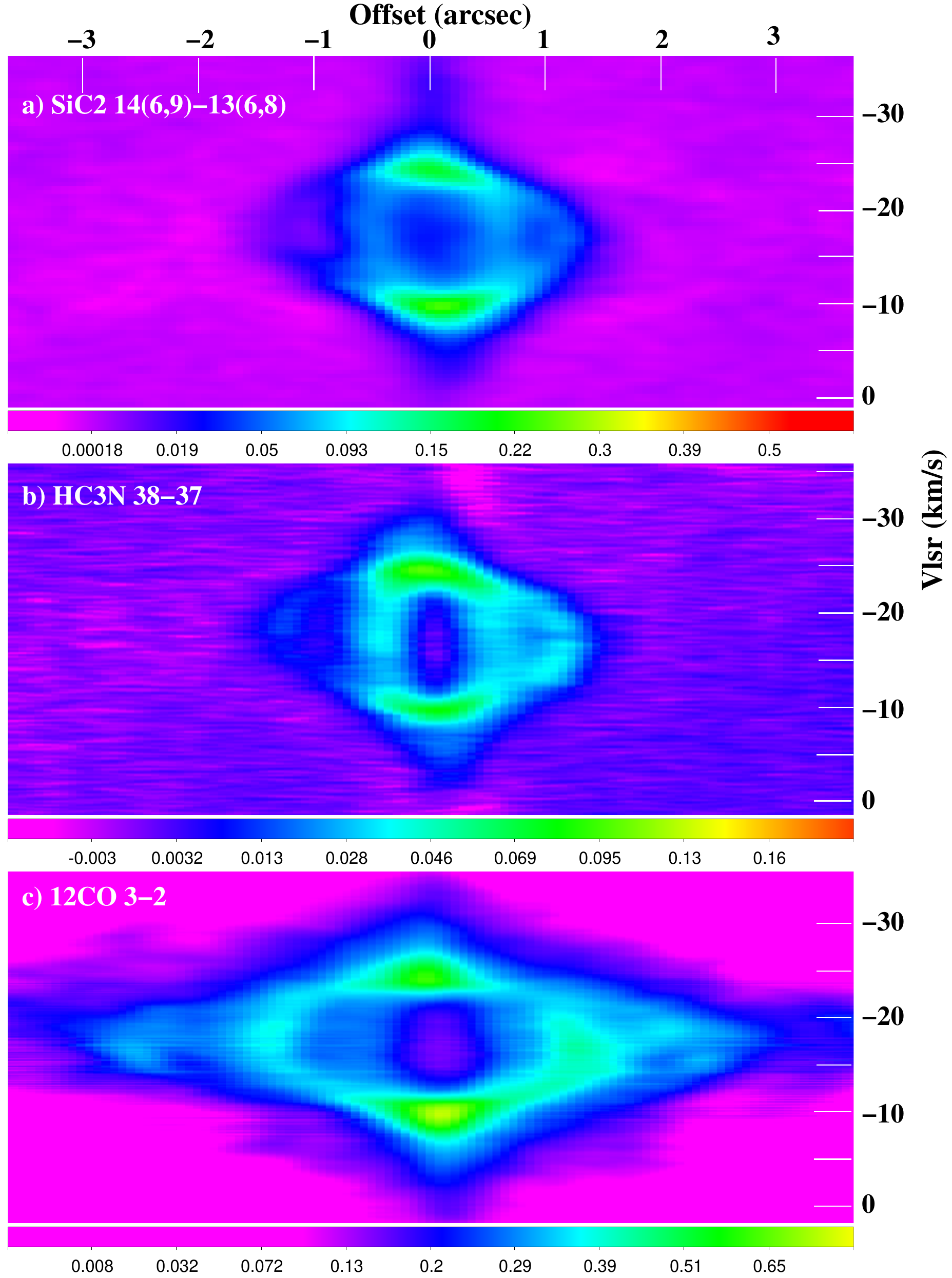}
\vspace{0.2cm}
\caption{Position-velocity intensity cuts taken along the major axis of the DUDE, for (a) \sicdo, (b) \hctresnhi, and (c)\codos~\jtres. The width of the cut is $3\farcs0$.  The intensity scale at the bottom of each panel is in Jy\,\kms\,beam$^{-1}$.
}
\label{sic2-hc3n-r0}
\end{figure}


\clearpage
\begin{thebibliography}{}

\bibitem[Akashi et al.(2015)]{2015MNRAS.453.2115A} Akashi, M., Sabach, E., Yogev, O., et al.\ 2015, \mnras, 453, 2115




\bibitem[Balick et al.(2014)]{2014ApJ...795...83B} Balick, B., Riera, A., Raga, A., et al.\ 2014, \apj, 795, 83

\bibitem[Balick \& Frank(2002)]{2002ARA&A..40..439B} Balick, B., \& Frank, A.\ 2002, \araa, 40, 439

\bibitem[Barnbaum et al.(1995)]{1995ApJ...450..862B} Barnbaum, C., Morris, M., \& Kahane, C.\ 1995, \apj, 450, 862
  


\bibitem[Brunner et al.(2019)]{2019A&A...621A..50B} Brunner, M., Mecina, M., Maercker, M., et al.\ 2019, \aap, 621, A50

\bibitem[Bujarrabal et al.(2013)]{2013A&A...557A.104B} Bujarrabal, V., Alcolea, J., Van Winckel, H., et al.\ 2013, \aap, 557, A104

\bibitem[Bujarrabal et al.(2001)]{2001A&A...377..868B} Bujarrabal, V., Castro-Carrizo, A., Alcolea, J., et al.\ 2001, \aap, 377, 868

\bibitem[Chiu et al.(2006)]{2006ApJ...645..605C} Chiu, P.-J., Hoang, C.-T., Dinh-V-Trung, et al.\ 2006, \apj, 645, 605

\bibitem[Corradi \& Schwarz(1993)]{1993A&A...268..714C} Corradi, R.~L.~M. \& Schwarz, H.~E.\ 1993, \aap, 268, 714

\bibitem[de Ruyter et al.(2006)]{2006A&A...448..641D} de Ruyter, S., van Winckel, H., Maas, T., et al.\ 2006, \aap, 448, 641

\bibitem[Dermine et al.(2013)]{dermine_et_al_2013} Dermine, T., Izzard, R.G., Jorissen, A., et al. 2013, \aap, 551, A50

\bibitem[Doan et al.(2020)]{2020A&A...633A..13D} Doan, L., Ramstedt, S., Vlemmings, W.~H.~T., et al.\ 2020, \aap, 633, A13

\bibitem[Draine(2006)]{2006ApJ...636.1114D} Draine, B.~T.\ 2006, \apj, 636, 1114

\bibitem[Draine \& Hensley(2012)]{2012ApJ...757..103D} Draine, B.~T., \& Hensley, B.\ 2012, \apj, 757, 103


\bibitem[Gielen et al.(2011)]{2011A&A...533A..99G} Gielen, C., Bouwman, J., van Winckel, H., et al.\ 2011, \aap, 533, A99

\bibitem[Gromadzki \& Miko{\l}ajewska(2009)]{2009A&A...495..931G} Gromadzki, M. \& Miko{\l}ajewska, J.\ 2009, \aap, 495, 931

\bibitem[Herman et al.(1986)]{1986A&A...167..247H} Herman, J., Burger, J.~H., \& Penninx, W.~H.\ 1986, \aap, 167, 247

\bibitem[Hirano et al.(2004)]{2004ApJ...616L..43H} Hirano, N., Shinnaga, H., Dinh-V-Trung, et al.\ 2004, \apjl, 616, L43

\bibitem[Homan et al.(2020)]{2020A&A...644A..61H} Homan, W., Montarg{\`e}s, M., Pimpanuwat, B., et al.\ 2020, \aap, 644, A61

\bibitem[Huang et al.(2016)]{2016ApJ...820..134H} Huang, P.-S., Lee, C.-F., Moraghan, A., et al.\ 2016, \apj, 820, 134

\bibitem[Huang et al.(2020)]{2020ApJ...889...85H} Huang, P.-S., Lee, C.-F., \& Sahai, R.\ 2020, \apj, 889, 85

\bibitem[Jura et al.(1988)]{1988A&A...201...80J} Jura, M., Kahane, C., \& Omont, A.\ 1988, \aap, 201, 80

\bibitem[Jura et al.(2001)]{2001ApJ...550L..71J} Jura, M., Webb, R.~A., \& Kahane, C.\ 2001, \apjl, 550, L71

\bibitem[Kahane et al.(1996)]{1996A&A...314..871K} Kahane, C., Audinos, P., Barnbaum, C., et al.\ 1996, \aap, 314, 871

\bibitem[Knapp et al.(1999)]{1999A&A...351...97K} Knapp, G.~R., Dobrovolsky, S.~I., Ivezi{\'c}, Z., et al.\ 1999, \aap, 351, 97

\bibitem[Knapp et al.(1997)]{1997A&A...326..318K} Knapp, G.~R., Jorissen, A., \& Young, K.\ 1997, \aap, 326, 318

\bibitem[Lagadec et al.(2005)]{2005A&A...433..553L} Lagadec, E., M{\'e}karnia, D., de Freitas Pacheco, J.~A., et al.\ 2005, \aap, 433, 553

\bibitem[Leal-Ferreira et al.(2013)]{2013A&A...554A.134L} Leal-Ferreira, M.~L., Vlemmings, W.~H.~T., Kemball, A., \& Amiri, N.\ 2013, \aap, 554, A134 


\bibitem[Liimets et al.(2018)]{2018A&A...612A.118L} Liimets, T., Corradi, R.~L.~M., Jones, D., et al.\ 2018, \aap, 612, A118

\bibitem[Lindegren et al.(2018)]{2018A&A...616A...2L} Lindegren, L., Hern{\'a}ndez, J., Bombrun, A., et al.\ 2018, \aap, 616, A2

\bibitem[Lindegren et al.(2021)]{2021A&A...649A...2L} Lindegren, L., Klioner, S.~A., Hern{\'a}ndez, J., et al.\ 2021, \aap, 649, A2

\bibitem[Massalkhi et al.(2019)]{2019A&A...628A..62M} Massalkhi, S., Ag{\'u}ndez, M., \& Cernicharo, J.\ 2019, \aap, 628, A62

\bibitem[McDonald et al.(2012)]{2012MNRAS.427..343M} McDonald, I., Zijlstra, A.~A., \& Boyer, M.~L.\ 2012, \mnras, 427, 343

\bibitem[Milam et al.(2009)]{2009ApJ...690..837M} Milam, S.~N., Woolf, N.~J., \& Ziurys, L.~M.\ 2009, \apj, 690, 837

\bibitem[Miszalski et al.(2018)]{2018PASA...35...27M} Miszalski, B., Manick, R., Miko{\l}ajewska, J., et al.\ 2018, \pasa, 35, e027

\bibitem[Morris(1987)]{1987PASP...99.1115M}Morris, M.\ 1987, \pasp, 99, 1115

\bibitem[Oomen et al.(2018)]{oomen_et_al_2018} Oomen, G.-M., Van Winckel, H., Pols, O., et al. 2018, \aap, 620, A85

\bibitem[O'Connor et al.(2000)]{2000ApJ...531..336O} O'Connor, J.~A., Redman, M.~P., Holloway, A.~J., et al.\ 2000, \apj, 531, 336


\bibitem[Rafikov(2016)]{2016ApJ...830....8R} Rafikov, R.~R.\ 2016, \apj, 830, 8

\bibitem[Reid \& Menten(1997)]{1997ApJ...476..327R} Reid, M.~J. \& Menten, K.~M.\ 1997, \apj, 476, 327

\bibitem[Sahai et al.(1999)]{1999AJ....118..468S} Sahai, R., Dayal, A., Watson, A.~M., et al.\ 1999, \aj, 118, 468

\bibitem[Sahai et al.(2008)]{2008ApJ...689.1274S} Sahai, R., Findeisen, K., Gil de Paz, A., \& S{\'a}nchez Contreras, C.\ 2008, \apj, 689, 1274-1278 

\bibitem[Sahai et al.(2003)]{2003Natur.426..261S} Sahai, R., Morris, M., Knapp, G.~R., Young, K., \& Barnbaum, C.\ 2003, \nat, 426, 261 

\bibitem[Sahai et al.(2011)]{2011AJ....141..134S} Sahai, R., Morris, M.~R., \& Villar, G.~G.\ 2011, \aj, 141, 134



\bibitem[Sahai et al.(2016)]{2016ApJ...827...92S} Sahai, R., Scibelli, S., \& Morris, M.~R.\ 2016, \apj, 827, 92

\bibitem[Sahai et al.(2009)]{2009ApJ...699.1015S} Sahai, R., Sugerman, B.~E.~K., \& Hinkle, K.\ 2009, \apj, 699, 1015 


\bibitem[Sahai et al.(2007)]{2007ApJ...658..410S} Sahai, R., S{\'a}nchez Contreras, C., Morris, M., \& Claussen, M.\ 2007, \apj, 658, 410 

\bibitem[Sahai \& Trauger(1998)]{1998AJ....116.1357S} Sahai, R., \& Trauger, J.~T.\ 1998, \aj, 116, 1357

\bibitem[Sahai et al.(2017)]{2017ApJ...841..110S} Sahai, R., Vlemmings, W.~H.~T., \& Nyman, L.-{\r{A}}.\ 2017, \apj, 841, 110

\bibitem[Sahai et al.(2006)]{2006ApJ...653.1241S} Sahai, R., Young, K., Patel, N.~A., et al.\ 2006, \apj, 653, 1241

\bibitem[Sahai \& Wannier(1988)]{1988A&A...201L...9S} Sahai, R., \& Wannier, P.~G.\ 1988, \aap, 201, L9 


\bibitem[Salas et al.(2019)]{2019MNRAS.487.3029S} Salas, J.~M., Naoz, S., Morris, M.~R., et al.\ 2019, \mnras, 487, 3029

\bibitem[Scibelli et al.(2019)]{2019ApJ...870..117S} Scibelli, S., Sahai, R., \& Morris, M.~R.\ 2019, \apj, 870, 117

\bibitem[Shirley(2015)]{2015PASP..127..299S} Shirley, Y.~L.\ 2015, \pasp, 127, 299


\bibitem[Soker(1998)]{1998MNRAS.299.1242S} Soker, N.\ 1998, \mnras, 299, 1242

\bibitem[Soker(2000)]{2000ApJ...540..436S} Soker, N.\ 2000, \apj, 540, 436

\bibitem[Soker(2015)]{2015ApJ...800..114S} Soker, N.\ 2015, \apj, 800, 114

\bibitem[Soker \& Rappaport(2000)]{2000ApJ...538..241S} Soker, N. \& Rappaport, S.\ 2000, \apj, 538, 241

\bibitem[Soker \& Zoabi(2002)]{2002MNRAS.329..204S} Soker, N. \& Zoabi, E.\ 2002, \mnras, 329, 204





\bibitem[S{\'a}nchez Contreras \& Sahai(2012)]{2012ApJS..203...16S} S{\'a}nchez Contreras, C. \& Sahai, R.\ 2012, \apjs, 203, 16


\bibitem[Trammell \& Goodrich(2002)]{2002ApJ...579..688T} Trammell, S.~R. \& Goodrich, R.~W.\ 2002, \apj, 579, 688

\bibitem[Van der Tak et al (2007)]{2007A&A} Van der Tak, F.F.S., Black, J.H., Schöier, F.L., Jansen, D.J., van Dishoeck, E.F., 2007, A\&A 468, 627


\bibitem[Velusamy et al.(2011)]{2011ApJ...741...60V} Velusamy, T., Langer, W.~D., Kumar, M.~S.~N., et al.\ 2011, \apj, 741, 60



\bibitem[Woodruff et al.(2009)]{2009ApJ...691.1328W} Woodruff, H.~C., Ireland, M.~J., Tuthill, P.~G., et al.\ 2009, \apj, 691, 1328


\end{thebibliography}
\end{document}